\documentclass[a4paper,12pt]{article}
\usepackage[utf8]{inputenc}
\usepackage[english]{babel}
\usepackage{amsmath, amssymb,graphicx}
\usepackage{braket}
\usepackage{caption}
\usepackage{subcaption}
\usepackage{ulem}
\usepackage{multicol}

\pdfoutput=1
\usepackage{jheppub}
\newcommand{\be}{\begin{equation}}
\newcommand{\ee}{\end{equation}}
\newcommand{\bea}{\begin{eqnarray}}
\newcommand{\eea}{\end{eqnarray}}

\definecolor{darkraspberry}{rgb}{0.53, 0.15, 0.34}

\definecolor{darkblue}{rgb}{0., 0, 1}

\definecolor{dgreen}{rgb}{0.,0.6,0.}

\newcommand{\cS}{\mathcal{S}}
\newcommand{\cA}{\mathcal{A}}

\newcommand{\fb}{\mathfrak{b}}

\title{Holographic Anisotropic Model for Light Quarks with
  Confinement-Deconfinement Phase Transition}

\author{Irina Ya. Aref'eva$^a$, Kristina Rannu$^b$ and Pavel Slepov$^a$}

\affiliation{$^a$Steklov Mathematical Institute, Russian Academy of
  Sciences,\\ Gubkina str. 8, 119991, Moscow, Russia\\
$^b$Peoples Friendship University of Russia,\\ Miklukho-Maklaya
str. 6, 117198, Moscow, Russia}

\emailAdd{arefeva@mi-ras.ru}
\emailAdd{rannu-ka@rudn.ru}
\emailAdd{slepov@mi-ras.ru}

\abstract{We present a five-dimensional anisotropic holographic model
  for light quarks supported by Einstein-dilaton-two-Maxwell
  action. This model generalizing isotropic holographic model with
  light quarks is characterized by a Van der Waals-like phase
  transition between small and large black holes. We compare the
  location of the phase transition for Wilson loops with the positions
  of the phase transition related to the background instability and
  describe the QCD phase diagram in the thermodynamic plane --
  temperature $T$ and chemical potential $\mu$. The Cornell potential
  behavior in this anisotropic model is also studied. The asymptotics
  of the Cornell potential at large  distances strongly depend on the
  parameter of anisotropy and orientation. There is also a nontrivial
  dependence of the Cornell potential on the boundary conditions of
  the dilaton field and parameter of anisotropy. With the help of the
  boundary conditions for the dilaton field one fits the results of
  the lattice calculations for the string tension as a function of
  temperature in isotropic case and then generalize to the anisotropic
  one.}

\keywords{AdS/QCD, holography, phase transition, Wilson loops, light
  quarks}

\begin{document}

\maketitle

\newpage

\section{Introduction}

Experimental research of the phase transitions structure for the quark
matter is one of important problems of modern collider facilities
\cite{1710.09425}. The phase diagram in the thermodynamical plane --
temperature $T$ and chemical potential $\mu$ -- has been only studied
experimentally for small $\mu$ and large $T$ values (RHIC, LHC) on the
one hand and for finite $\mu$ and small $T$ values (SPS) on the
other. The study of the phase diagram in between these two selected
cases is one of the main goals of FAIR and NICA projects.

According to results of heavy ion collision (HIC) experiments at RHIC
and LHC the quark gluon-plasma (QGP) should exist at large
temperatures and densities. Temperature of the 
confinement/deconfinement phase transition most likely depends on
chemical potential, i.e. the phase transition can be displayed on the
$(\mu,T)$-plane. Under some circumstances QGP behaves like an almost
viscous liquid with an initial spatial anisotropy
\cite{Strickland:2013uga}, therefore phase transition in anisotropic
QCD should be considered.

Perturbative methods are not suited for the QGP studies. Several
calculations have been performed within the lattice approach
\cite{Philipsen:2010, Boyda:2017dyo, Philipsen:2019rjq, Ratti}, but
lattice calculations cannot provide full phase transition picture in
$(\mu,T)$-plane because of so-called sign problems. It is holographic
duality \cite{Solana, IA, DeWolf} that opens up an alternative
approach to the QCD phase transitions' researches. Among other things,
this approach has a natural framework to deal with spatial anisotropy
\cite{Mateos:2011ix,Mateos:2011tv,Rebhan:2011vd,Giataganas:2012, Cheng:2014sxa,Cheng:2014qia,Jain:2014vka,AG, AGG, Giataganas:2017koz,ARS-2019qfthep}. Note that anisotropic
lattice calculations have been performed in \cite{Forcrand2018}.

Holographical QCD (HQCD) as a phenomenological model has to describe
QCD at all energy scales. That means to reproduce the usual QCD
results obtained by perturbative theory at short distances and Lattice
QCD results at large distances (confinement etc.). The other purpose
of HQCD concerns intermediate energy scales. It has to give
theoretical results, that are in agreement with experiments, as well
as to predict new results  especially in extremal conditions such as
hight density or large chemical potential.

HQCD is formulated as 5-dimensional theory, where the 5-dim background
usually is a deformed version of a 5-dimensional Schwarzschild-AdS
or Reissner-Nordstr\"om  Schwarzschild-AdS space time. A scalar
(dilaton) field's dynamics describes the running coupling in 4-dim 
quantum theory, and the 5-th coordinate playes a role of an energy
scale.

Different isotropic holographic QCD models are distinguished by the
choice of the warp factor \cite{Kiritsis, 1301.0385, yang2015,
  Yang-2017, Dudal:2017max, Dudal:2018ztm, Mahapatra:2019uql,
  Ebrahim:2020qif, He:2020fdi, Zhou:2020ssi}. Models with different
warp factors describe different phenomenological models. In the
isotropic case there are special warp factors that describe the QCD
for light or heavy quarks \cite{yang2015, Yang-2017}. It is
interesting to consider a more complicated version of the warp factor
that describes model for both light and heavy quarks. It happens that
to describe chiral phase transition one has to modify holographic
model essentially by introducing additional scalar fields
\cite{Fang:2015ytf, Ballon-Bayona:2020qpq}.

\begin{figure}[t!]
  \centering
  \includegraphics[scale=0.57]{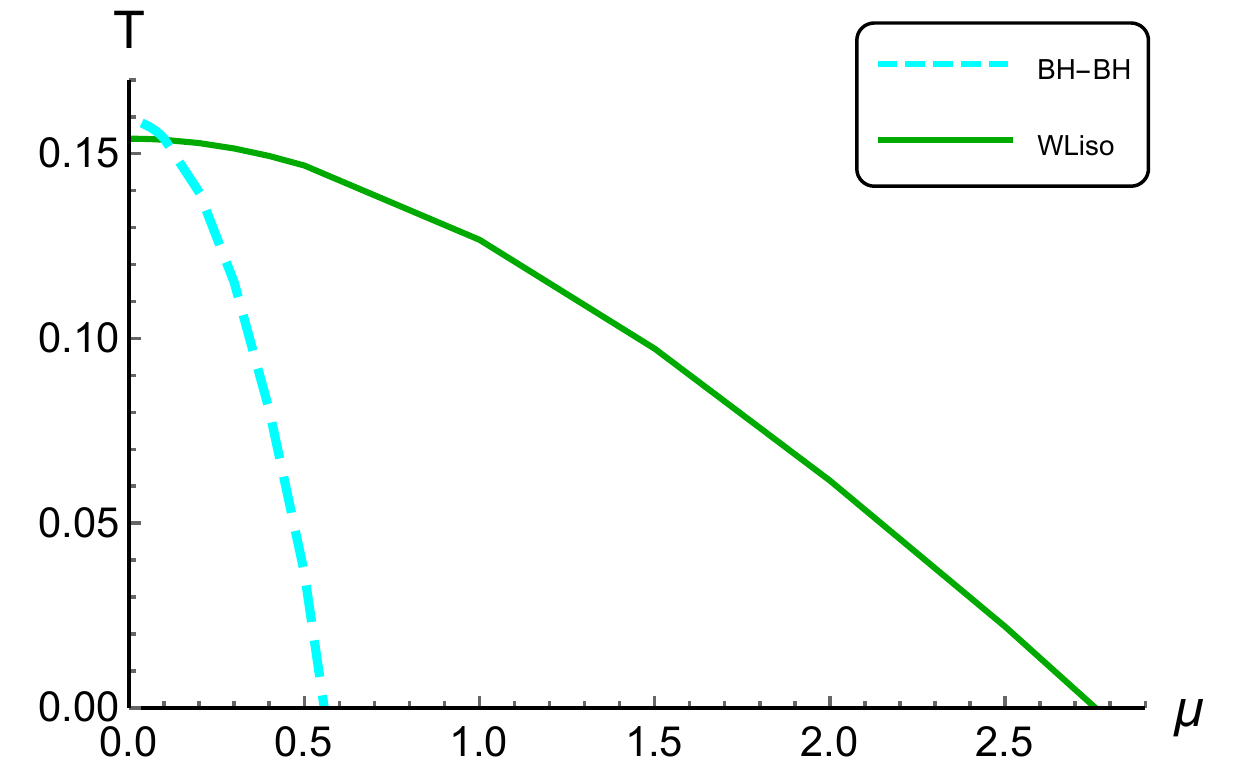} \quad
  \includegraphics[scale=0.57]{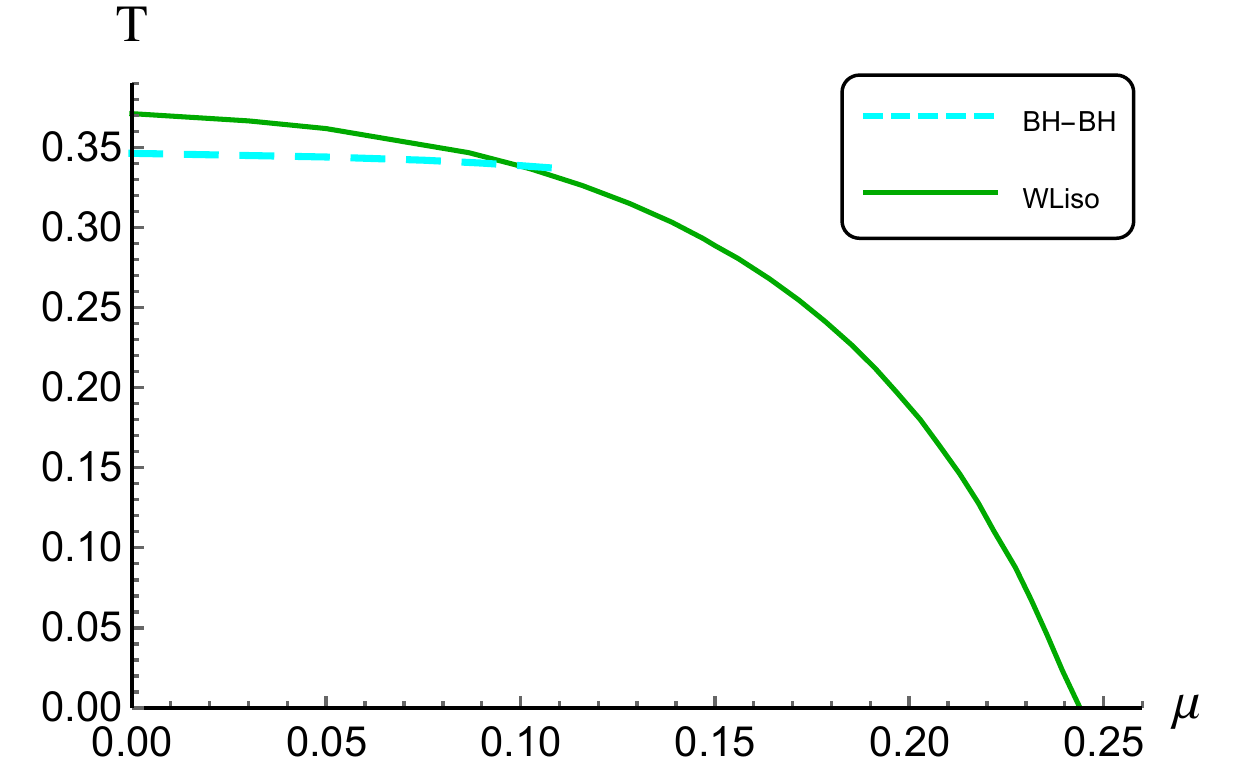} \\
  A \hspace{220pt} B
  \caption{Holographic QCD phase diagrams for light (A) and heavy (B)
    quarks in the isotropic case \cite{yang2015, Yang-2017}. Here
   first-order Hawking-Page-like phase transitions (BH-BH) are
   indicated by dashed lines. Wilson loop (crossover) phase
   transitions (WLiso) are indicated by solid lines.}
  \label{Hybrid1} 
\end{figure}

\begin{figure}[t!]
  \centering
  \includegraphics[scale=0.57]{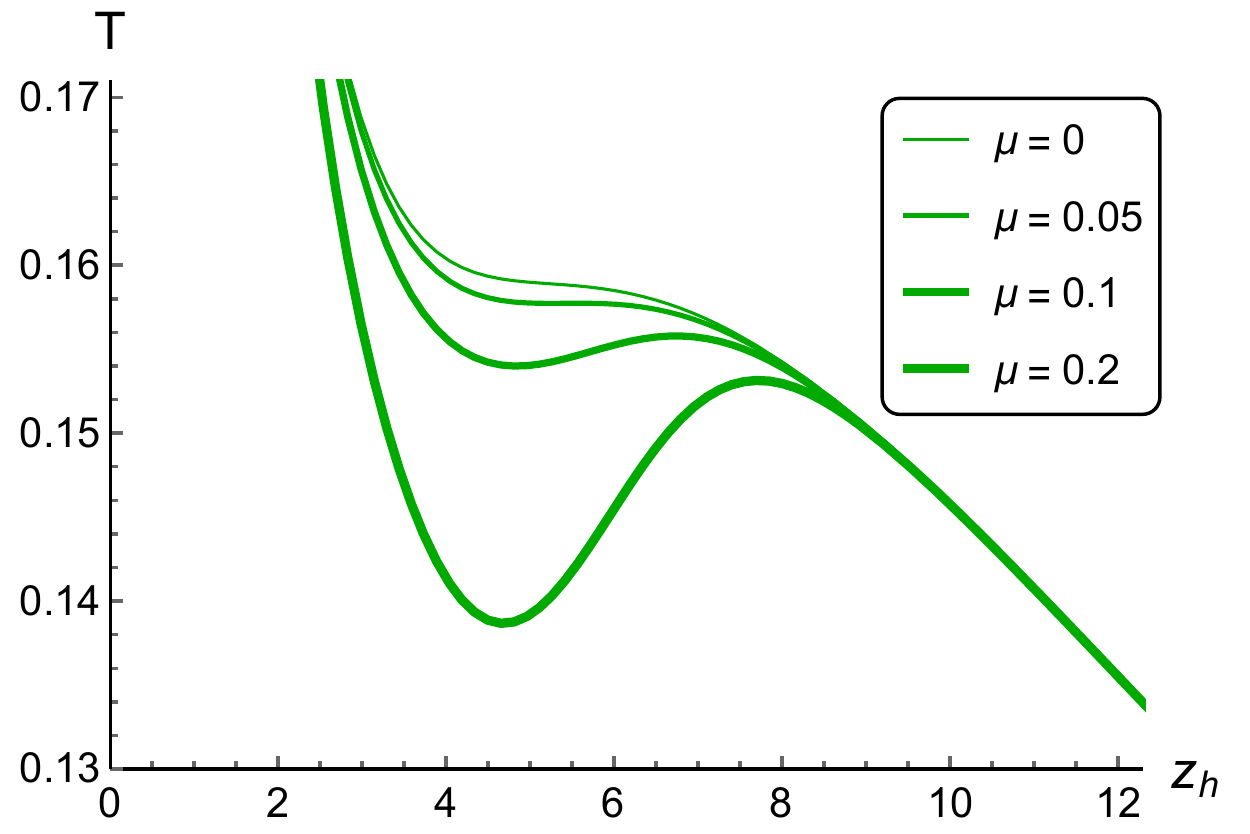} \quad
  \includegraphics[scale=0.57]{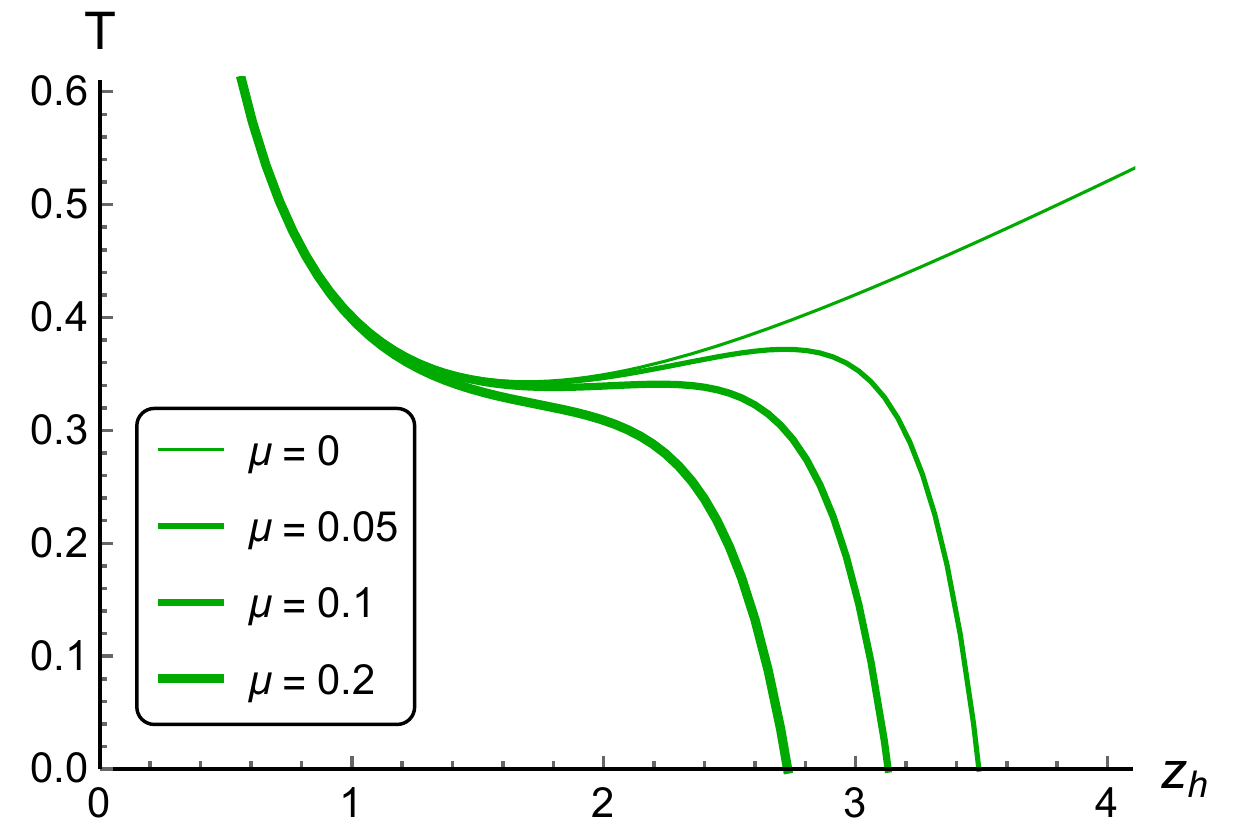} \\
  A \hspace{220pt} B
  \caption{Temperature as function of horizon for different $\mu$ in
    isotropic model for light~(A) and heavy (B) quarks.}
  \label{t-zh-h-l}
\end{figure}

Qualitative look of the confinement/deconfinement phase transition on
$(\mu,T)$-plane for different quarks' mass in isotropic media is
displayed in Fig.\ref{Hybrid1}. Phase transition of the light quarks
is supposed to have a crossover for small chemical potentials and a
first-order phase transition for large chemical potentials
(Fig.\ref{Hybrid1}.A). This picture was obtained in
\cite{Yang-2017}. Phase transition of the heavy quarks, on the 
opposite, has a first-order phase transition for small chemical
potentials and a crossover for large chemical potentials
(Fig.\ref{Hybrid1}.B). Difference in the behavior of holographic
models for heavy and light quarks is caused by difference in
dependences of temperature on the horizon size in these models
(Fig.\ref{t-zh-h-l}). Here and below chemical potential and
temperature are taken in GeV units and $z_h$ in GeV$^{-1}$.

There are several reasons to consider anisotropic versions of the
holographic models mentioned above: to reproduce the experimental data
for the energy dependence of the total multiplicity \cite{AG}, to
describe inverse magnetic catalysis \cite{Bohra:2019ebj,
  Gursoy:2018ydr, Gursoy:2020, He:2020fdi,Ballon-Bayona:2020xtf} or to
take into account anisotropic geometry of colliding ions.

In \cite{AR-2018} the anisotropic holographic model for heavy quarks
was studied. A peculiar feature of the model is the relation between
anisotropy of the background and anisotropy of the colliding heavy
ions geometry. In \cite{AR-2018} anisotropy is described by a special
parameter $\nu$, and it's value of about $4.5$ gives the dependence of
the produced entropy on energy in accordance with the experimental
data for the energy dependence of the total multiplicity of particles
produced in heavy ion collisions \cite{Alice}. Isotropic holographic
models had not been able to recover the experimental multiplicity
dependence on energy (\cite{AG} and refs therein). As shown in
\cite{ARS-2019plb}, the solution \cite{AR-2018} describes smeared
confinement/deconfinement phase transitions. This model also indicates
the relations of the fluctuations of the multiplicity, i.e. the
entanglement entropy, with the phase transitions \cite{APS}.

The purpose of this paper is to perform similar investigation for the
model describing the light quarks. As in case of the heavy quarks
\cite{AR-2018}, the anisotropic model for light quarks considered in
this work is described by the parameter $\nu$
again. 

We also study the Cornell potential behavior in this anisotropic model
and discuss a nontrivial dependence of the Cornell potential on
boundary conditions of the dilaton field and parameter of
anisotropy. Particular forms of $\sigma(T)$-function and their
connection to the boundary condition for the scalar field are
investigated. As a result we suggest a boundary definition that allows
to fit string tension behavior from Lattice QCD \cite{Bali2001}.

Holographic calculations for heavy and light quarks models are
essentially different since solutions for the heavy quarks model can
be expressed explicitly, unlike the model of light quarks, where the
solution is presented in quadratures. Therefore the  generalization to
anisotropic model  of the light quarks is a more complicated task.

The paper is organized as follows. In Sect.~\ref{Model} we present the
action and the ansatz that solves the EOM for the anisotropic model
with symmetry in transversal directions. In Sect.~\ref{Solution} we
briefly describe the solutions of the considered model. The
thermodynamics of the background is described in
Sect.~\ref{Thermodynamics}. In Sect.~\ref{Wilson loop} we compare the 
position of the phase transition for Wilson loops with the positions
of phase transitions related to the background instability in the
thermodynamical plane -- temperature $T$ and chemical potential
$\mu$. We end the paper with the discussion of future directions of
research on the subject.

\section{Model}  

\subsection{Metric and EOM} \label{Model}

We take the action in Einstein frame
\begin{gather}
  \cS = \cfrac{1}{16\pi G_5} \int d^5x \ \sqrt{-g} \left[ 
    R - \cfrac{f_1(\phi)}{4} \ F_{(1)}^2 
    - \cfrac{f_2(\phi)}{4} \ F_{(2)}^2
    - \cfrac{1}{2} \ \partial_{\mu} \phi \partial^{\mu} \phi
    - V(\phi) \right], \label{eq:1.01}  \\
  \begin{split}
    F_{\mu\nu}^{(1)} = \partial_{\mu} A_{\nu} - \partial_{\nu} A_{\mu}
    \quad &\Rightarrow \quad A_{\mu}^{(1)} = A_t (z) \delta_\mu^0, \\
    F_{\mu\nu}^{(2)} = q \ dy^1 \wedge dy^2 \quad &\Rightarrow \quad
    F_{23}^{(2)} = q,
  \end{split}\label{eq:1.02} 
\end{gather}
where $\phi = \phi(z)$ is the scalar field, $f_1(\phi)$ and
$f_2(\phi)$ are the coupling functions associated with the Maxwell
fields $A_{\mu}$ and $F_{\mu\nu}^{(2)}$ correspondingly, $q$ is the
constant and $V(\phi)$ is the scalar field potential. Thus
\eqref{eq:1.01} is the same action that was used in \cite{AR-2018}.

To consider action \eqref{eq:1.01} let us take the ansatz in the
following view:
\begin{gather}
  ds^2 = \cfrac{L^2}{z^2} \ \fb(z) \left[
    - \ g(z) dt^2 + dx^2
    + \left( \cfrac{z}{L} \right)^{2-\frac{2}{\nu}} dy_1^2
    + \left( \cfrac{z}{L} \right)^{2-\frac{2}{\nu}} dy_2^2
    + \cfrac{dz^2}{g(z)} \right], \label{eq:1.20} \\
  \fb(z) = e^{2{\cA}(z)}, \quad {\cA}(z) = - \, a \ln (b z^2 +
  1), \label{eq:1.04}
\end{gather}
where $L$ is the AdS-radius, $\fb(z)$ is the warp factor, ${\cA}(z)$
is its half-power, $g(z)$ is the blackening function and $\nu$ is the
parameter of anisotropy. Following \cite{Yang-2017} we choose
\eqref{eq:1.04} for the warp factor to get the solution for the
light quarks. Therefore the EOM simplifies to:
\begin{gather}
    \phi'' + \phi' \left( \cfrac{g'}{g} + \cfrac{3 \fb'}{2 \fb} - 
      \cfrac{\nu + 2}{\nu z} \right)
    + \left( \cfrac{z}{L} \right)^2 \cfrac{\partial f_1}{\partial
      \phi} \ \cfrac{(A_t')^2}{2 \fb g}
    - \left( \cfrac{L}{z} \right)^{2-\frac{4}{\nu}} \cfrac{\partial
      f_2}{\partial \phi} \ \cfrac{q^2}{2 \fb g} 
    - \left( \cfrac{L}{z} \right)^2 \cfrac{\fb}{g} \ \cfrac{\partial
      V}{\partial \phi} = 0,
    \label{eq:1.21} \\
  A_t'' + A_t' \left( \cfrac{\fb'}{2 \fb} + \cfrac{f_1'}{f_1} +
    \cfrac{\nu - 2}{\nu z} \right) = 0, \label{eq:1.22} \\
  g'' + g' \left(\cfrac{3 \fb'}{2 \fb} - \cfrac{\nu + 2}{\nu z}
  \right)
  - \left( \cfrac{z}{L} \right)^2 \cfrac{f_1 (A_t')^2}{\fb} =
  0, \label{eq:1.23} \\
  \fb'' - \cfrac{3 (\fb')^2}{2 \fb} + \cfrac{2 \fb'}{z}
  - \cfrac{4 \fb}{3 \nu z^2} \left( 1 - \cfrac{1}{\nu} \right)
  + \cfrac{\fb \, (\phi')^2}{3} = 0, \label{eq:1.24} \\
  2 g' \ \cfrac{\nu - 1}{\nu} + 3 g \ \cfrac{\nu - 1}{\nu} \left(
    \cfrac{\fb'}{\fb} - \cfrac{4 \left( \nu + 1 \right)}{3 \nu z}
  \right)
  + \left( \cfrac{L}{z} \right)^{1-\frac{4}{\nu}} \cfrac{L \, q^2 \,
    f_2}{\fb} = 0, \label{eq:1.25} \\
  \begin{split}
    \cfrac{\fb''}{\fb} + \cfrac{(\fb')^2}{2 \fb^2}
    + \cfrac{3 \fb'}{\fb} \left( \cfrac{g'}{2 g}
      - \cfrac{\nu + 1}{\nu z} \right)
    - \cfrac{g'}{3 z g}  \left( 5 + \cfrac{4}{\nu} \right)
    &+ \cfrac{8}{3 z^2} \left( 1 + \cfrac{3}{2 \nu} + \cfrac{1}{2
        \nu^2} \right) + \\
    &+ \cfrac{g''}{3 g}
    + \cfrac{2}{3} \left( \cfrac{L}{z} \right)^2 \cfrac{\fb V}{g} =
    0.
  \end{split} \label{eq:1.26}
\end{gather}

Excluding anisotropy and normalizing to the AdS-radius, i.e. putting
$L = 1$, $\nu = 1$ and $f_2 = 0$ into
\eqref{eq:1.21}--\eqref{eq:1.26}, one can get the expressions that 
fully coincide with the EOM (2.12)--(2.16) from \cite{Yang-2017}. We
consider the general form of the boundary conditions:
\begin{gather}
  A_t(0) = \mu, \quad A_t(z_h) = 0, \label{eq:1.17} \\
  g(0) = 1, \quad g(z_h) = 0, \label{eq:1.18} \\
   \phi(z_0) = 0, \label{eq:1.19}
\end{gather}
where $z_h$ is a size of horizon and $z_0$ is the boundary condition
point located between 0 and $z_h$ $(0\leq z_0\leq z_h)$. The case of
$z_0=0$ corresponds to \cite{Yang-2017} and $z_0=z_h$ corresponds to
\cite{AR-2018}. The choice of the boundary condition for the scalar
field  discussed in details in Sect.\ref{scalar}.

In this paper we assume $a = 4.046$, $b = 0.01613$, $c = 0.227$ to
make our solution agree with results from \cite{Yang-2017} in the
isotropic case. These values are due to the mass spectrum of $\rho$
meson with its excitations and to the lattice results for the phase
transition temperature. We use the same values of $a$, $b$ and $c$ for
anisotropic case, as we still do not have anisotropic lattice data for
the spectrum.


\subsection{Solution} \label{Solution}

To solve EOM \eqref{eq:1.21}--\eqref{eq:1.26} we need to determine the
form of the coupling function $f_1$. Choosing it we base on our
previous experience in anisotropic heavy quarks model \cite{AR-2018} 
and also follow the proposition for the isotropic light quarks model
\cite{Yang-2017}, that reproduces the Regge spectrum: 
\begin{gather}
  f_1 = e^{-cz^2-{\cA}(z)} \ z^{-2+\frac{2}{\nu}}
  , \label{eq:1.27}
\end{gather}
i.e. $f_1^{iso}/f_1^{aniz}$ is the same as for ``heavy quark model''
in \cite{AR-2018}.
  
Solving \eqref{eq:1.22} with coupling function \eqref{eq:1.27} and
boundary conditions \eqref{eq:1.17} gives the same answer as in
\cite{AR-2018, yang2015, Yang-2017}:
\begin{gather}
  A_t = \mu \ \cfrac{e^{cz^2} - e^{cz_h^2}}{1 -
    e^{cz_h^2}}. \label{eq:1.28}
\end{gather}
Note that to obtain $A_t$ given by \eqref{eq:1.28} in this work we take the
simplest form of the coupling function $f_1$. This choice isn't the
only possible one (see for example \cite{Yang-2020}), but comparison
of advantages and disadvantages of different forms of $f_1$ is not the
subject of the current discussion.

\subsubsection{Blackening function $g(z)$}

\begin{figure}[b!]
  \centering
  \includegraphics[scale=0.42]{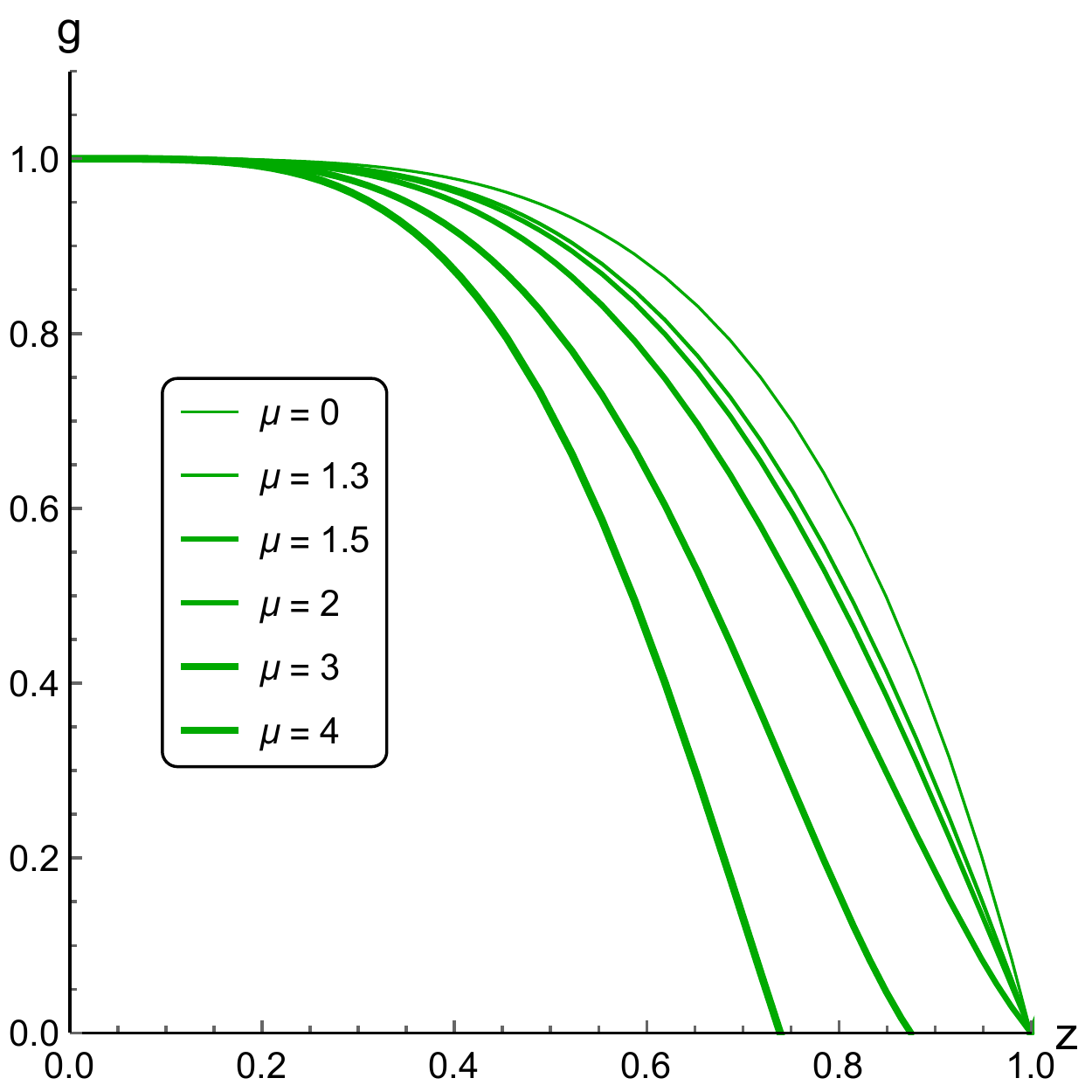} \qquad
  \includegraphics[scale=0.42]{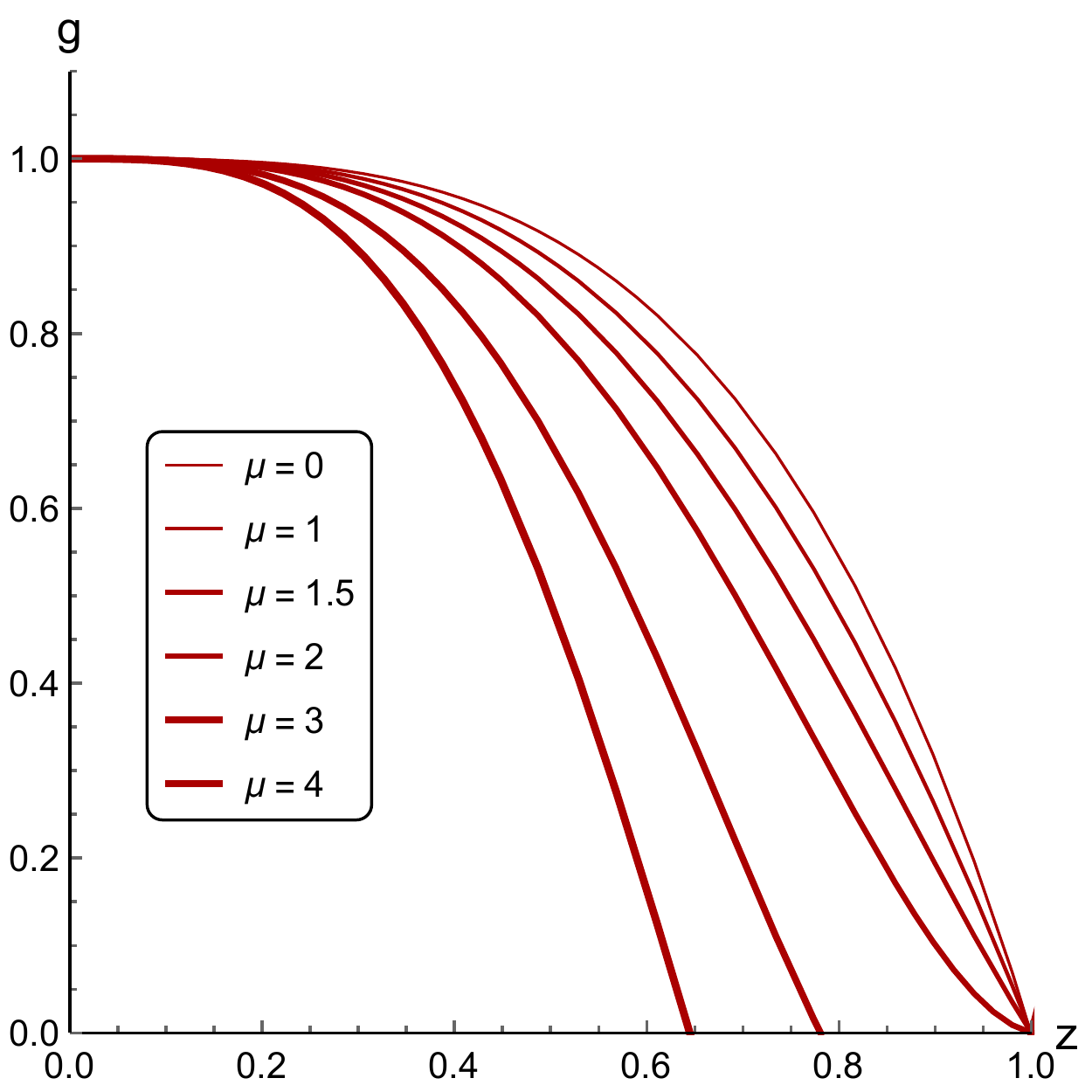} \\
  A \hspace{180pt} B \\
  \includegraphics[scale=0.42]{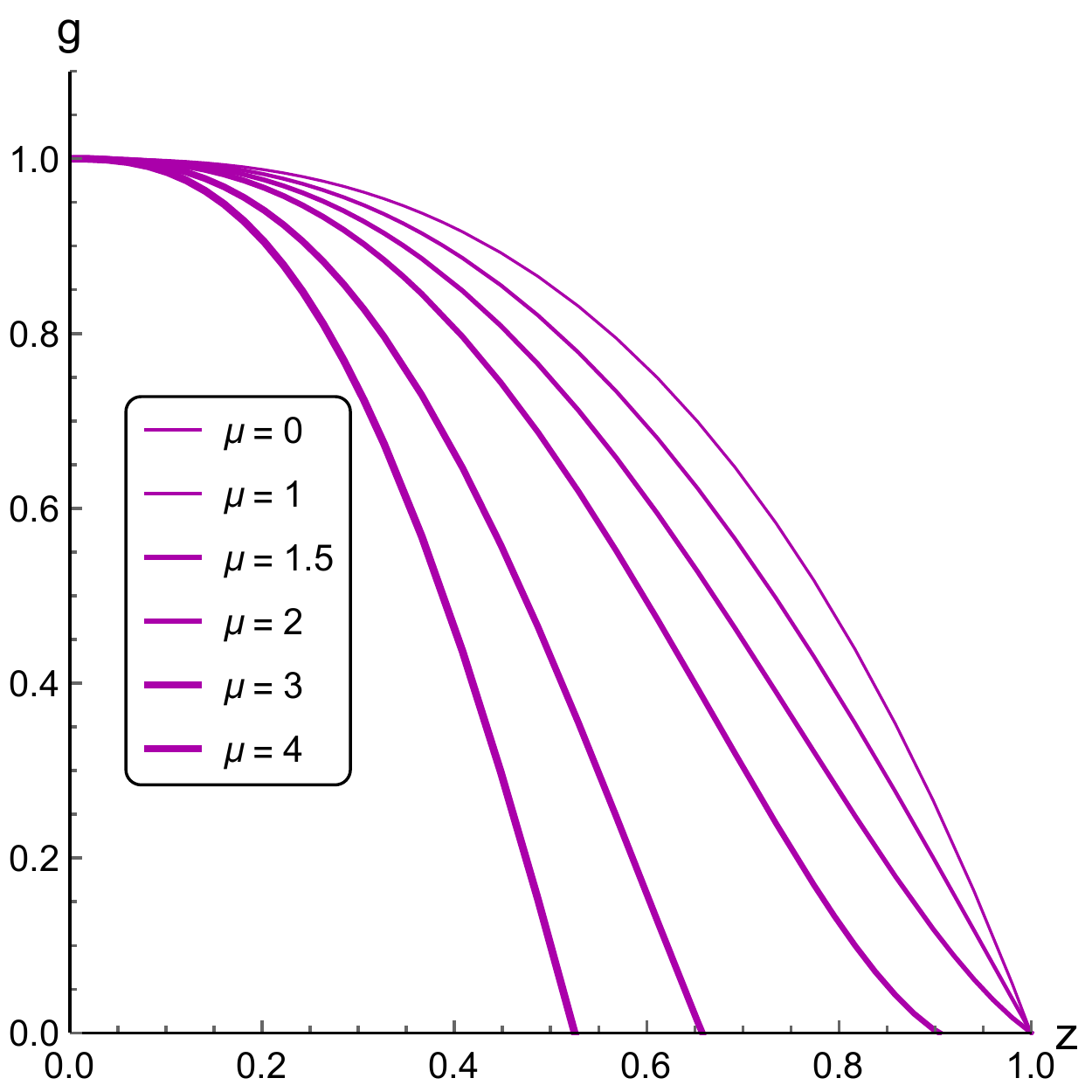} \qquad
  \includegraphics[scale=0.42]{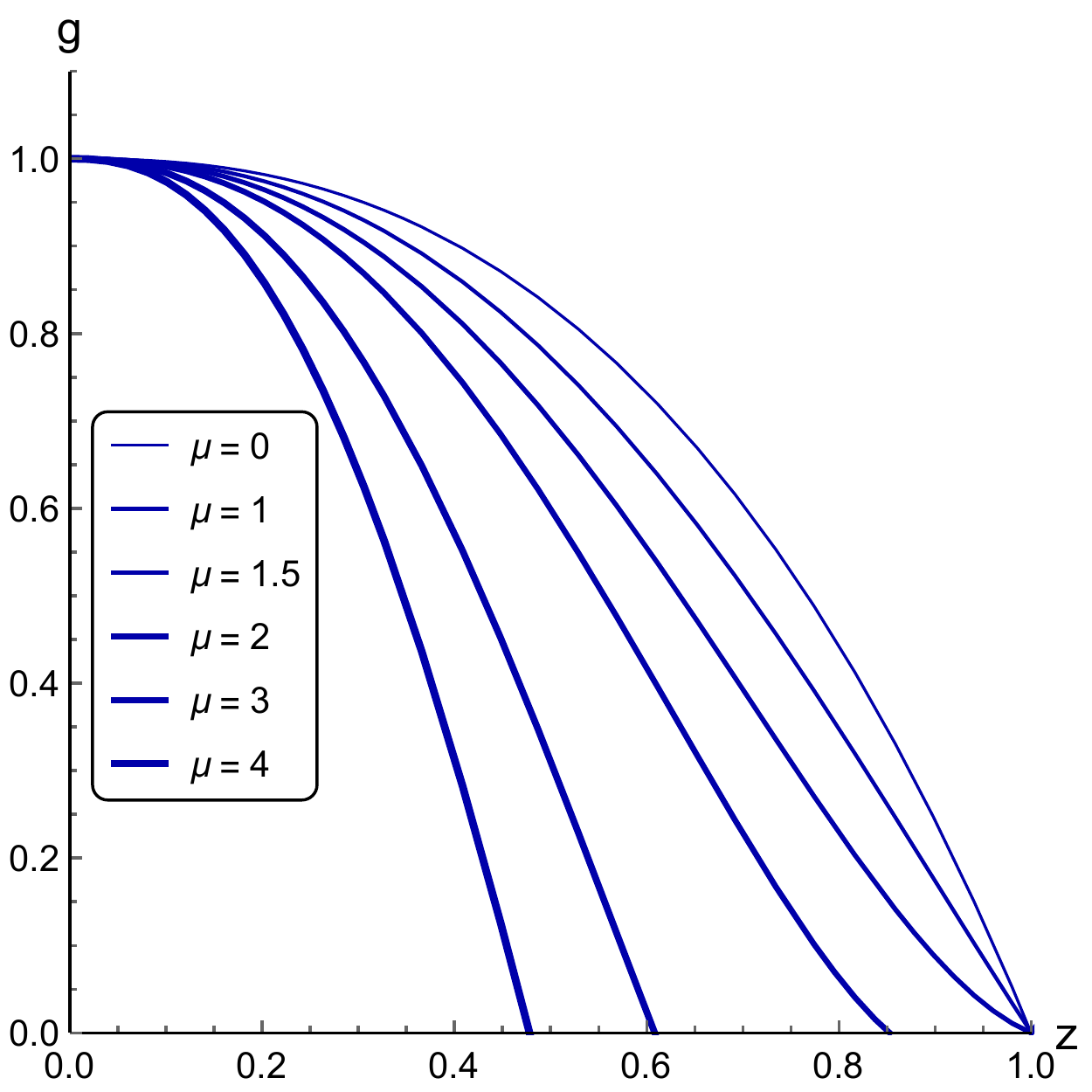} \\
  C \hspace{180pt} D
  \caption{Blackening function $g(z)$ for different $\mu$ in isotropic
    (A) and anisotropic cases for $\nu = 1.5$ (B), $\nu = 3$  (C),
    $\nu = 4.5$ (D); $a = 4.046$, $b = 0.01613$, $c = 0.227$, $z_h =
    1$.}
  \label{Fig:gznu}
\end{figure}
Solving \eqref{eq:1.23} with coupling function \eqref{eq:1.27} and
boundary conditions \eqref{eq:1.18} gives

\begin{gather}
  \begin{split}
    g &=  1 - \cfrac{
      \int_0^z \left(1 + b \, \xi^2 \right)^{3a} \xi^{1+\frac{2}{\nu}}
      \, d \xi}{
      \int_0^{z_h} \left(1 + b \, \xi^2 \right)^{3a}
      \xi^{1+\frac{2}{\nu}} \, d \xi}
    + \cfrac{2 \mu^2 c}{L^2 \left( 1 - e^{cz_h^2} \right)^2}
    \int_0^z e^{c\xi^2} \left(1 + b \, \xi^2 \right)^{3a}
    \xi^{1+\frac{2}{\nu}} \, d \xi \ \times \\
    &\times \left[ 1 -
      \cfrac{
        \int_0^z \left(1 + b \, \xi^2 \right)^{3a}
        \xi^{1+\frac{2}{\nu}} \, d \xi}{
        \int_0^{z_h} \left(1 + b \, \xi^2 \right)^{3a}
        \xi^{1+\frac{2}{\nu}} \, d \xi} \
      \cfrac{
        \int_0^{z_h} e^{c\xi^2} \left(1 + b \, \xi^2 \right)^{3a}
        \xi^{1+\frac{2}{\nu}} \, d \xi}{
        \int_0^z  e^{c\xi^2} \left(1 + b \, \xi^2 \right)^{3a}
        \xi^{1+\frac{2}{\nu}} \, d \xi} \right].
  \end{split}\label{eq:1.29}
\end{gather}

On Fig.\ref{Fig:gznu} we see $g(z)$ behavior for different chemical
potentials and anisotropy parameter values. For zero $\mu$ blackening
function monotonously decreases (Fig.\ref{Fig:gznu}.A), nonzero
chemical potentials lead to the appearence of the second horizon that
is decreasing with increasing $\mu$. For small $\mu$ the second
horizon doesn't matter, but at some moment it becomes lesser than the
fixed one and continues decreasing while increasing $\mu$. From this
moment it starts to play the main role and the fixed horizon loses
actual influence. For larger $\nu$ we also have lesser second (moving)
horizon values.

\begin{figure}[h!]
  \centering
  \includegraphics[scale=0.46]{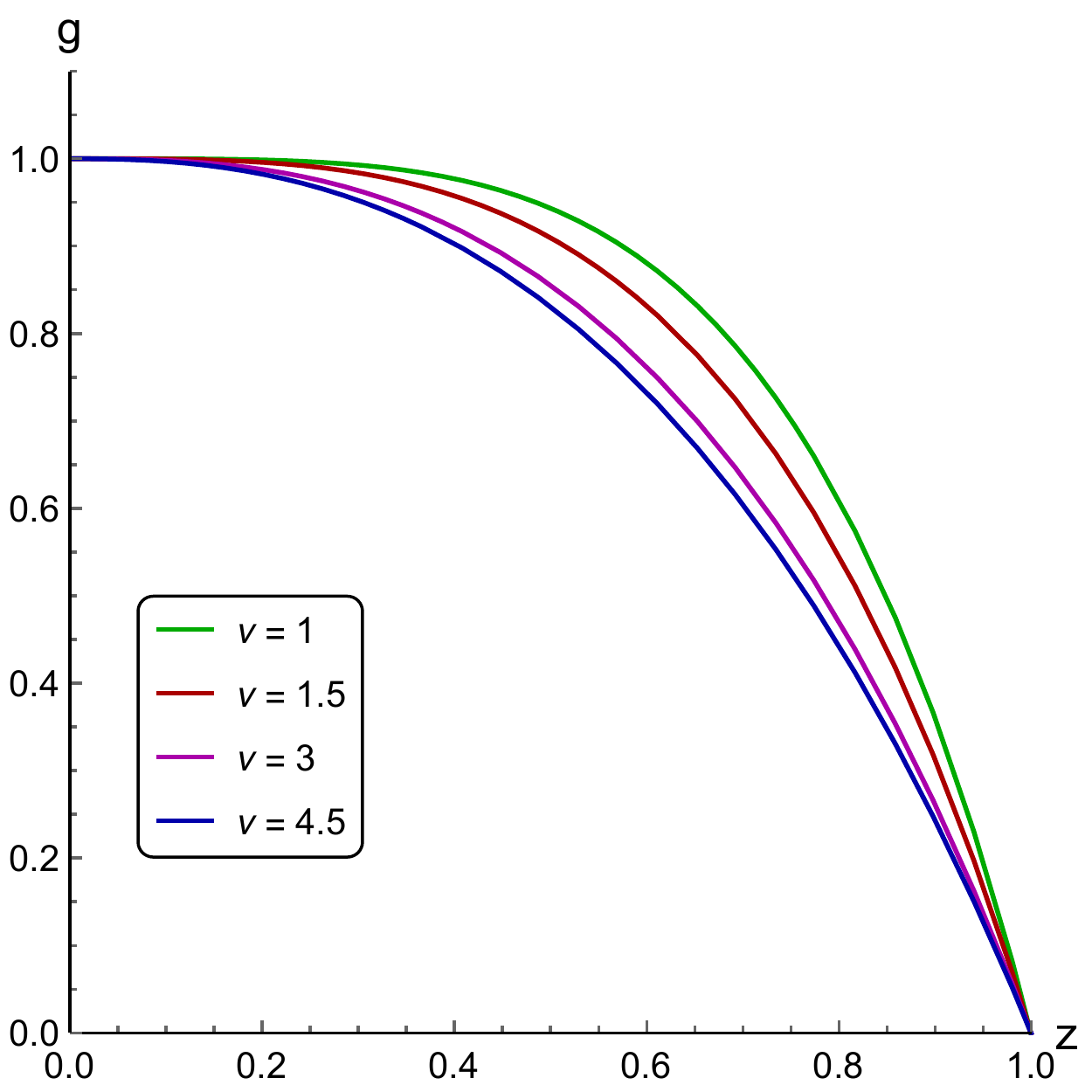} \qquad
  \includegraphics[scale=0.46]{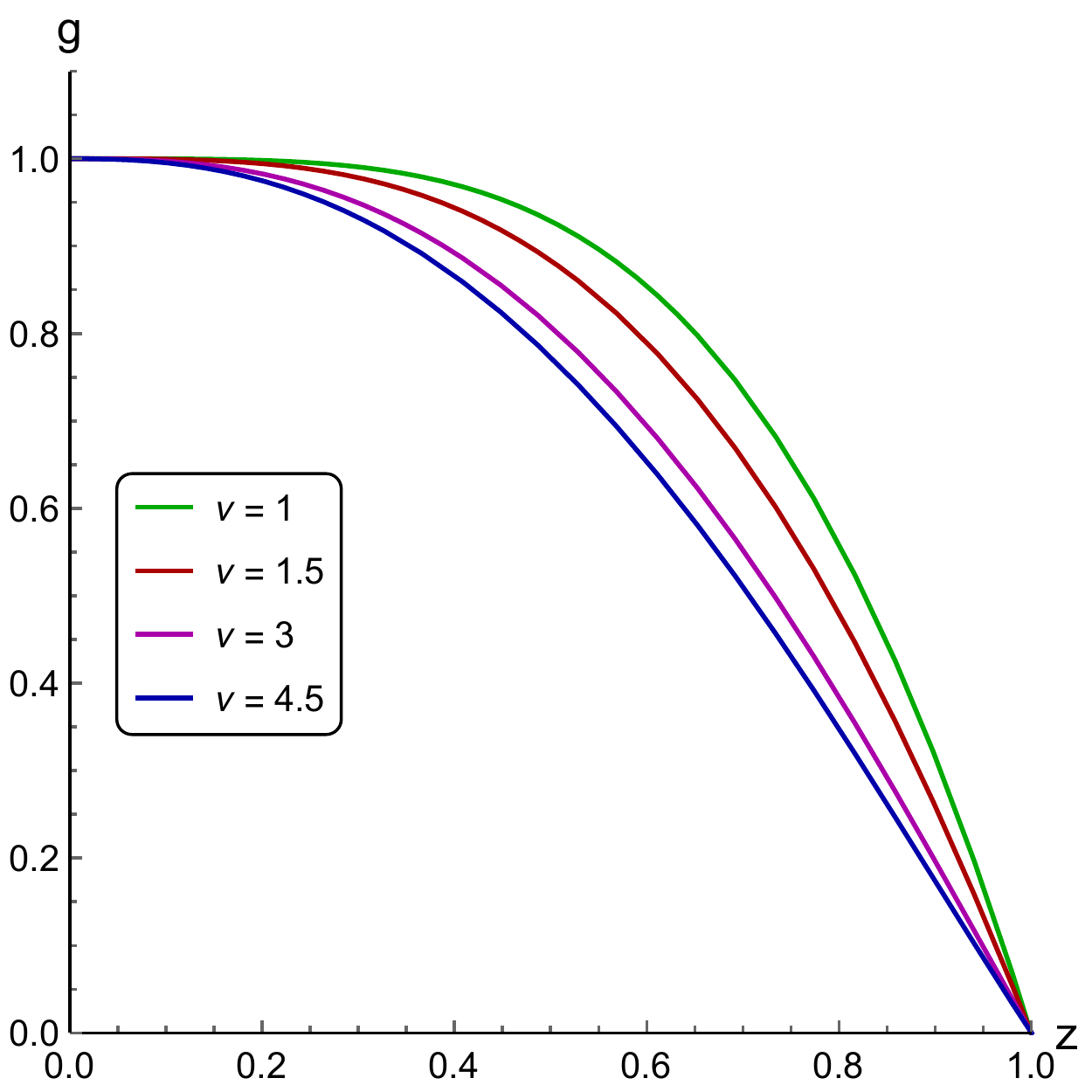} \\
  A \hspace{220pt} B \\
  \includegraphics[scale=0.46]{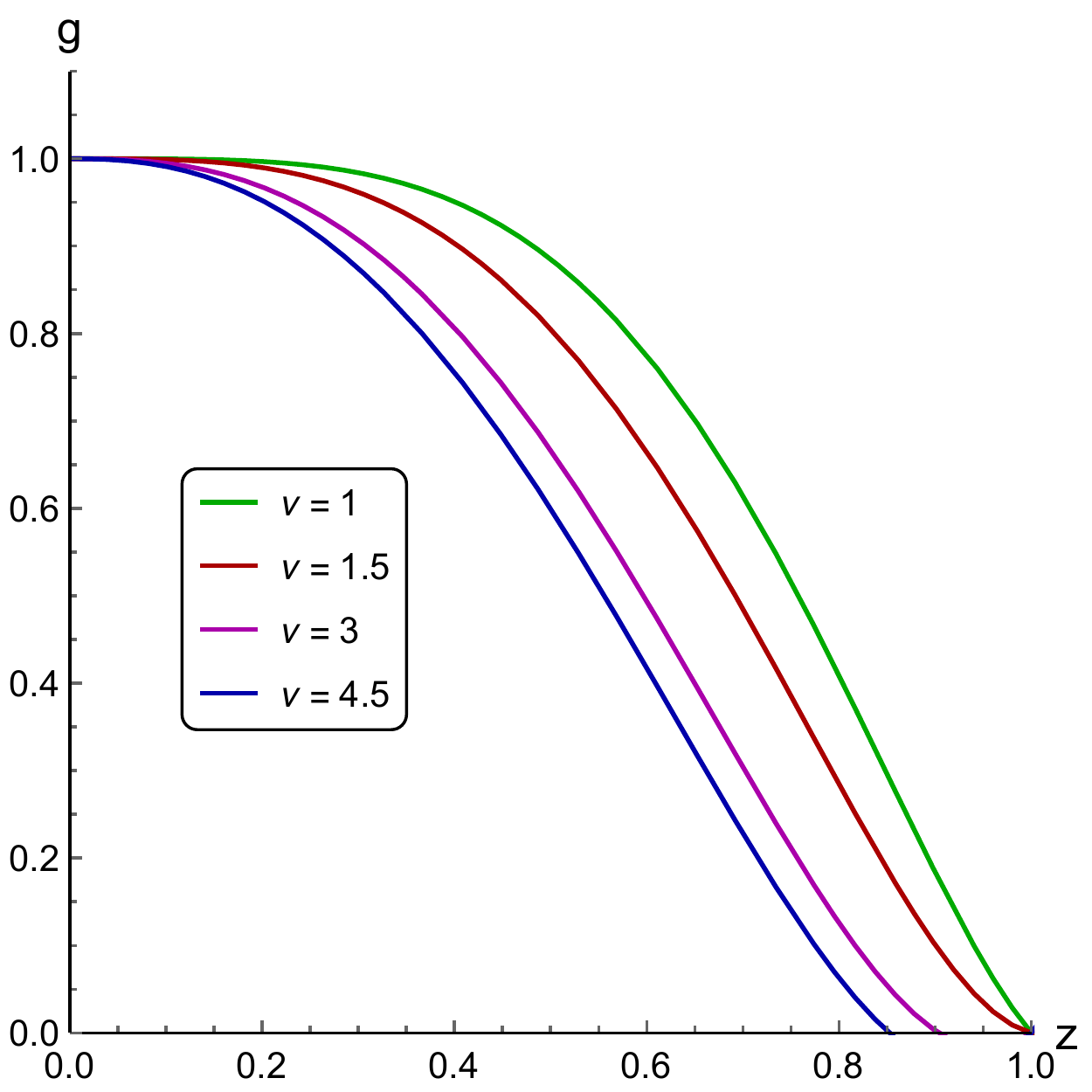} \qquad
  \includegraphics[scale=0.46]{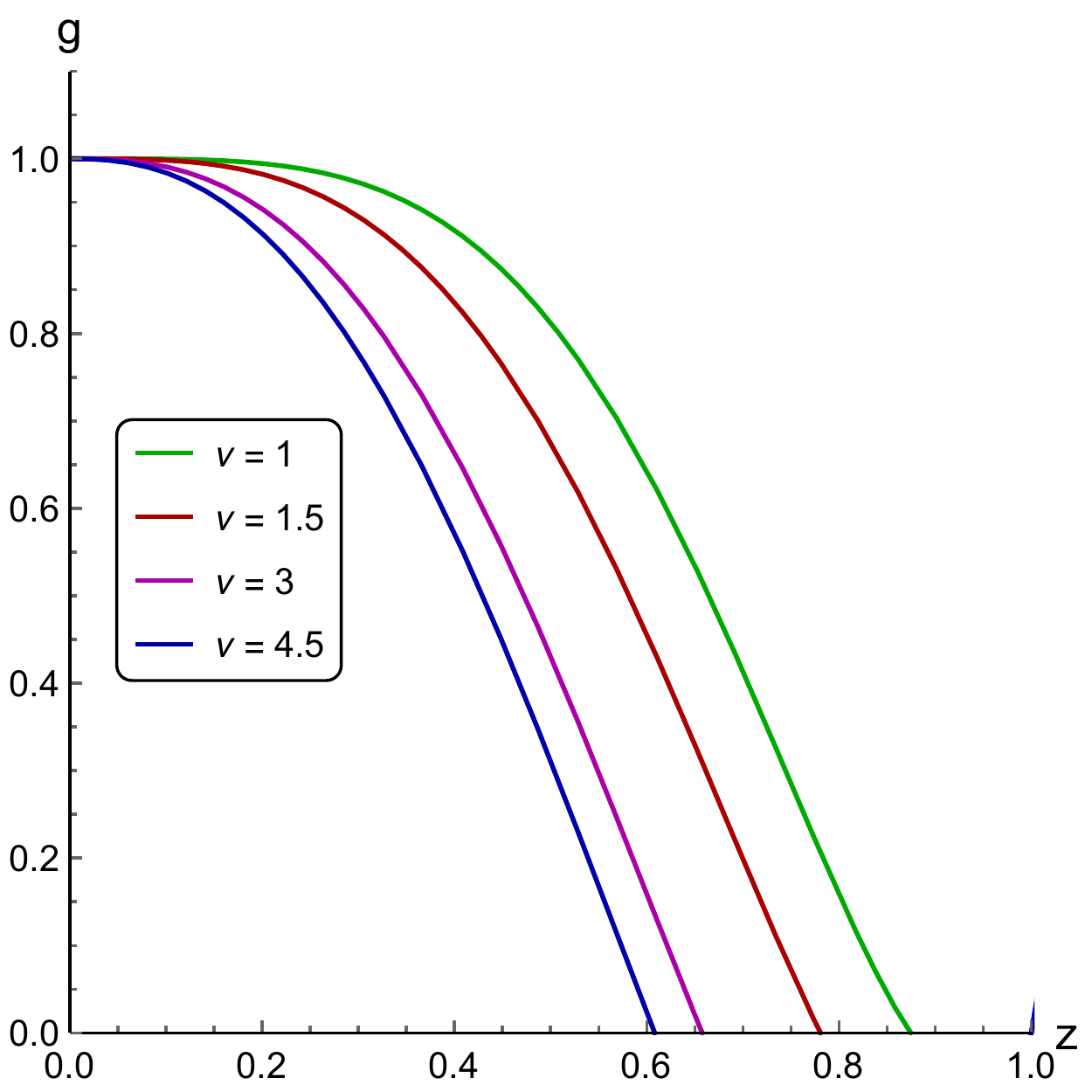} \\
  C \hspace{220pt} D
  \caption{Blackening function $g(z)$ in isotropic and anisotropic
    cases ($\nu = 1, \ 1.5, \ 3, \ 4.5$) for $\mu = 0$ (A), $\mu = 1$
    (B), $\mu = 2$ (C) and $\mu = 3$ (D); $a = 4.046$, $b = 0.01613$,
    $c = 0.227$, $z_h = 1$.}
  \label{Fig:gzmu}
\end{figure}

On Fig.\ref{Fig:gzmu} blackening function curves for different $\nu$
and fixed chemical potential are displayed. For zero chemical potential
(Fig.\ref{Fig:gzmu}.A) the blackening function decreases faster near
the horizon for lesser $\nu$, and this tendency continues for non-zero
$\mu$ values in one form or another. The second (non-fixed) horizon is 
lesser for larger $\nu$ values (Fig.\ref{Fig:gzmu}.C,D).

Solution (2.31) from \cite{Yang-2017} satisfies our equation
\eqref{eq:1.23} for $\nu = 1$ and $f_1 = e^{-cz^2-{\cA}(z)}$ and can
be obtained from \eqref{eq:1.29} putting $L = \nu = 1$.

\begin{figure}[t!]
  \centering
  \includegraphics[scale=0.36]{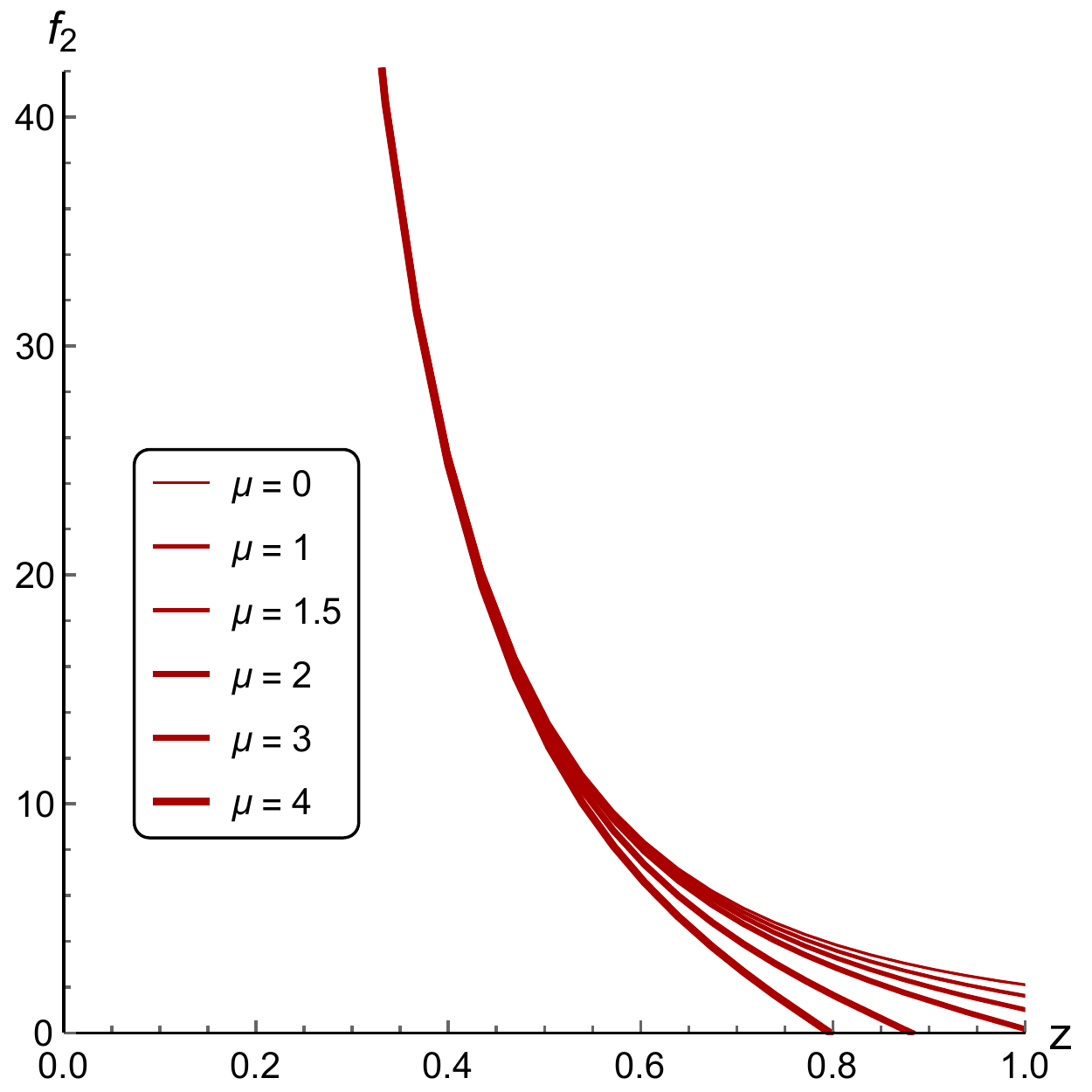} \quad
  \includegraphics[scale=0.36]{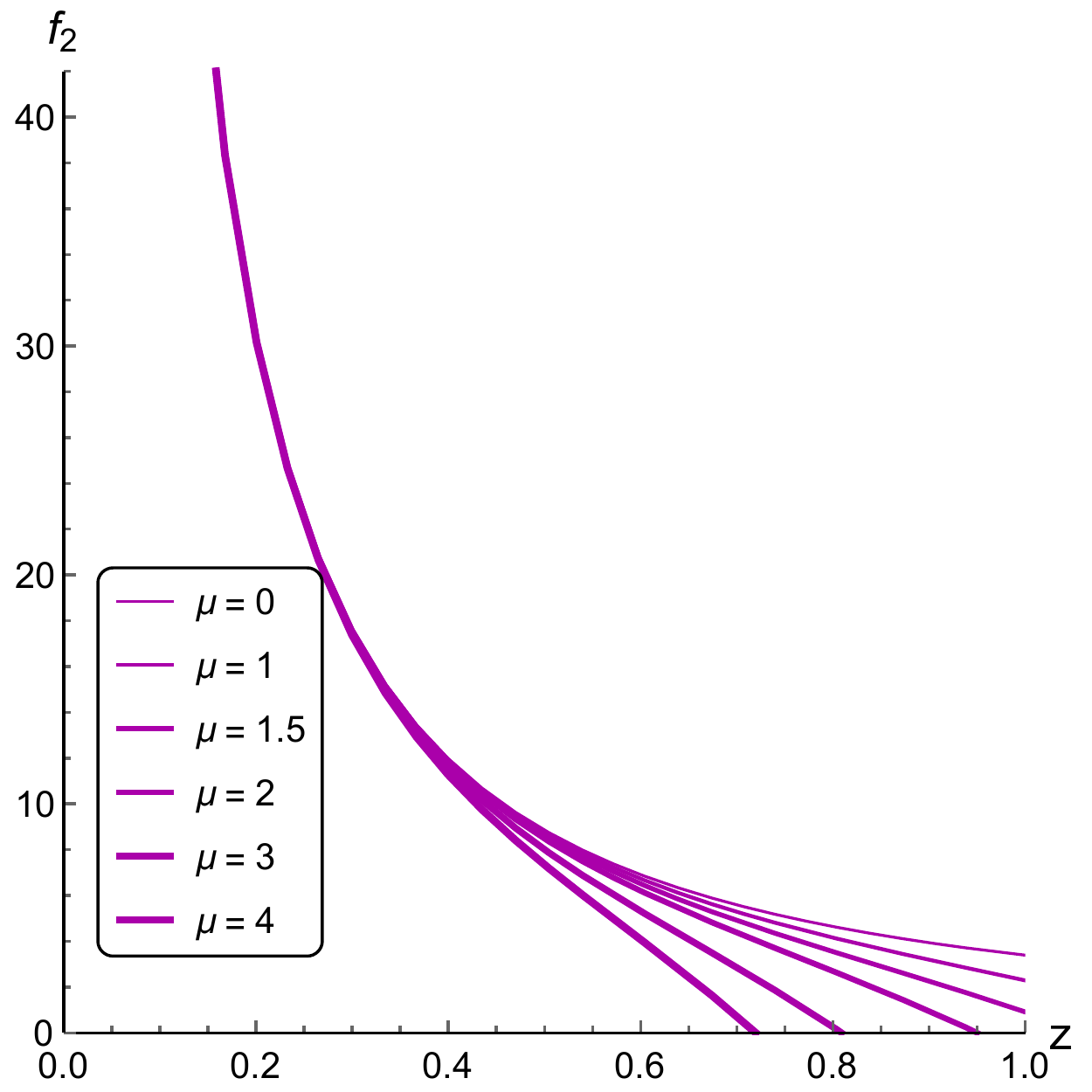} \quad
  \includegraphics[scale=0.36]{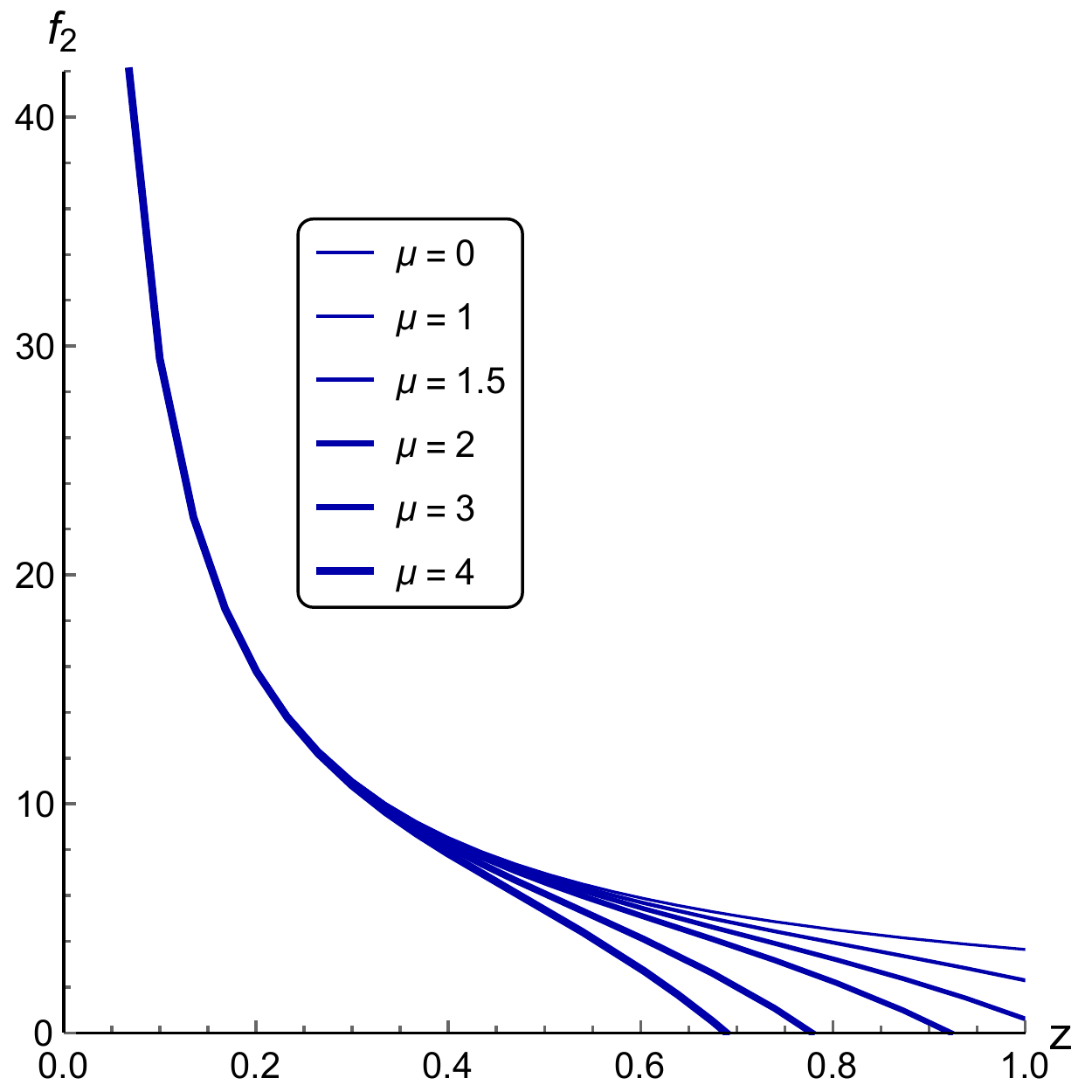} \\
  A \hspace{125pt} B \hspace{125pt} C
  \caption{Coupling function $f_2(z)$ for different $\mu$ in
    anisotropic cases $\nu = 1.5$ (A), $\nu = 3$ (B), $\nu = 4.5$ (C);
    $a = 4.046$, $b = 0.01613$, $c = 0.227$, $z_h = 1$ and $q = 1$.\\}
  \label{Fig:f2znumu}
  \centering
  \includegraphics[scale=0.36]{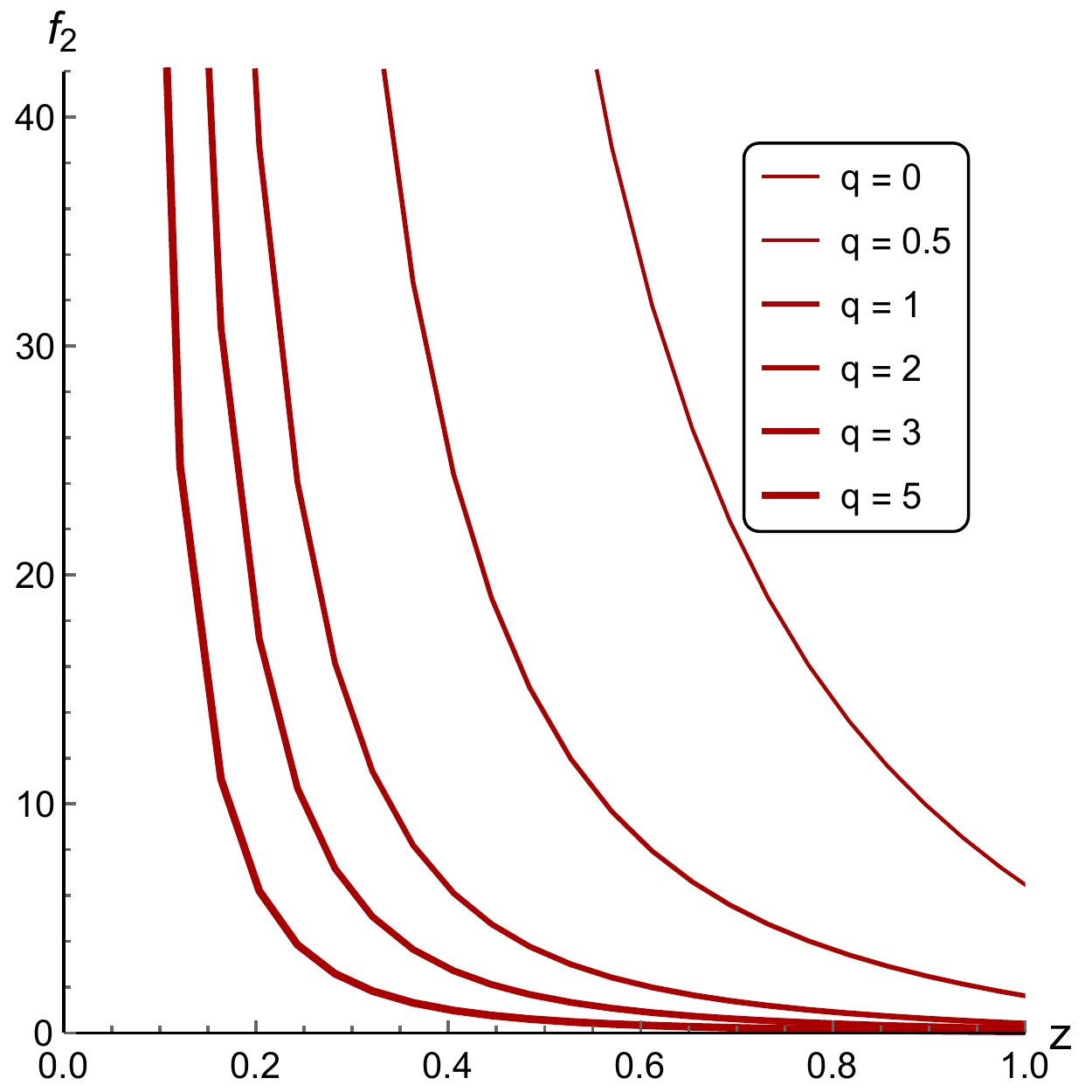} \quad
  \includegraphics[scale=0.36]{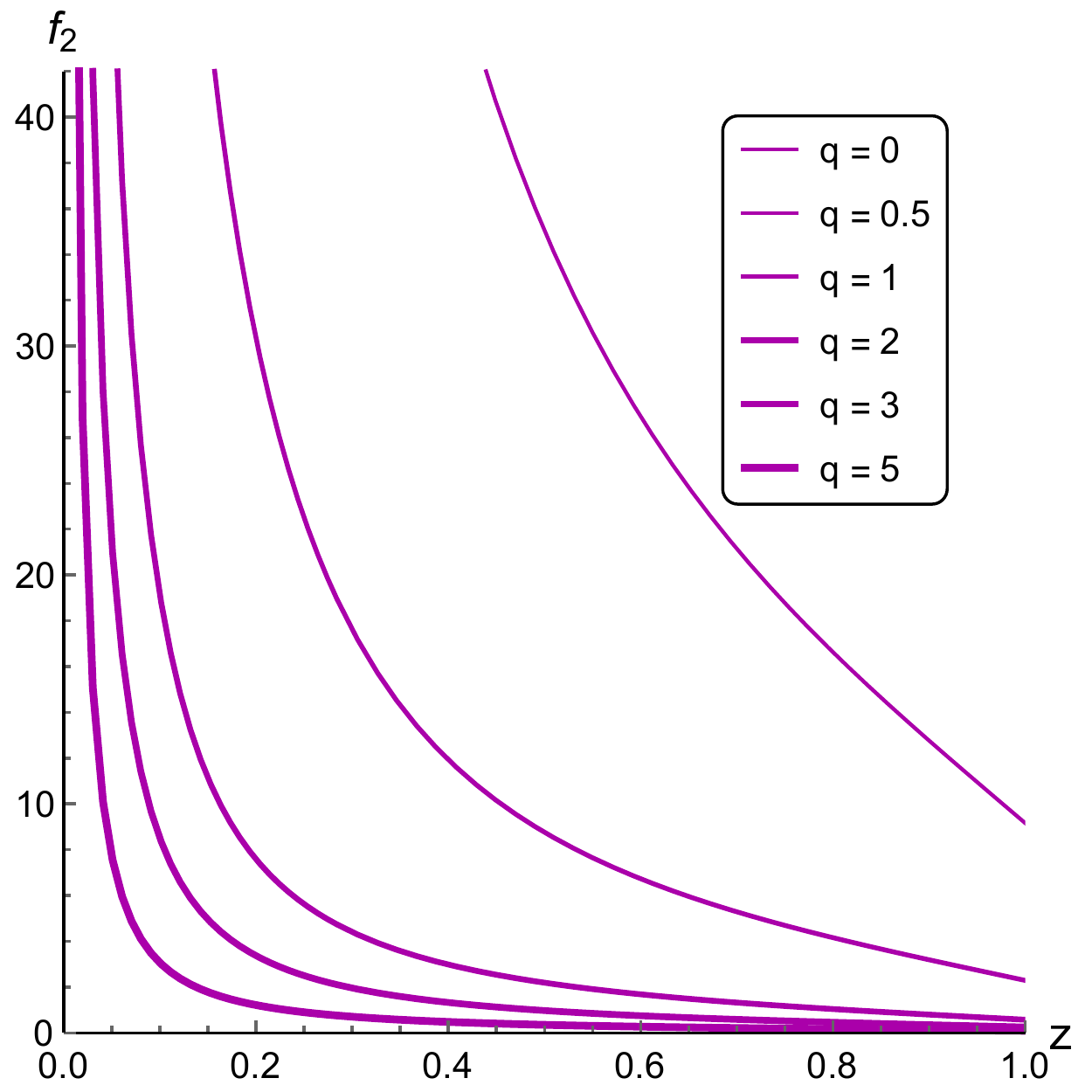} \quad
  \includegraphics[scale=0.36]{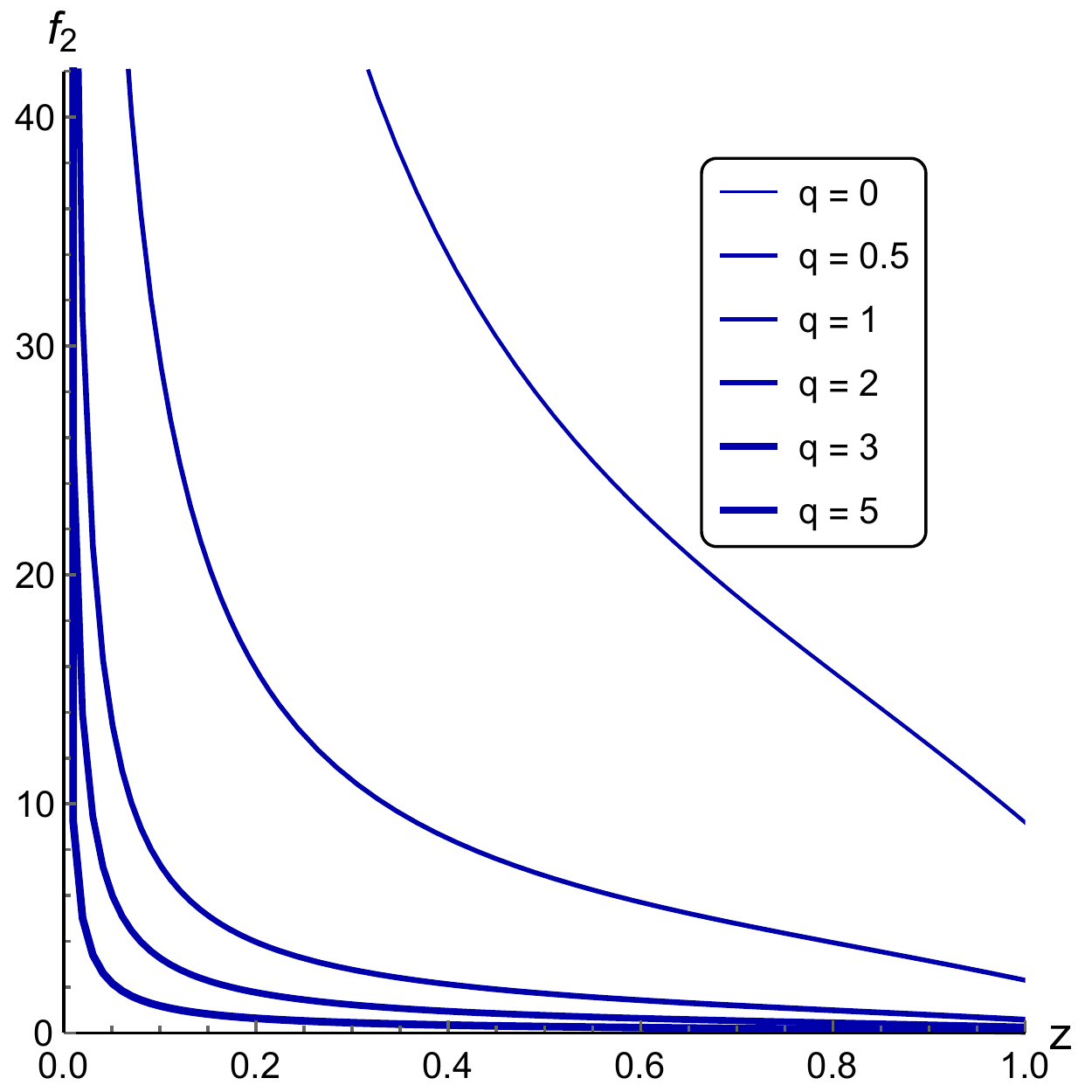} \\
  A \hspace{125pt} B \hspace{125pt} C
  \caption{Coupling function $f_2(z)$ for different $q$ in anisotropic
    cases $\nu = 1.5$ (A), $\nu = 3$ (B), $\nu = 4.5$ (C); $a =
    4.046$, $b = 0.01613$, $c = 0.227$, $z_h = 1$ and $\mu = 1$.\\}
  \label{Fig:f2znuq}
  \includegraphics[scale=0.36]{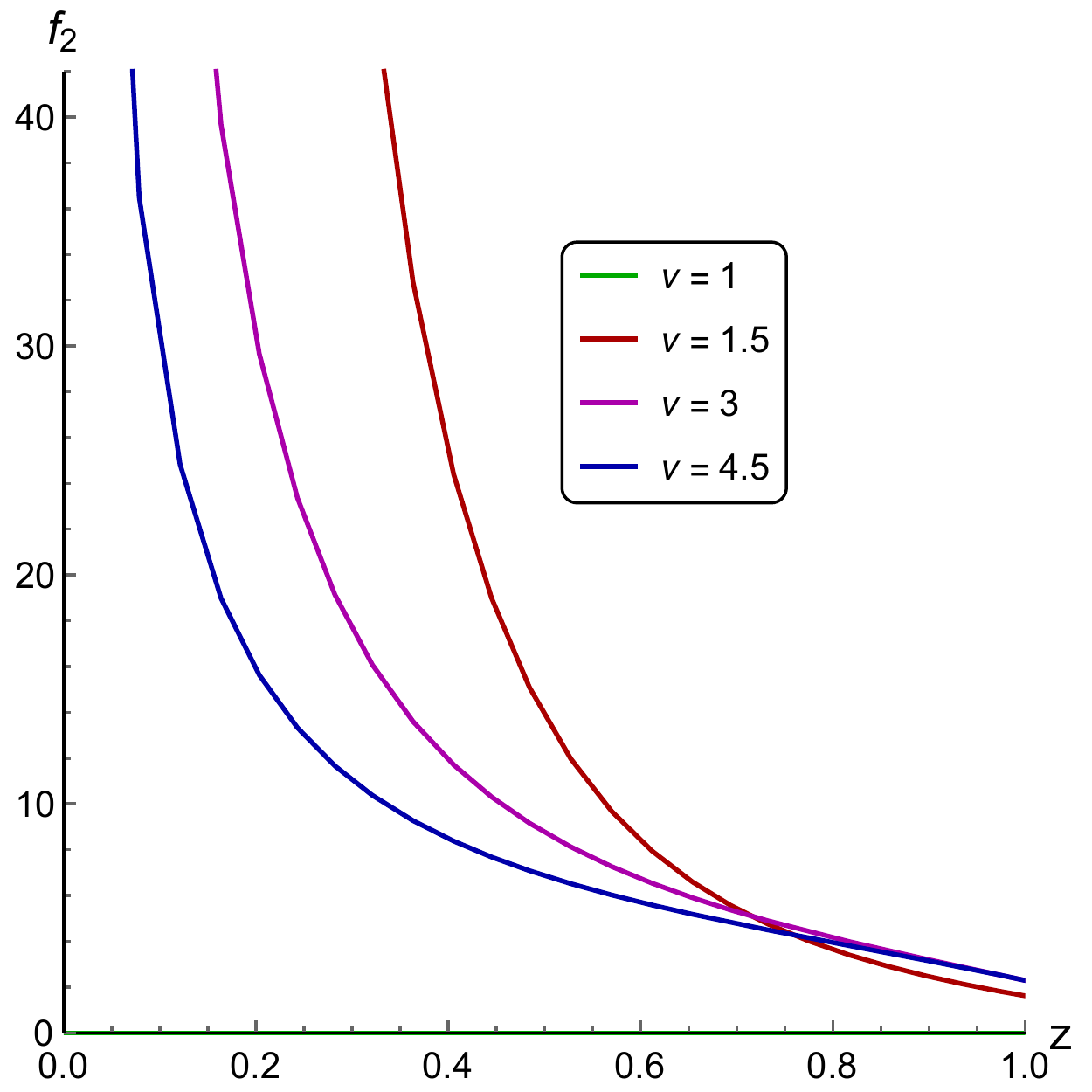} \quad
  \includegraphics[scale=0.36]{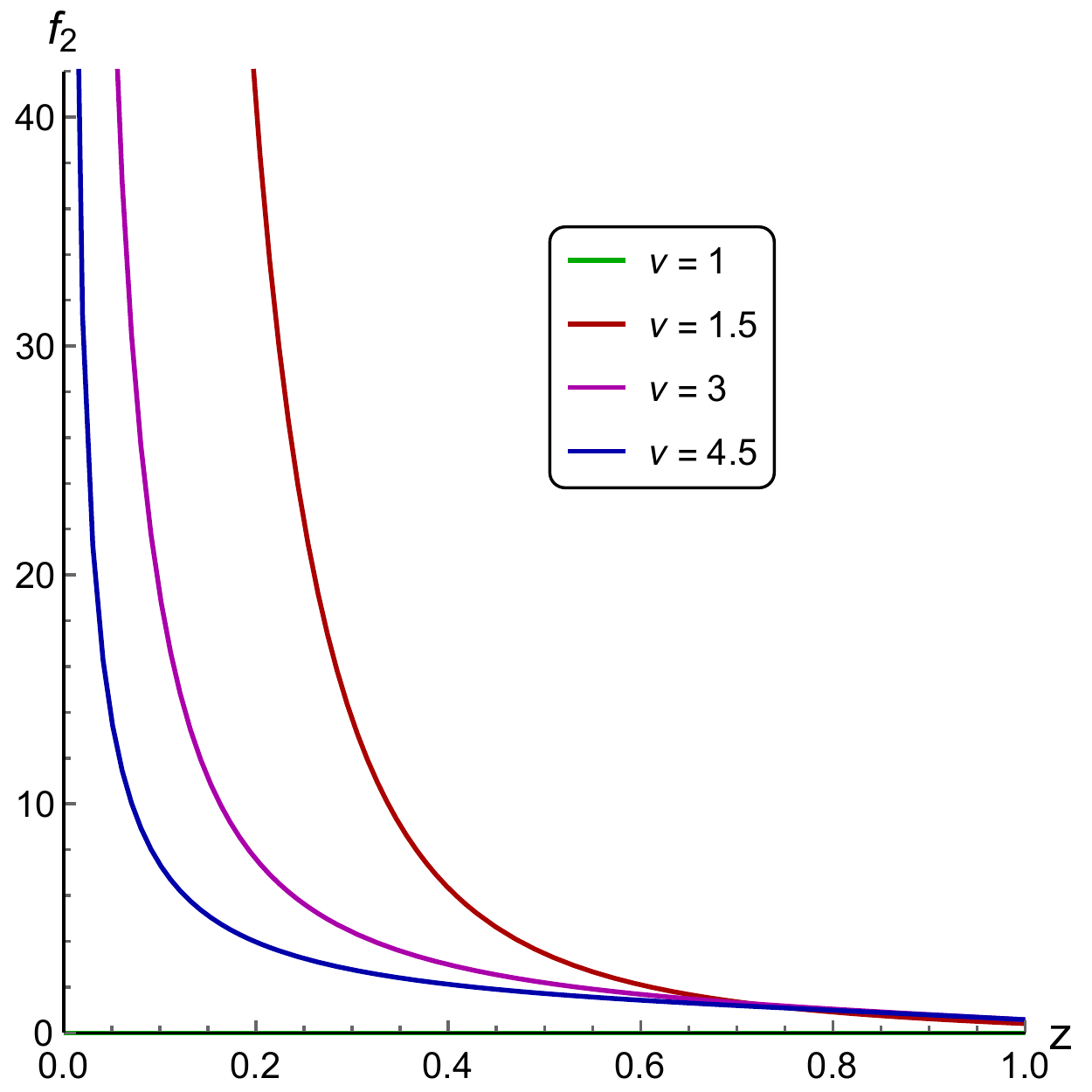} \quad
  \includegraphics[scale=0.36]{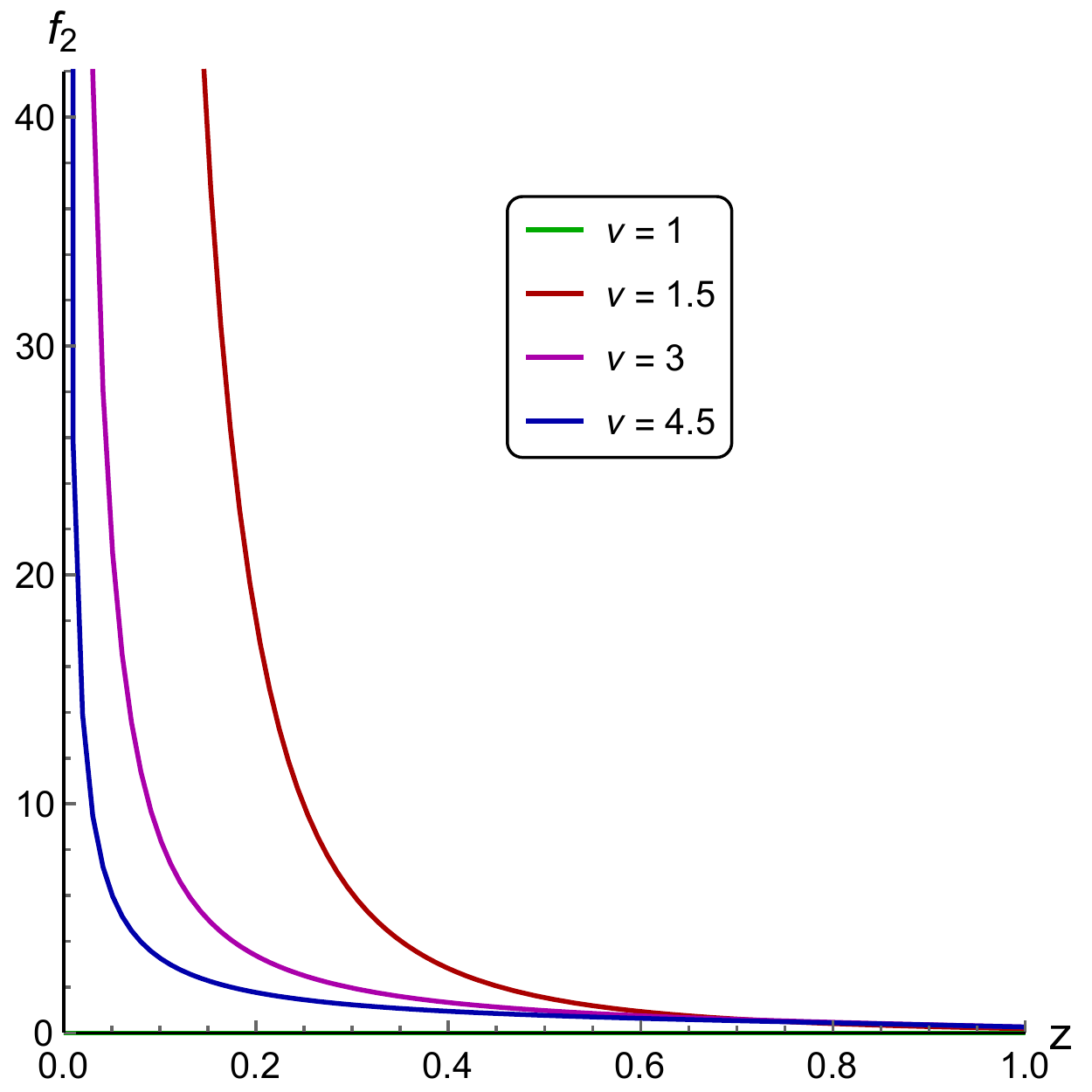} \\
  A \hspace{125pt} B \hspace{125pt} C
\caption{Coupling function $f_2(z)$ in isotropic and anisotropic
    cases ($\nu = 1, \ 1.5, \ 3, 4.5$) for $q = 1$ (A), $q = 2$ (B),
    $q = 3$ (C); $a = 4.046$, $b = 0.01613$, $c = 0.227$, $z_h = 1$
    and $\mu = 1$.}
  \label{Fig:f2zqnu}
\end{figure}

\newpage 

\subsubsection{Coupling function $f_2(z)$}

Soving \eqref{eq:1.25} and \eqref{eq:1.26} we get
\begin{gather}
  f_2 = - \left( \cfrac{z}{L} \right)^{1-\frac{4}{\nu}}
  \cfrac{e^{2{\cA}}}{L \, q^2} \left[
    2 g' \, \cfrac{\nu - 1}{\nu} 
    + 6 g \, \cfrac{\nu - 1}{\nu} \left(
      {\cA}' - \cfrac{2 \, (\nu + 1)}{3 \nu z} \right)
  \right]. \label{eq:1.32}
\end{gather}

Fig.\ref{Fig:f2znumu} shows coupling function $f_2(z)$ behavior for
different chemical potential and anisotropy parameter values. The
coupling function tends to zero without chemical potential, and the
larger $\mu$ and $\nu$ we have the faster $f_2(z)$ decreases. This can
also be seen from Fig.\ref{Fig:f2zqnu}. On the opposite, the larger
charge $q$ makes $f_2(z)$ to decrease more slowly for the fixed
chemical potential value (Fig.\ref{Fig:f2znuq}). In isotropic case
$f_2(z) \equiv 0$ (Fig.\ref{Fig:f2zqnu}).

Appropriate solutions require positive $f_2$ values. As we can see
from plots of Fig.\ref{Fig:gzmu}-\ref{Fig:f2zqnu}, the considered region
limited by the horizon with lesser $z$ fullfills this requirement.

Note, that the function $f_1$ is taken at hoc, while $f_2$ is found
from EOM. The reason for this is the following. In spite of the fact
that $f_1$ and $f_2$ are included into the action symmetrically, their
roles are quite different. Coupling function $f_1$ provides the
chemical potential and is related with the Regge spectrum, while $f_2$
specifies the anisotropy. We use the electric ansatz for the first
Maxwell field and the magnetic one for the second like in
\cite{AGG}. Technically one can fix $f_2$ and find $f_1$ from EOM,
but we do not have restrictions and requirements for the form of
$f_2$-function.

\subsubsection{Scalar field $\phi(z)$}\label{scalar} 

We should reproduce the correct behavior of the string tension on
temperature. The condition \eqref{eq:1.19} for $z_0 = z_h$ can give
the behavior considered in \cite{ARS-2019qfthep} (for heavy quarks
and, as it would be seen, for light quarks situation is the
same). Solving \eqref{eq:1.24} with boundary condition
\eqref{eq:1.19} gives
\begin{gather}
  \phi = \int_{z_0}^z \cfrac{2 \sqrt{\nu - 1 + \left( 2 (\nu - 1) + 9
        a \nu^2 \right) b \, \xi^2 + \left( \nu - 1 + 3 a (1 + 2 a)
        \nu^2 \right) b^2 \, \xi^4 }}{(1 + b \xi^2) \, \nu \, \xi} \ d
  \xi. \label{eq:1.33}
\end{gather}
\begin{figure}[t!]
  \centering
  \includegraphics[scale=0.37]{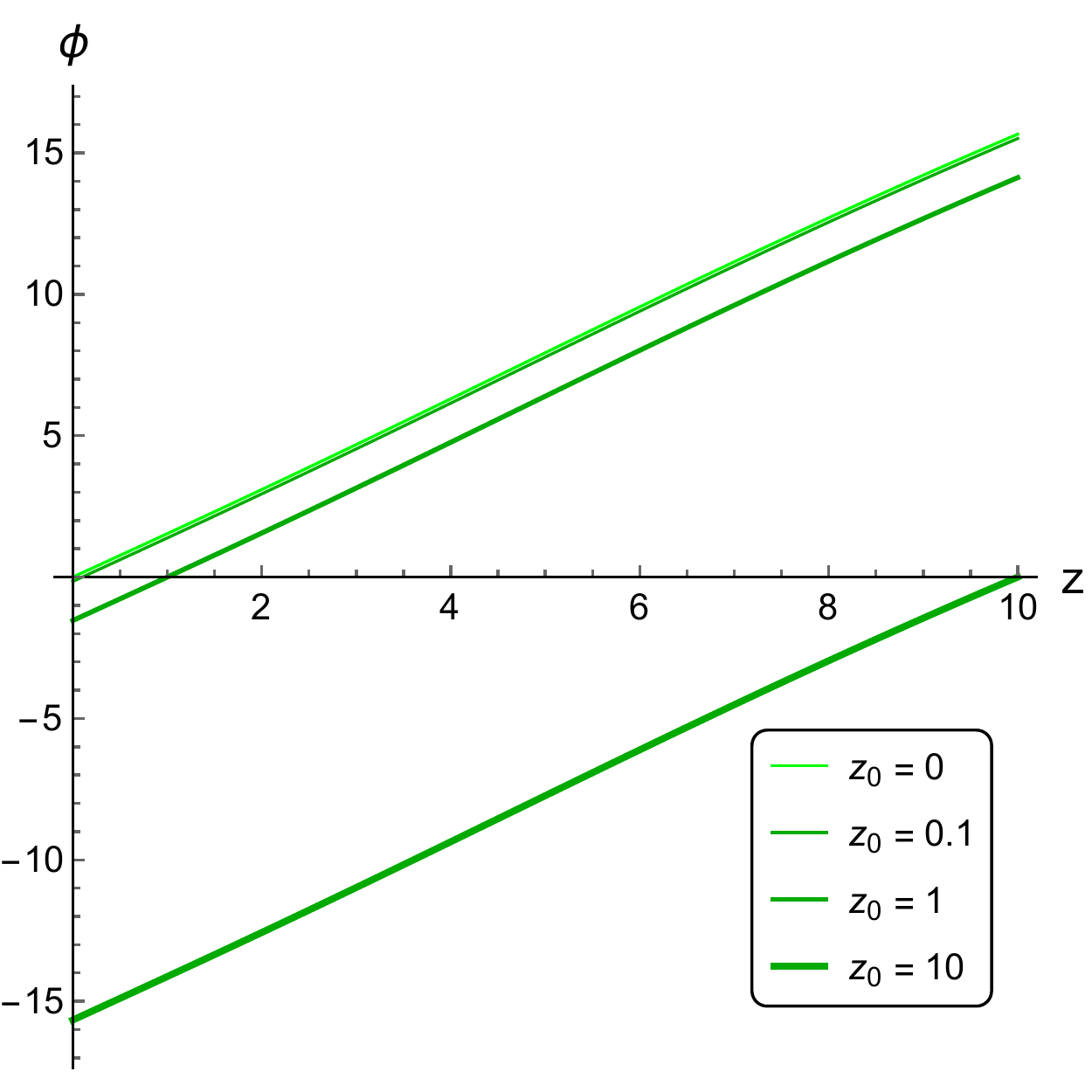} \qquad
  \includegraphics[scale=0.37]{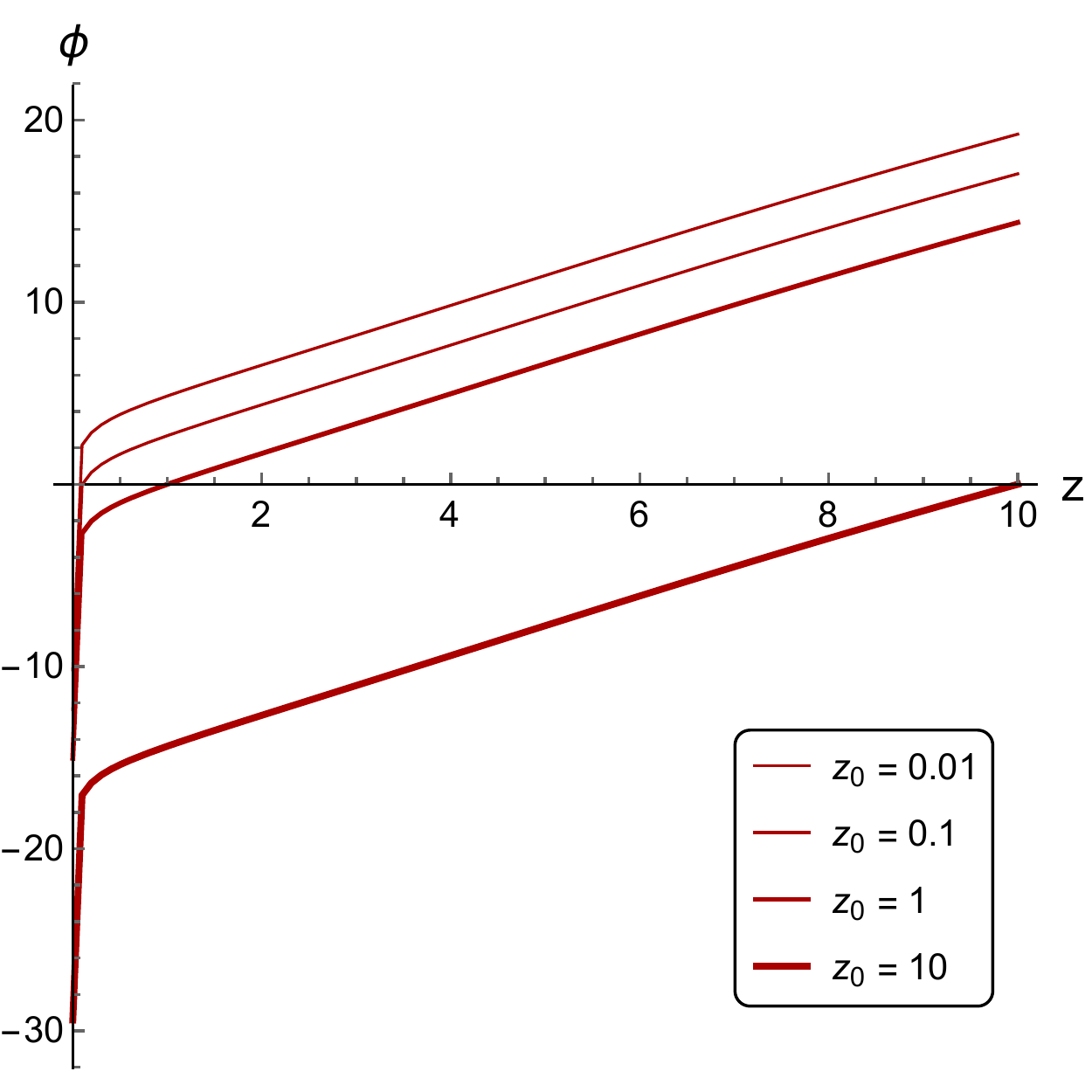} \\
  \ A \hspace{150pt} B \\
  \includegraphics[scale=0.37]{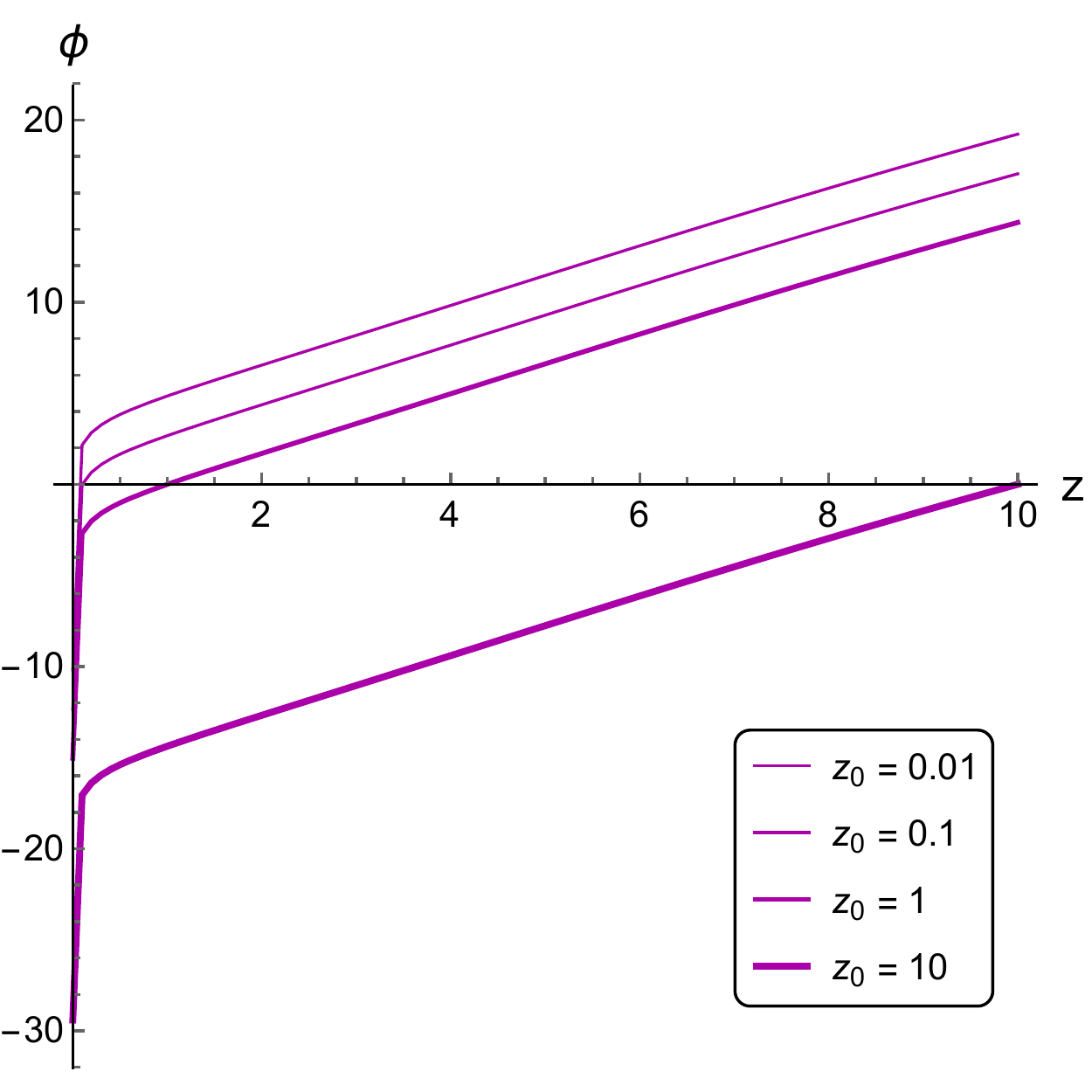} \qquad
  \includegraphics[scale=0.37]{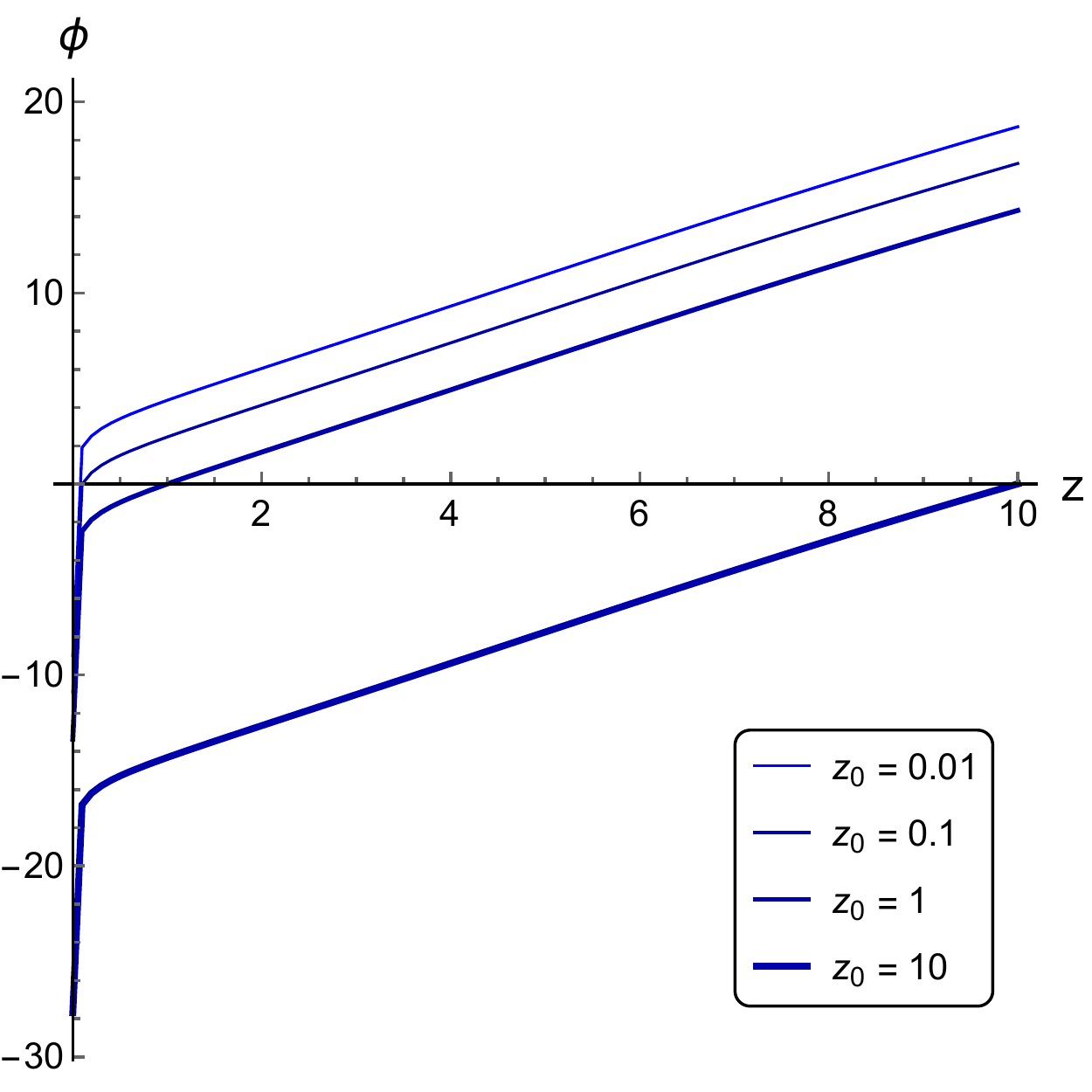} \\
  \ C \hspace{150pt} D
  \caption{Scalar field $\phi(z)$ in isotropic (A) and anisotropic
    cases for $\nu = 1.5$ (B), $\nu = 3$ (C) and $\nu = 4.5$ (D) for
    $z_0 = 0, \ 0.01, \ 0.1, \ 1, \ 10$; $a = 4.046$, $b = 0.01613$,
    $c = 0.227$.}
  \label{Fig:phiznu}
\end{figure}

In the anisotropic case for $z_0 = 0$ the dilaton field has a
logarithmic divergence $\phi(z) \sim \int_{0}^{z} dz/z$. There are no
divergences in the isotropic case for the dilaton field and the
expression is reduced as:
\begin{gather}
  \phi_{iso} = \int_{z_0}^z \cfrac{2 \sqrt{ 9a b  + \left( 3 a  + 6
        a^2 \right) b^2 \, \xi^2 }}{(1 + b \xi^2) } \ d \xi. 
\end{gather}
Note that boundary conditions influence on the temperature dependence
of string tension (i.e. on the coefficient in the linear term of
Cornell potential). String tension should decrease while temperature
increases and drop to zero after the confinement/deconfinement phase
transition \cite{Digal:2005ht, Cardoso:2011hh, Bicudo:2010hg}. To keep
this behavior on the one hand and avoid divergences in anisotropic
cases on the other hand we generalize boundary condition for dilaton
field as $\phi(z_0) = 0$ \cite{He:2010ye}, where $z_0$ can be a
function of $z_h$. Conditions $z_0 = z_h$ \cite{AR-2018} or $z_0 = 0$
\cite{Yang-2017} are it's particular cases.

As we can see from Fig.\ref{Fig:phiznu}.A, even $z_0 = 0.1$ can be
hardly distinguished from the $z_0 = 0$ in the isotropic case. This
allows to assume that the results for sufficiently small $z_0$
reproduce the proper behavior of the scalar field. On the other hand
the difference between various $z_0$ cases can be used to fit the
experimental data in the future. The dilaton field's dependence on $z$
is a monotonically increasing function (Fig.\ref{Fig:phiznu}). This
function has finite negative value near $z = 0$ in isotropic case and
decreases quickly in anisotropic cases (Fig.\ref{Fig:phiznu}.B-D). We
can also see that in this case the $\phi(z)$ behavior depends on the
anisotropy parameter $\nu$ rather weakly.

\begin{figure}[h!]
  \centering
  \includegraphics[scale=0.5]{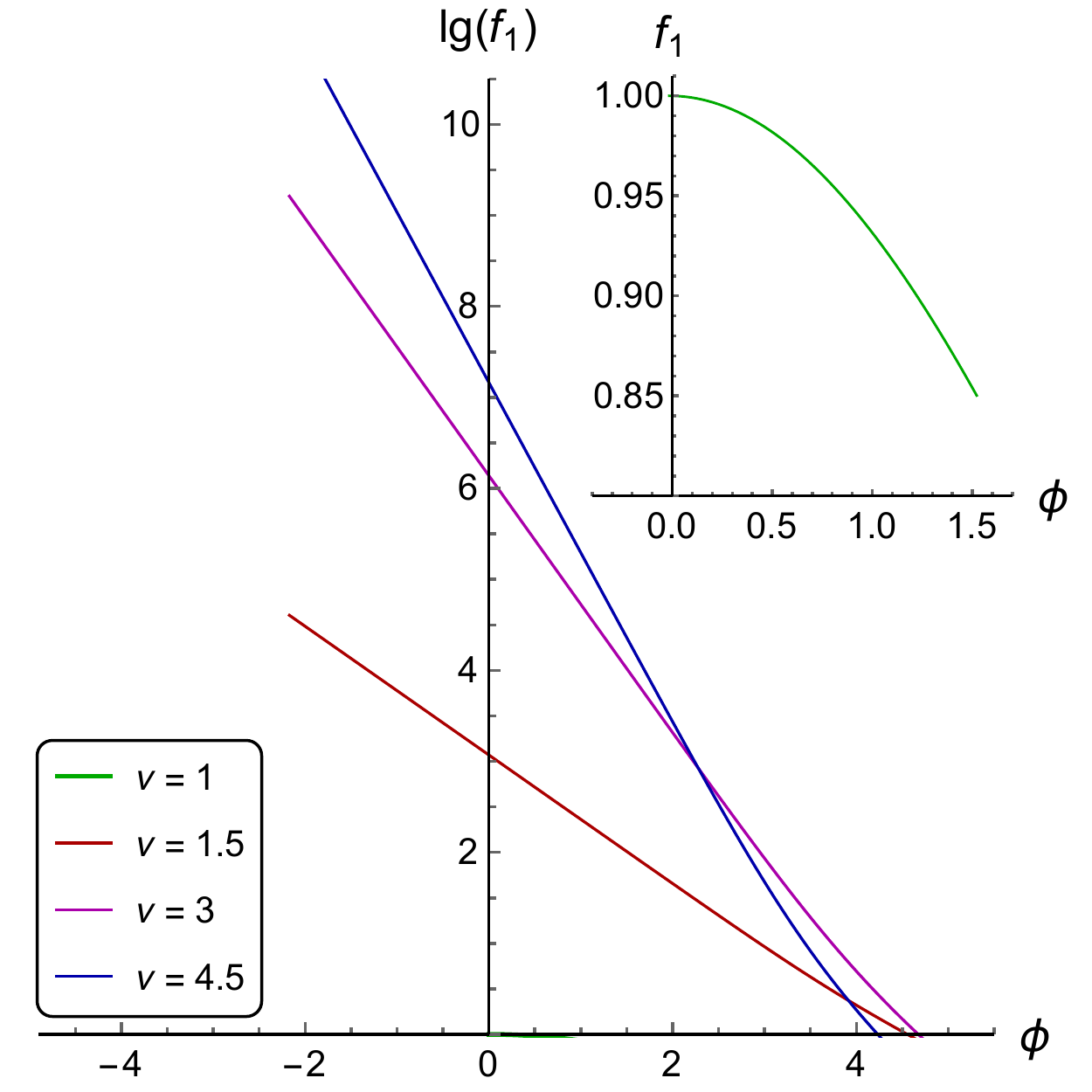} \qquad
  \includegraphics[scale=0.5]{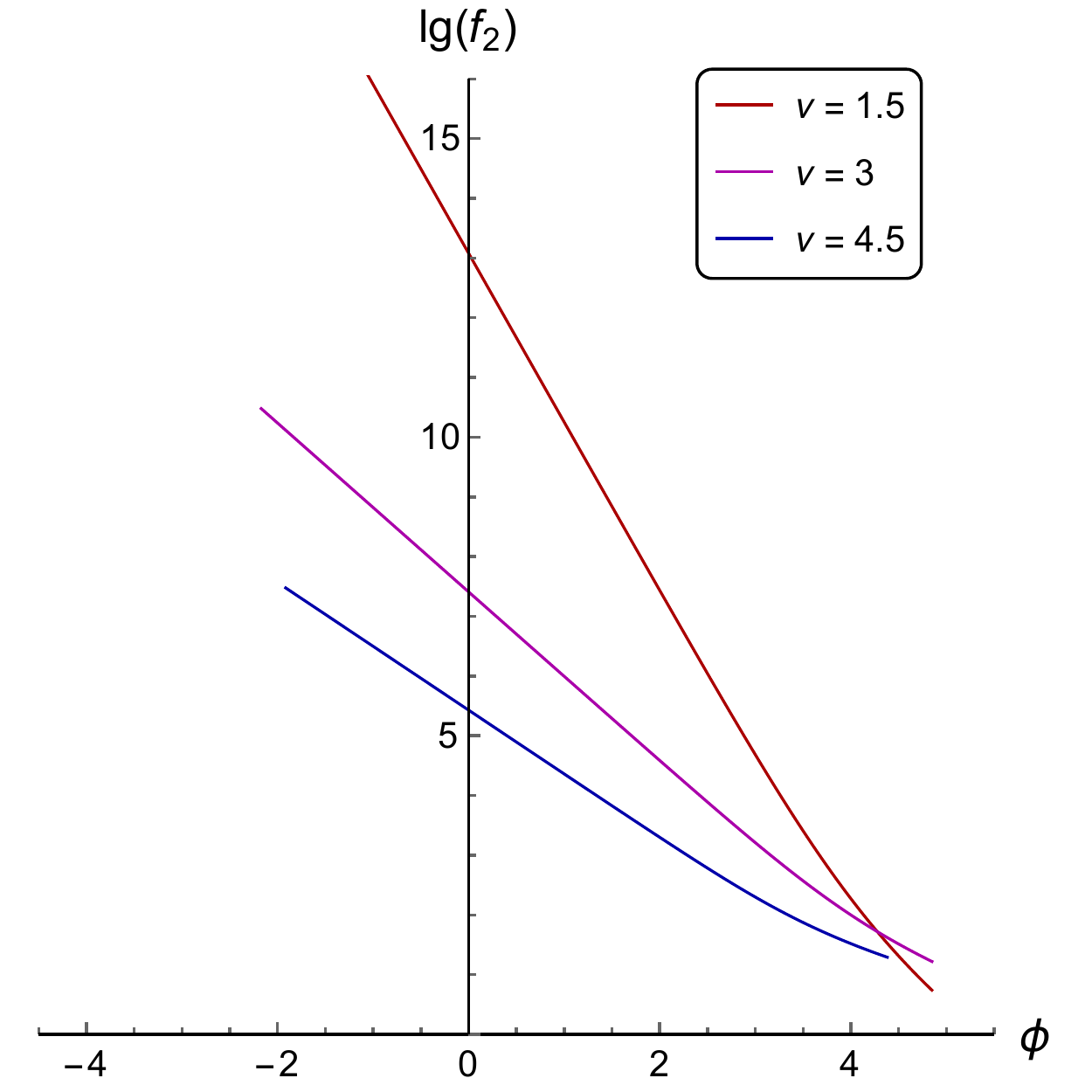} \\
  \ \\
  \includegraphics[scale=0.5]{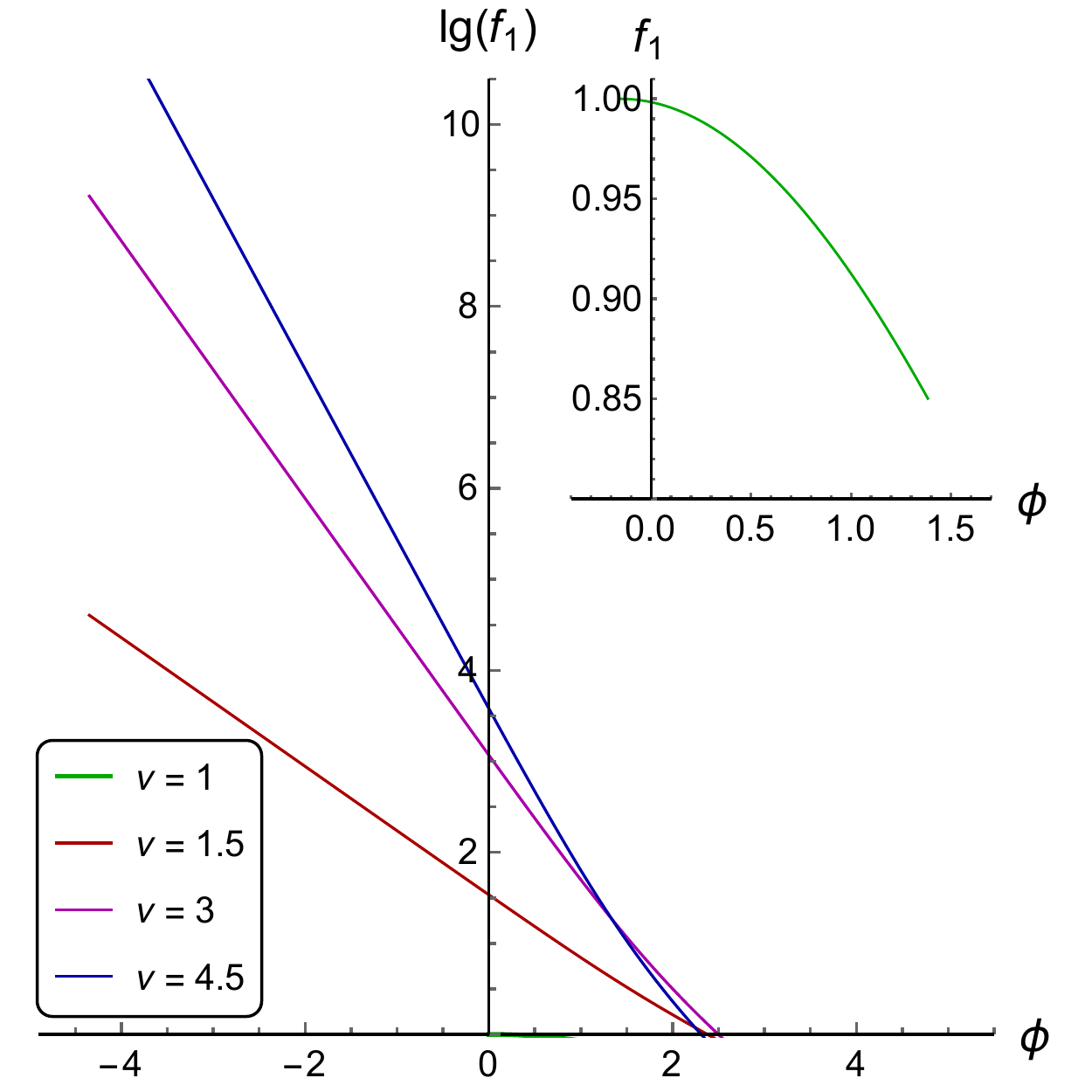} \qquad
  \includegraphics[scale=0.5]{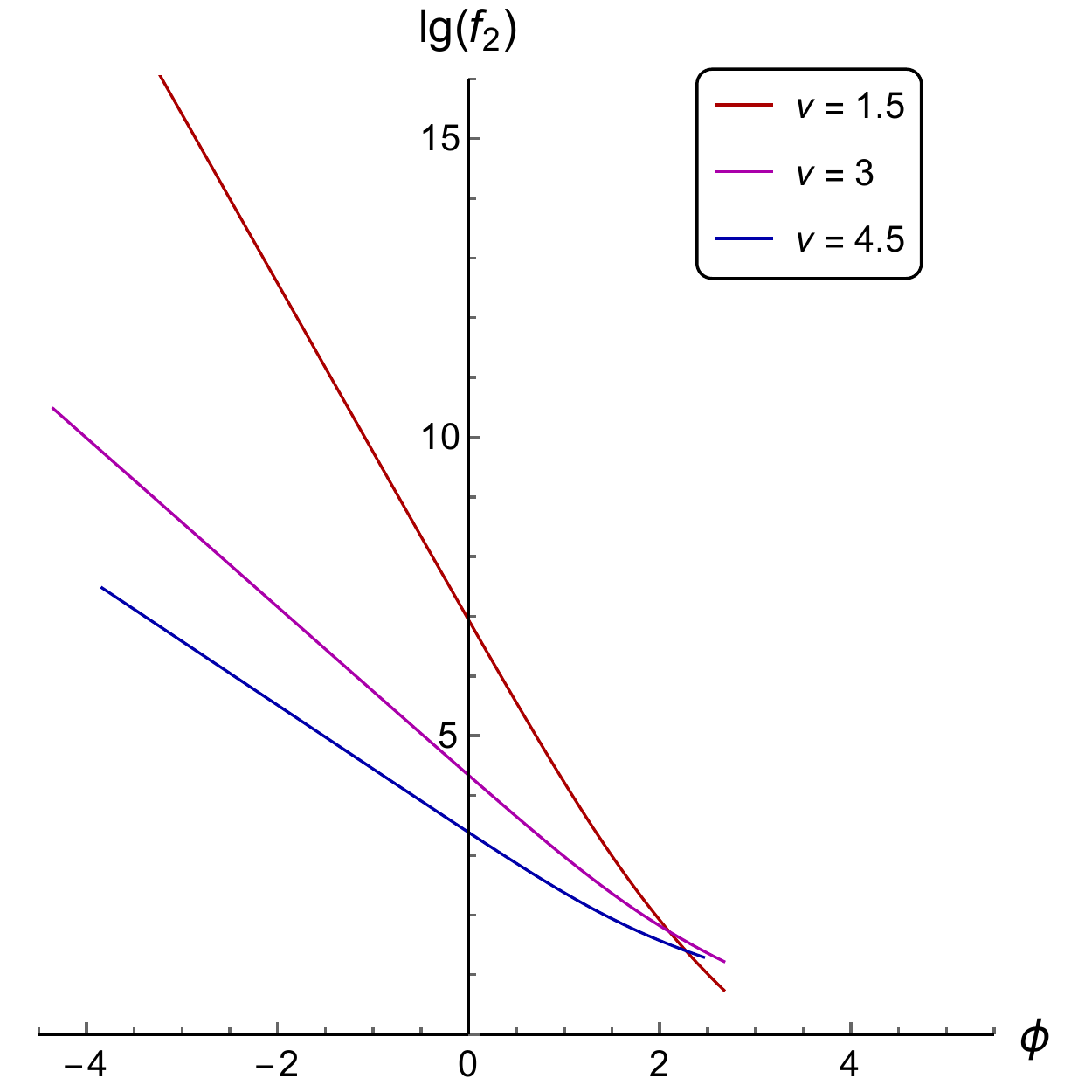} \\
  A \hspace{200pt} B\\
 \caption{Coupling functions $f_1(\phi)$ (A) and $f_2(\phi)$ (B) in
    logarithmic scale for different anisotropy $\nu$, $z_0 = 0.01$
    (1-st line) and $z_0 = 0.1$ (2-nd line), $a = 4.046$, $b =
    0.01613$, $c = 0.227$ and $q = 1$.}
  \label{Fig:f1f2phi}
\end{figure}

Since the function $\phi(z)$ is monotonic, we reconstruct the form of
$f_1$ and $f_2$ as functions of $\phi$ (Fig.\ref{Fig:f1f2phi}). Both
functions decrease quickly and monotonically while increasing
$\phi$. The first (top) line of Fig.\ref{Fig:f1f2phi} presents
$f_1(\phi)$ (Fig.\ref{Fig:f1f2phi}.A) and $f_2(\phi)$
(Fig.\ref{Fig:f1f2phi}.B) for $z_0 = 0.01$ and different
anisotropies in logarythmic scale (except $f_1(\phi)$ for $\nu =
1$). The second (bottom) line show the same plots for $z_0 =
0.1$. The increasing of the boundary $z_0$ has rather weak effect on
the shape of the coupling functions and mainly shifts them to the
left, to lower $\phi$-values.

\subsubsection{Scalar potential $V(\phi)$}

\begin{figure}[t!]
  \centering
  \includegraphics[scale=0.35]{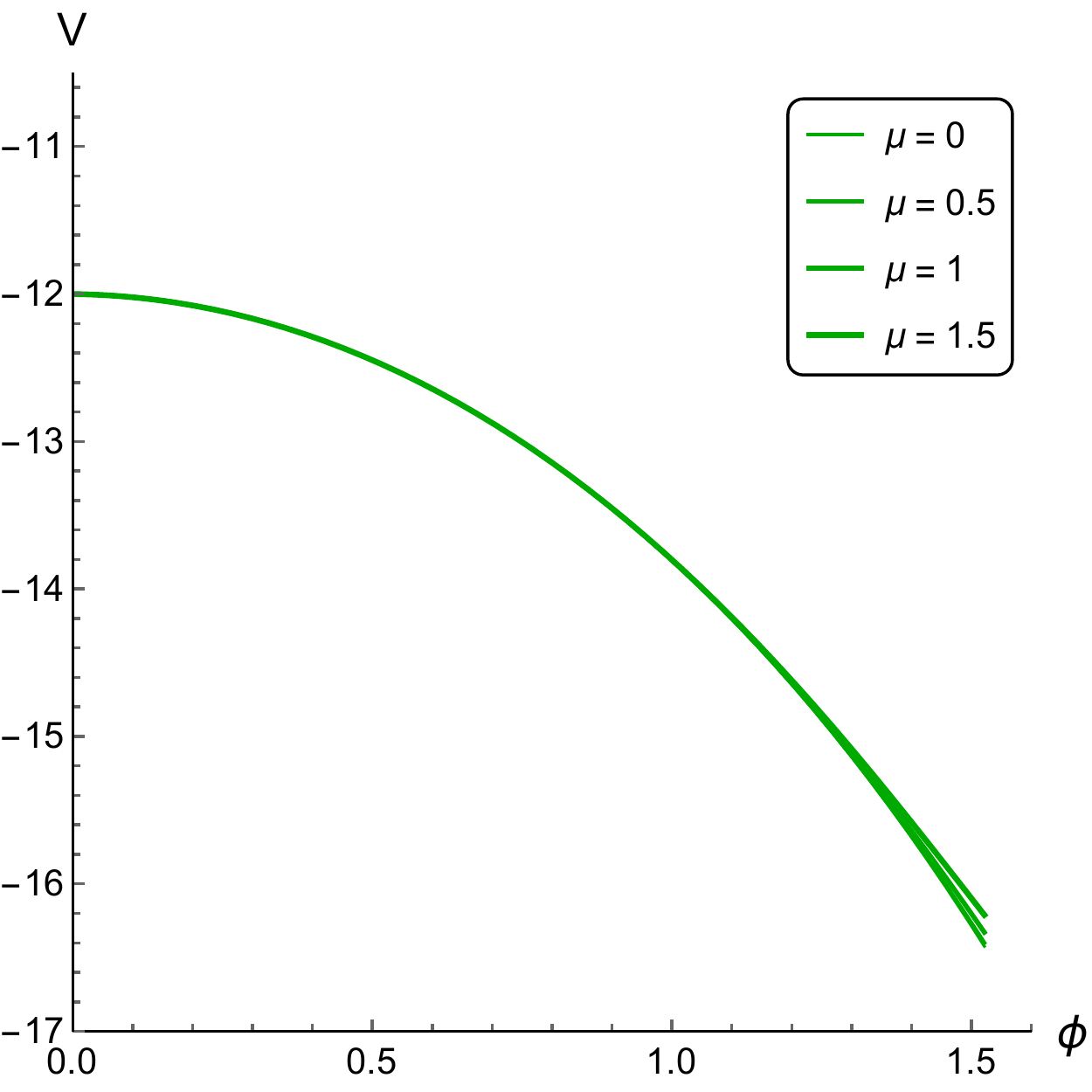} \quad
  \includegraphics[scale=0.35]{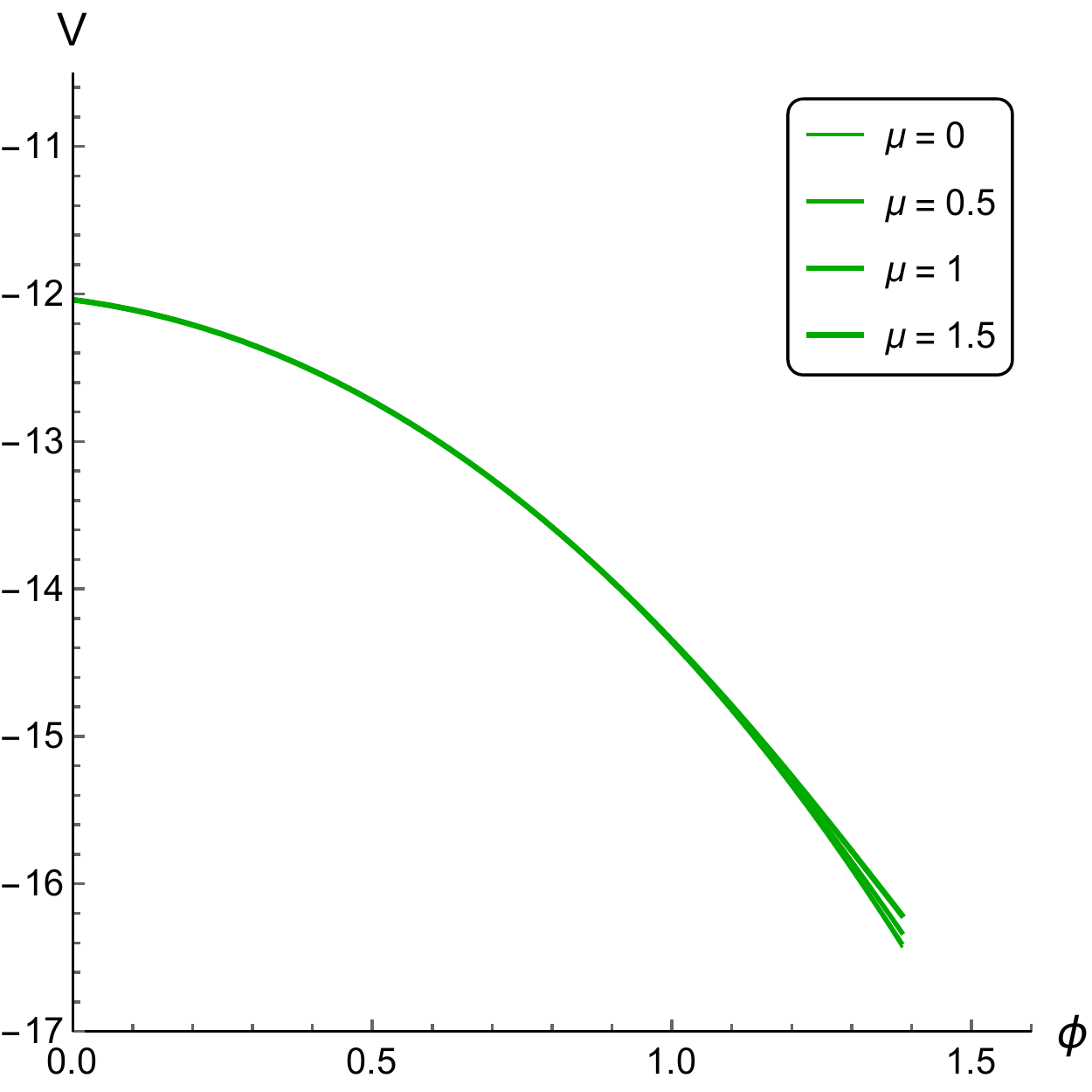} \quad
  \includegraphics[scale=0.35]{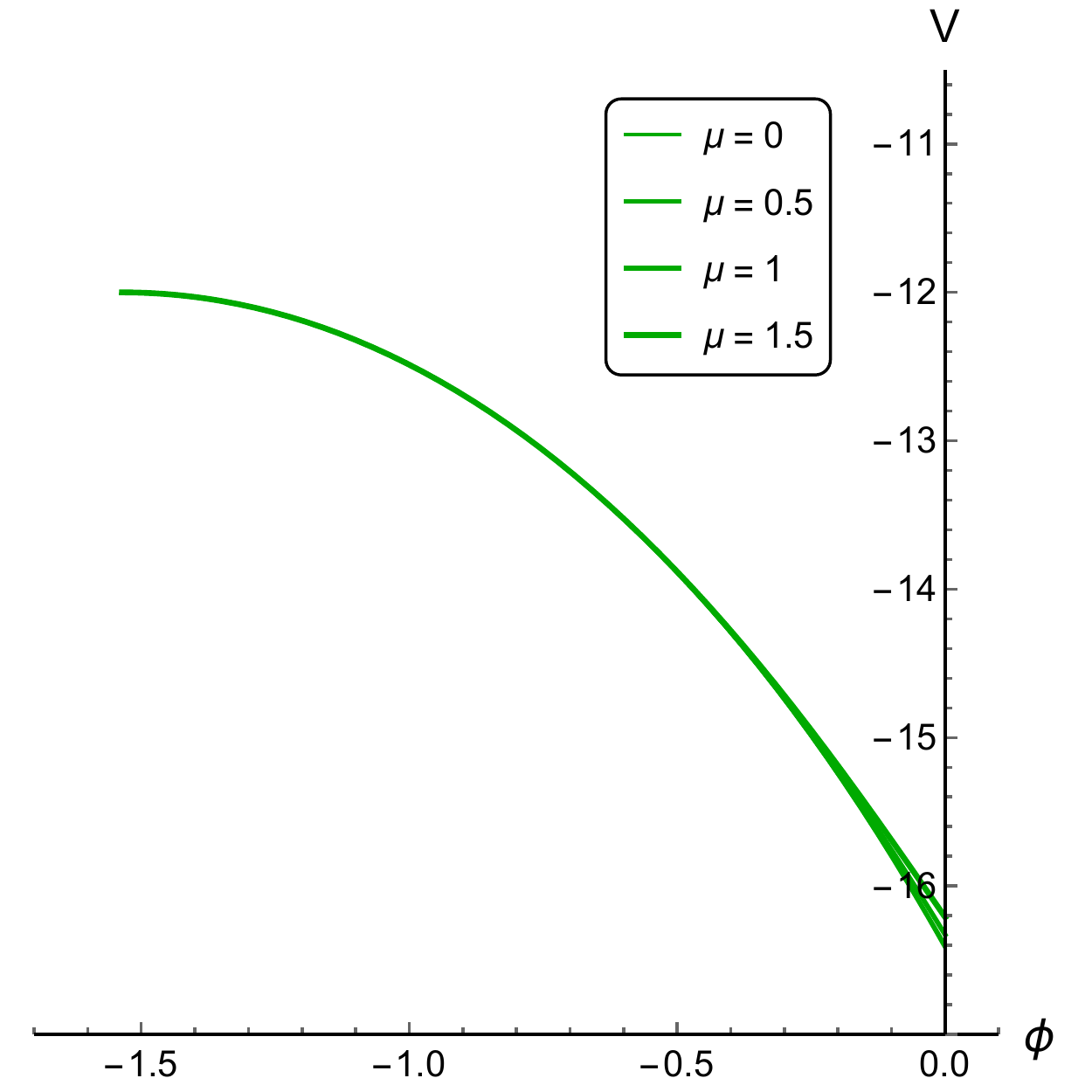} \\
  \includegraphics[scale=0.35]{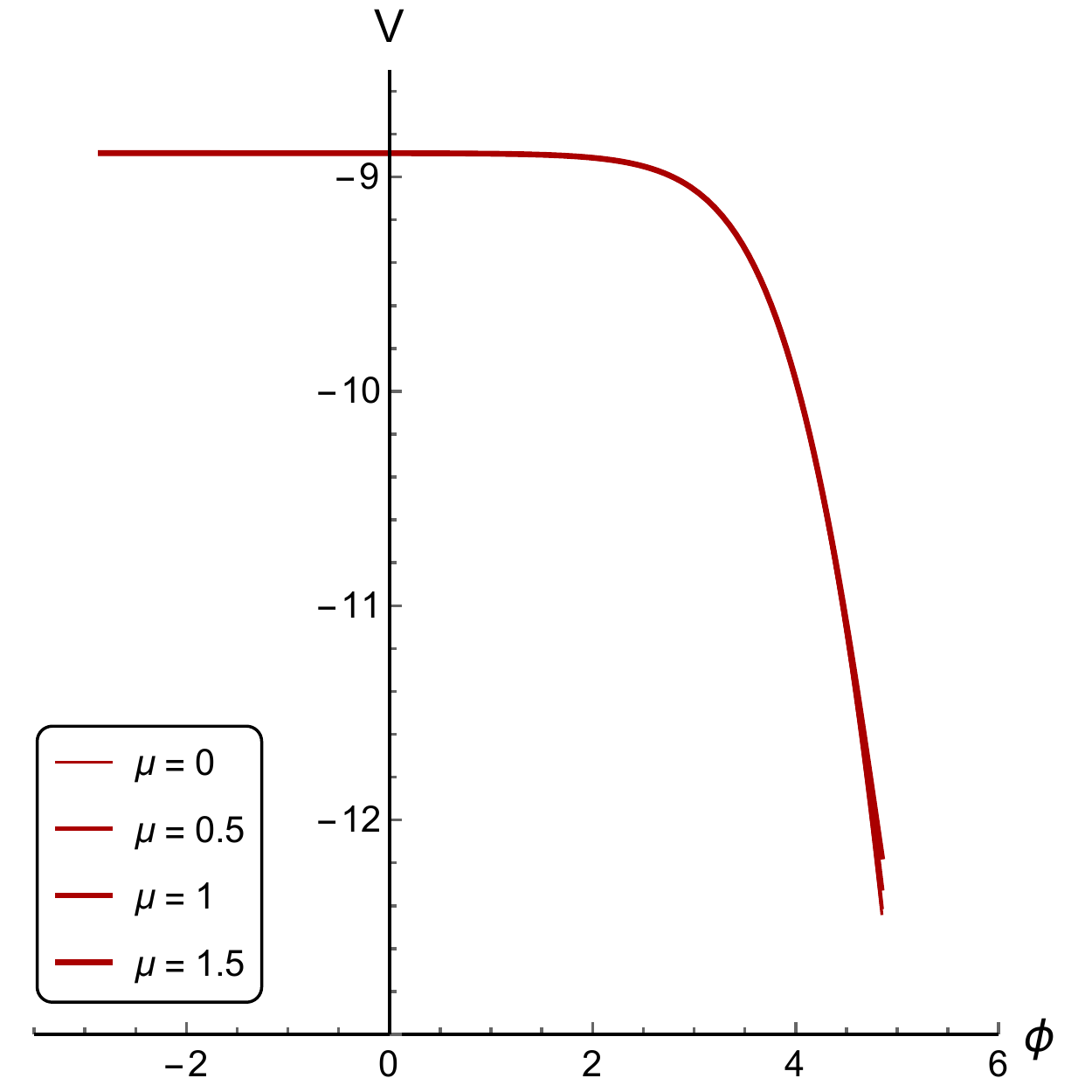} \quad
  \includegraphics[scale=0.35]{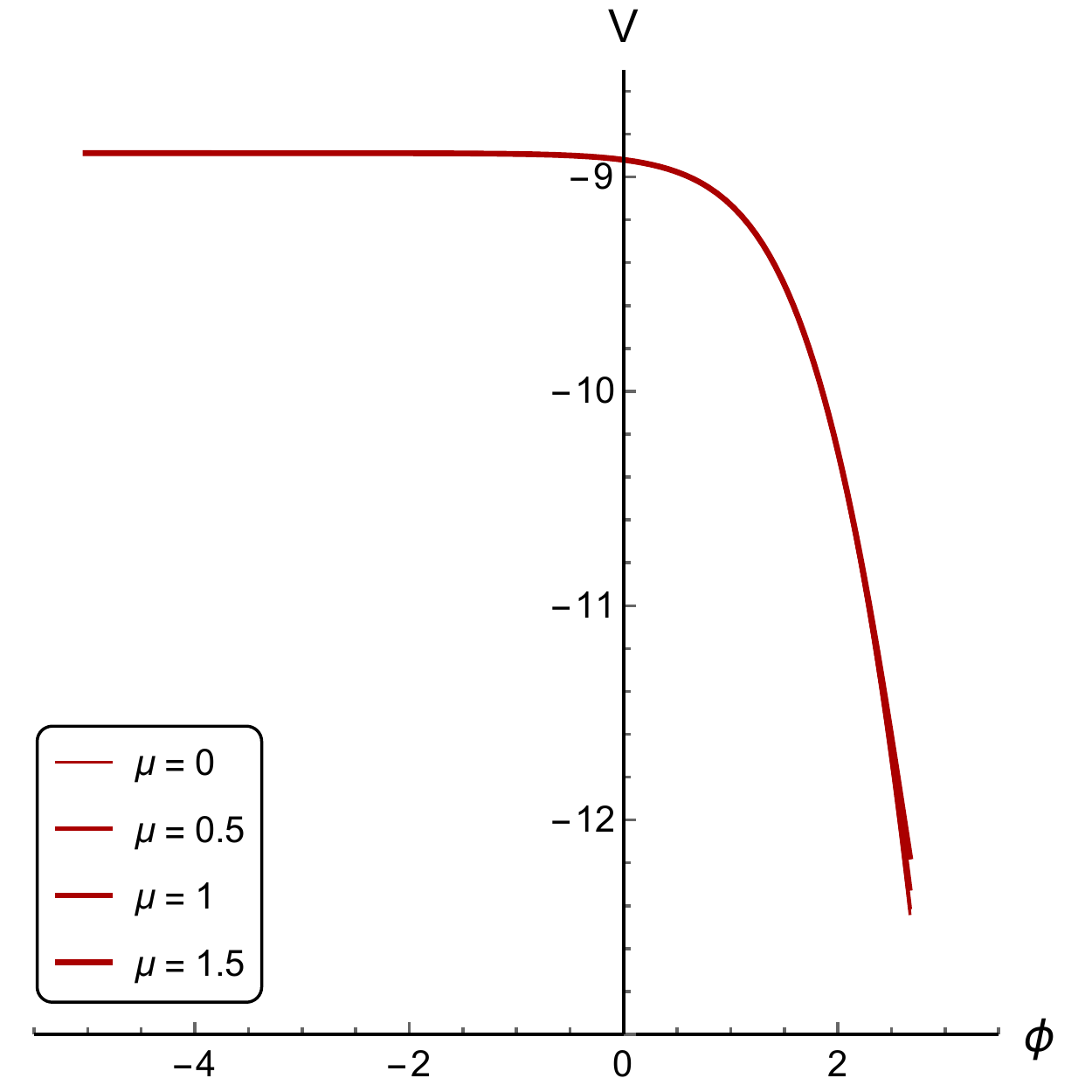} \quad
  \includegraphics[scale=0.35]{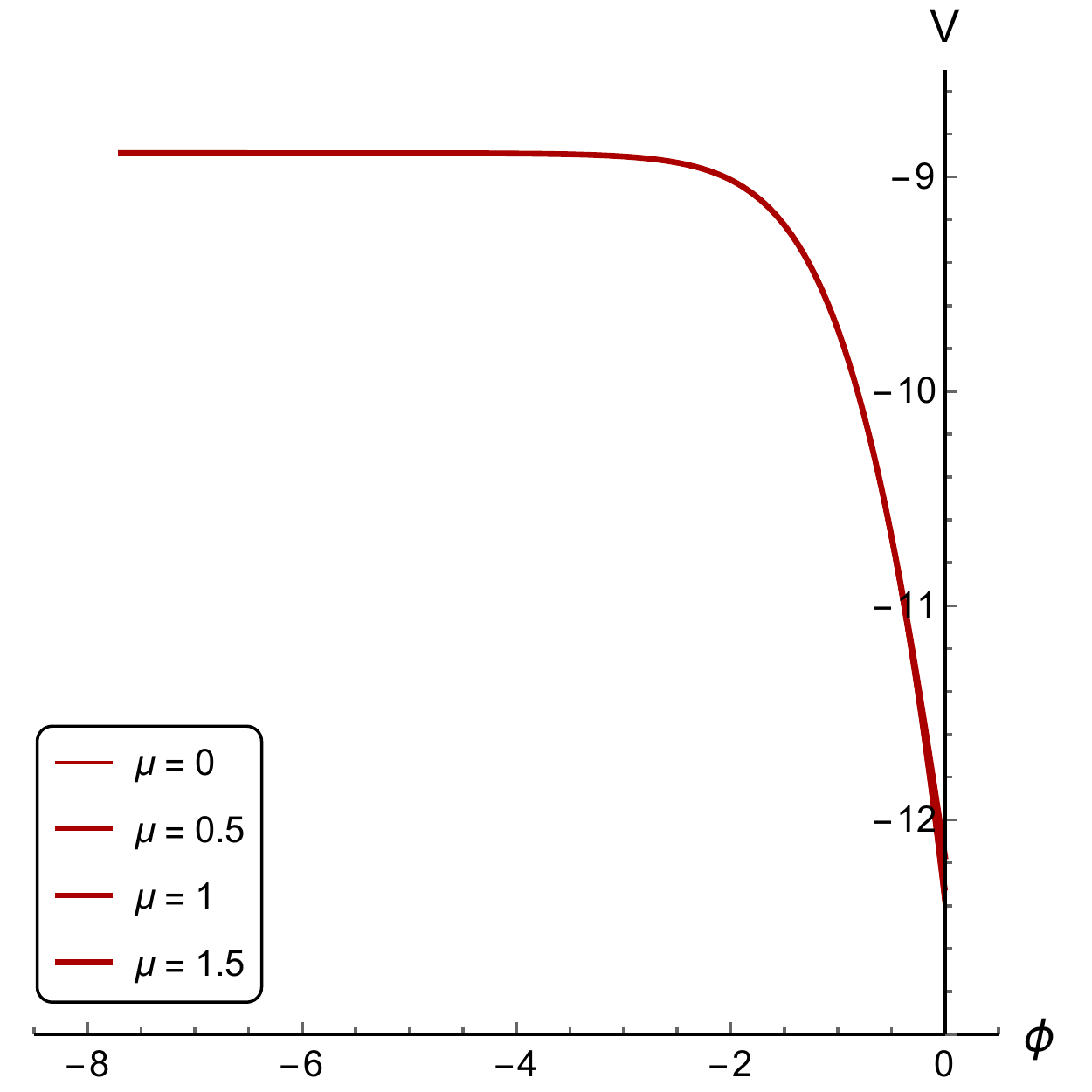} \\
  \includegraphics[scale=0.35]{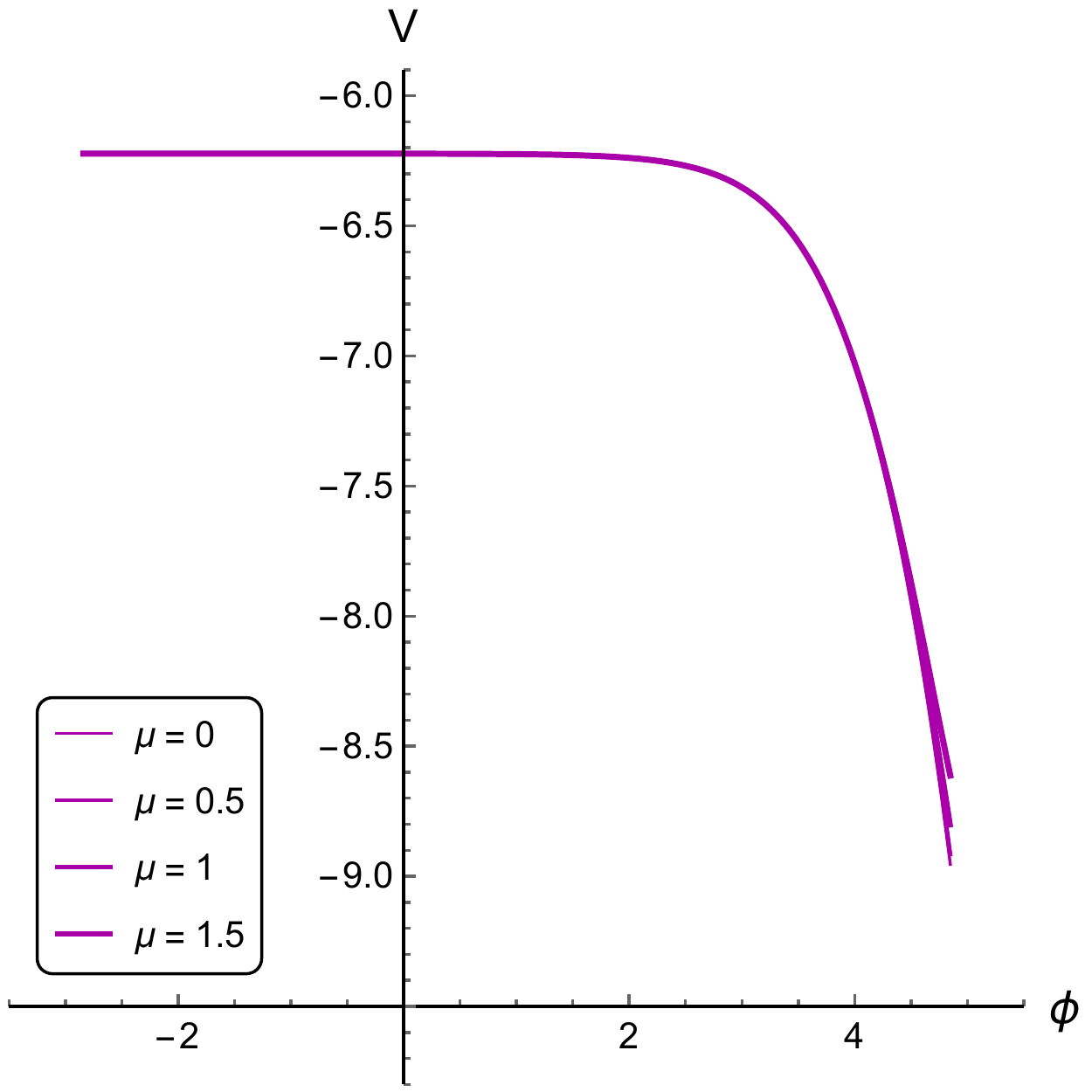} \quad
  \includegraphics[scale=0.35]{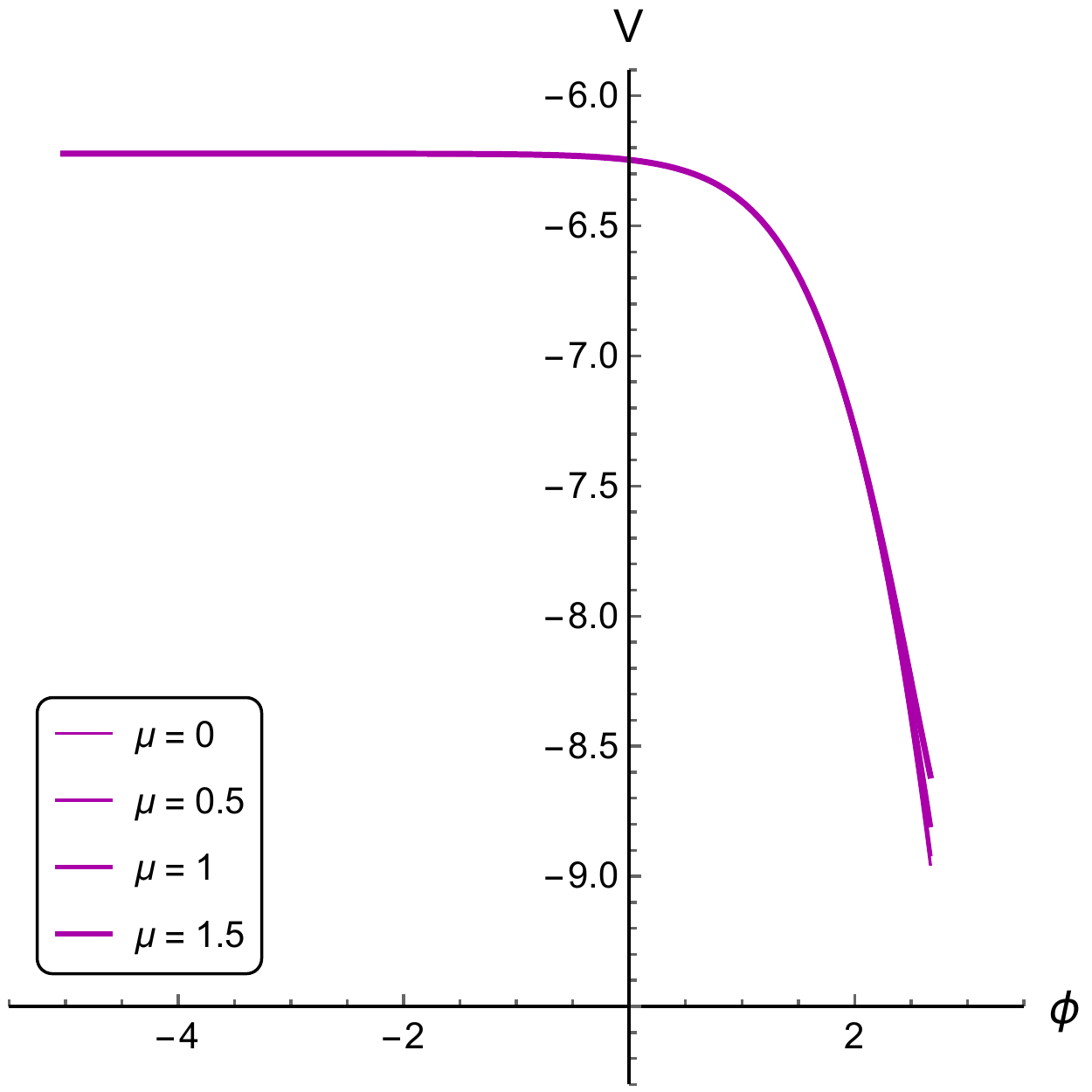} \quad
  \includegraphics[scale=0.35]{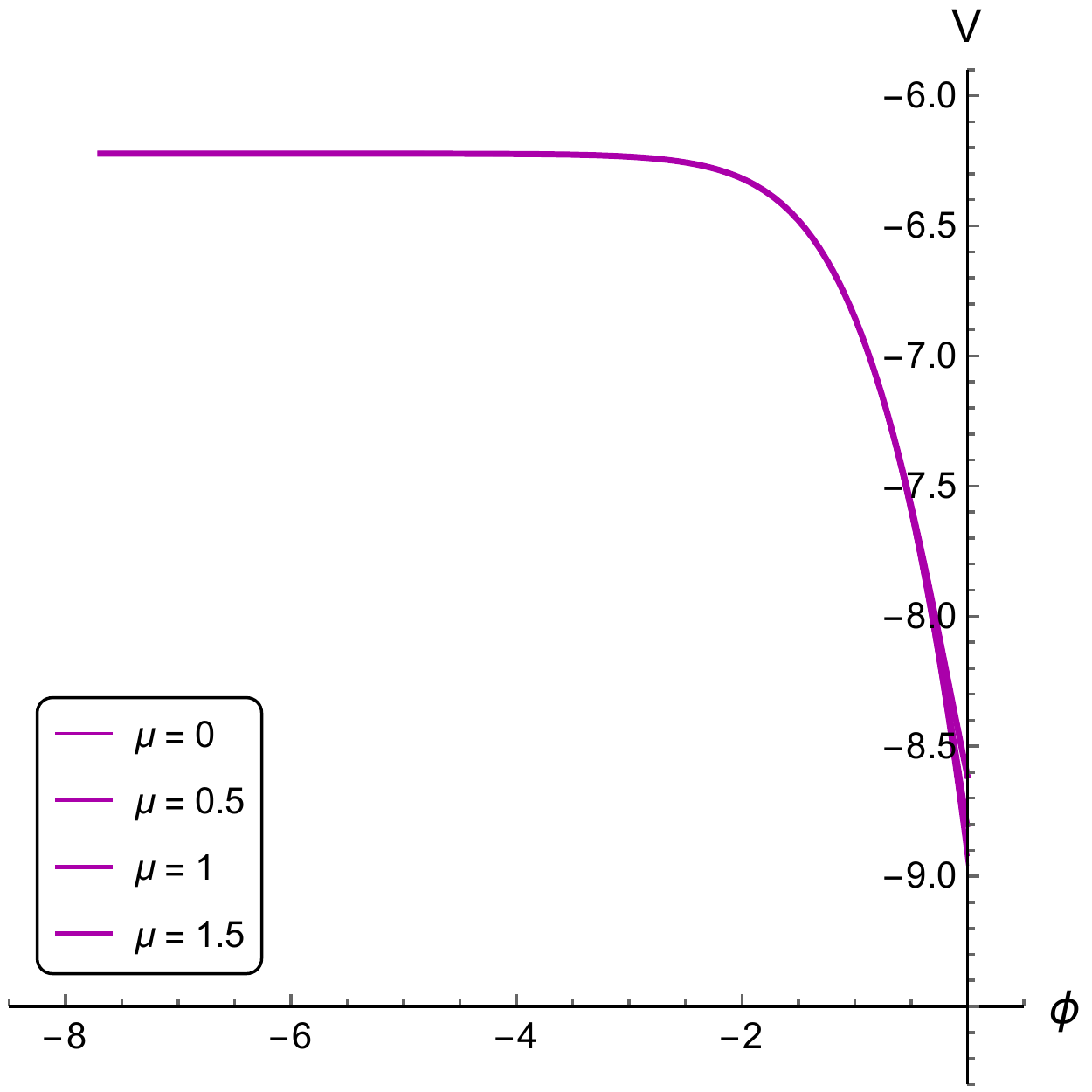} \\
  \includegraphics[scale=0.35]{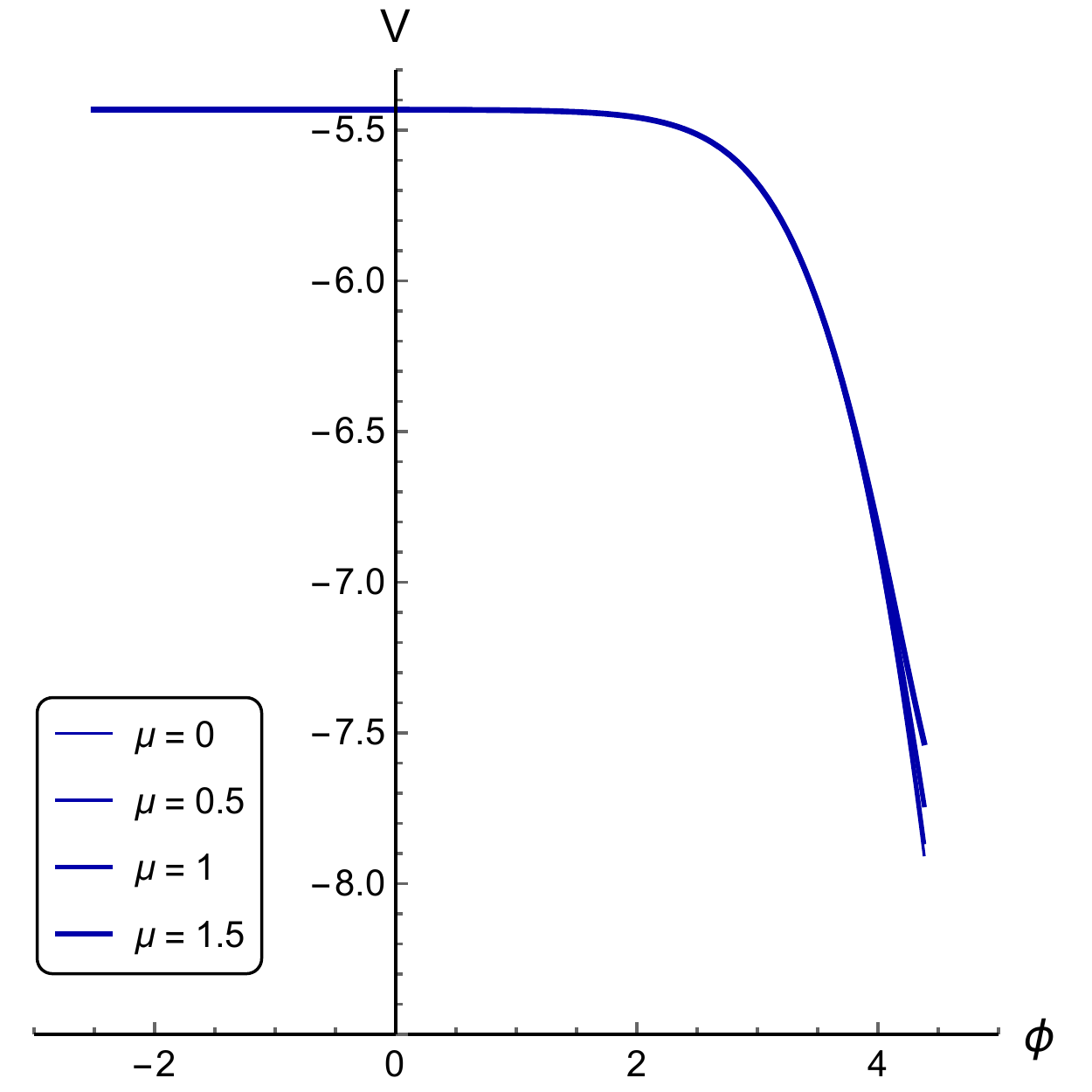} \quad
  \includegraphics[scale=0.35]{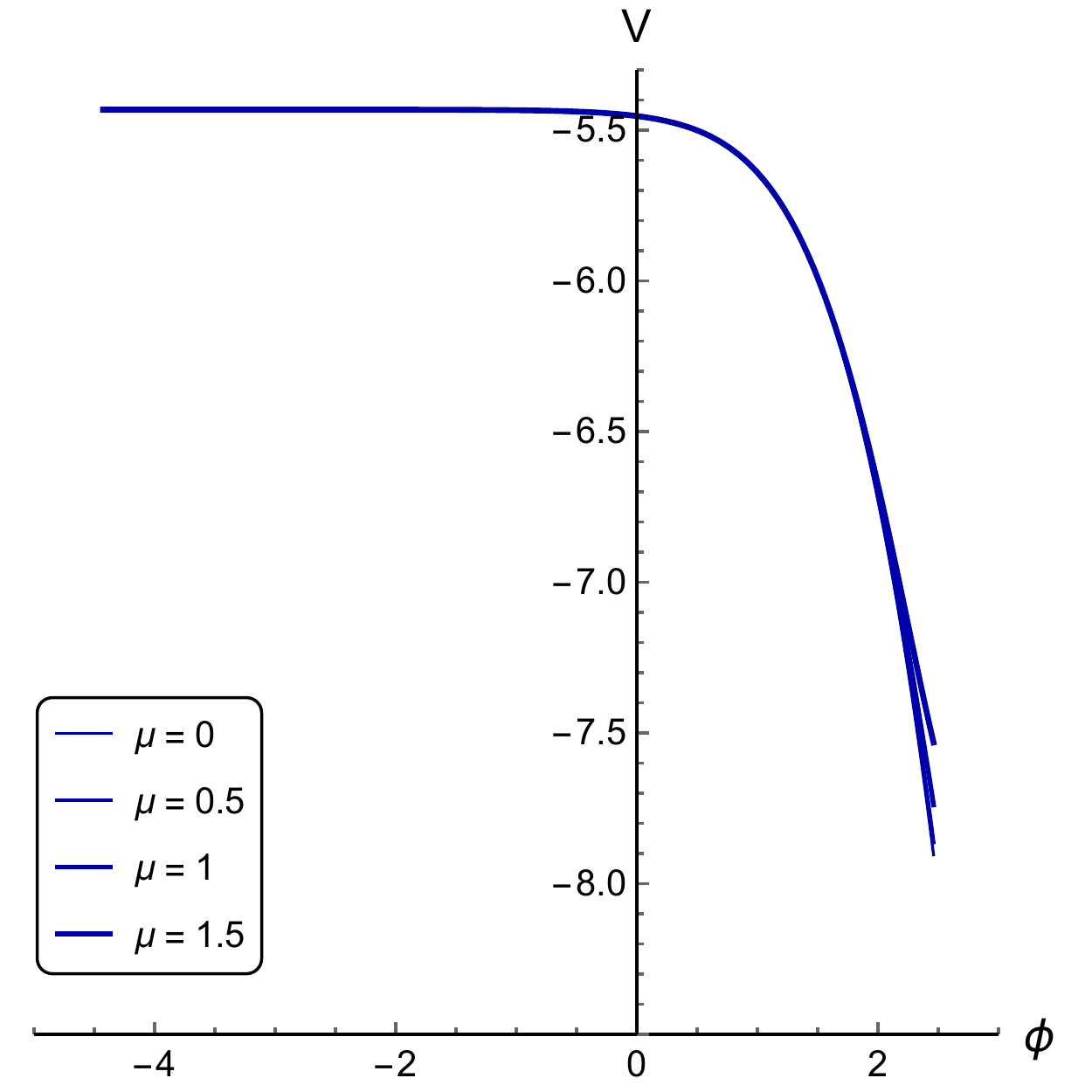} \quad
  \includegraphics[scale=0.35]{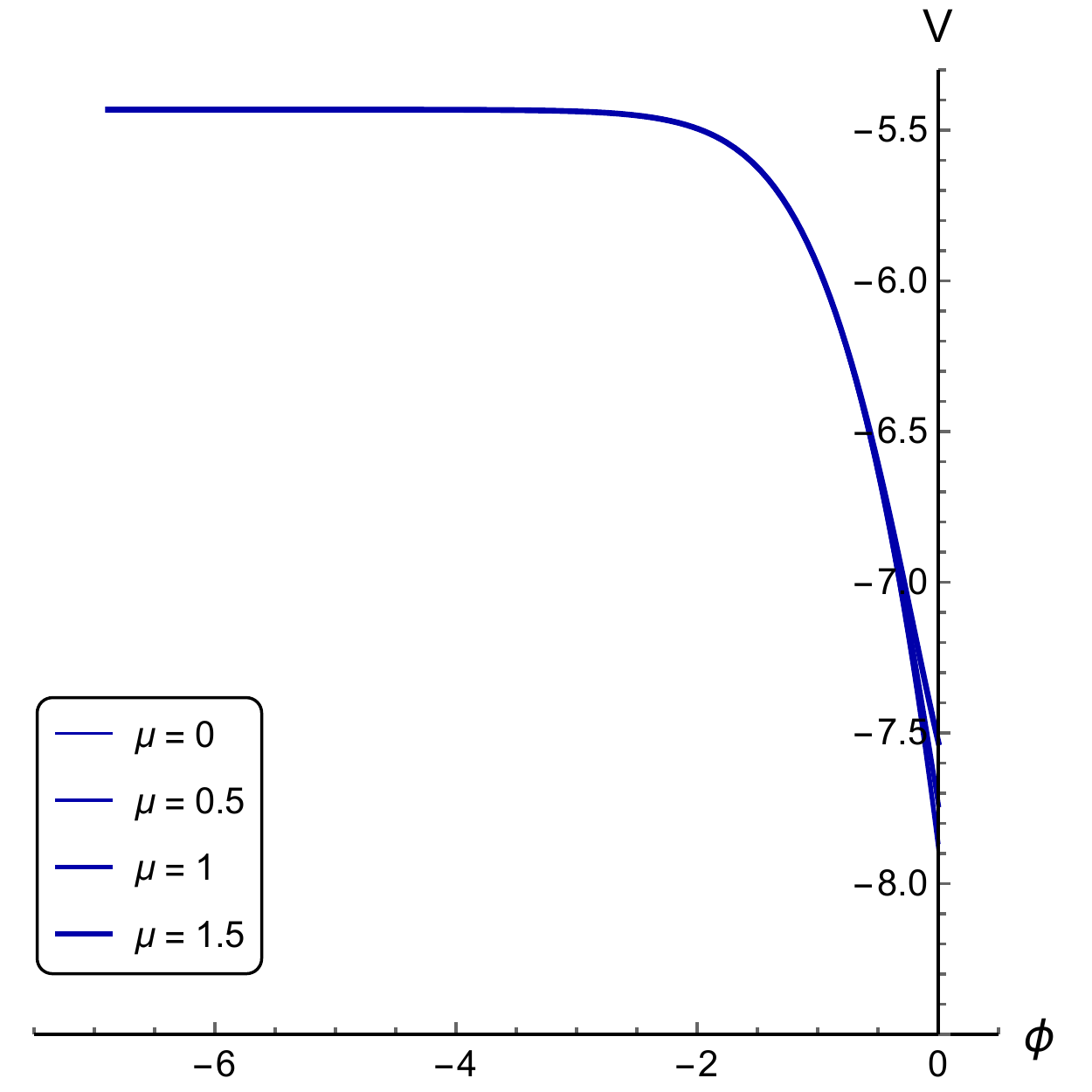} \\
  A \hspace{130pt} B \hspace{130pt} C
  \caption{Scalar field potential $V(\phi)$ for different $\mu$ in
    isotropic (1-st line) and anisotropic cases for $\nu = 1.5$  (2-nd
    line), $\nu = 3$ (3-rd line) and $\nu = 4.5$ (4-th line) for
    boundary $z_0 = 0.01$ (A), $z_0 = 0.1$ (B) and $z_0 = 1$ (C); $a =
    4.046$, $b = 0.01613$, $c = 0.227$ and $z_h = 1$. Vertical lines
    show the boundary of valid scalar field values $(\phi(z_h))$; for
    plots from the column C the axis $V(\phi)$ serves as such a
    boundary, because these are plots for $\phi(z_h) = 0$.}
  \label{Fig:Vphiz0nu}
\end{figure}

Solving \eqref{eq:1.25} and \eqref{eq:1.26} we get
\begin{gather}
  \begin{split}
    V = - \, \cfrac{3 g \, z^2 e^{-2{\cA}}}{L} &\left[
      {\cA}'' + 3 {\cA}'^2
      + \left( \cfrac32 \ \cfrac{g'}{g} - \cfrac{3 (\nu + 1)}{\nu z}
      \right) {\cA}' \right. - \\
    &\qquad \qquad \quad 
    - \left. \cfrac{1}{\nu z} \left( \cfrac{4 + 5 \nu}{6} \
        \cfrac{g'}{g} - \cfrac{2 (\nu + 1) \left( 2 \nu + 1 \right)}{3 
          \nu z} \right) + \cfrac{g''}{6 g} \right],
  \end{split}\label{eq:1.35}
\end{gather}
that transforms into (2.32) from \cite{Yang-2017} for $L = \nu =
1$. We can see that potential $V(z)$ \eqref{eq:1.35} does not depend
on the boundary conditions or the dilaton field, as equation
\eqref{eq:1.26} inculdes the dilaton field's derivative, not the
dilaton itself. But $V(\phi)$ behavoir does depend on them.

On Fig.\ref{Fig:Vphiz0nu} $V(\phi,\mu)$-curves for different
boundaries $z_0$ are presented. Scalar potential is plotted till the
horizon, i.e. for $\phi(\epsilon \to 0) \le \phi \le \phi(z_h)$. We
can see that the increasing of the boundary $z_0$ doesn't actually
affect the shape of $V(\phi)$, just shifts the whole curve to the left, to
lower $\phi$-values. Chemical potential influence on the scalar field
is visible, but weak near the horizon only and therefore can be
neglected.


\begin{figure}[b!]
  \centering
  \includegraphics[scale=0.5]{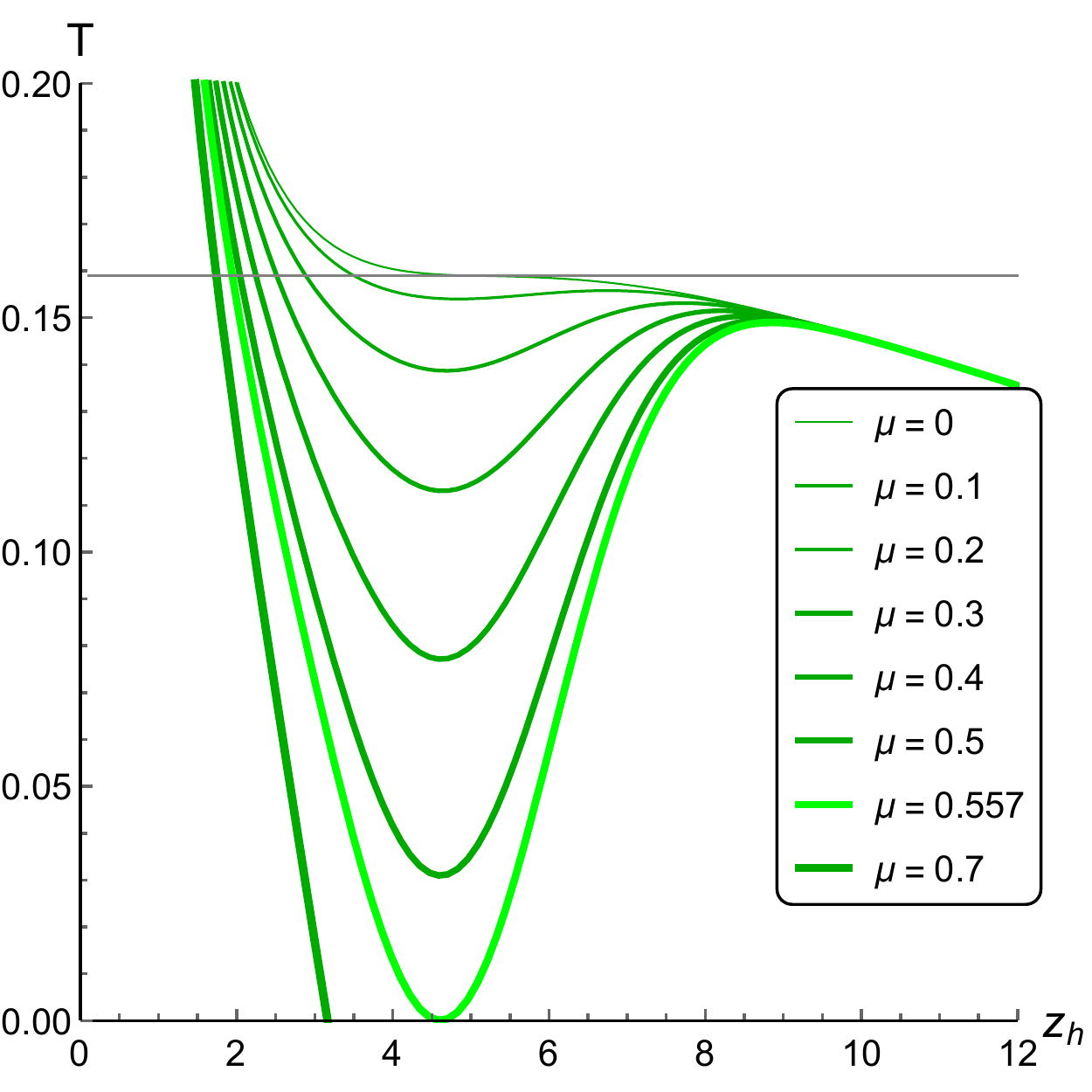} \qquad
  \includegraphics[scale=0.5]{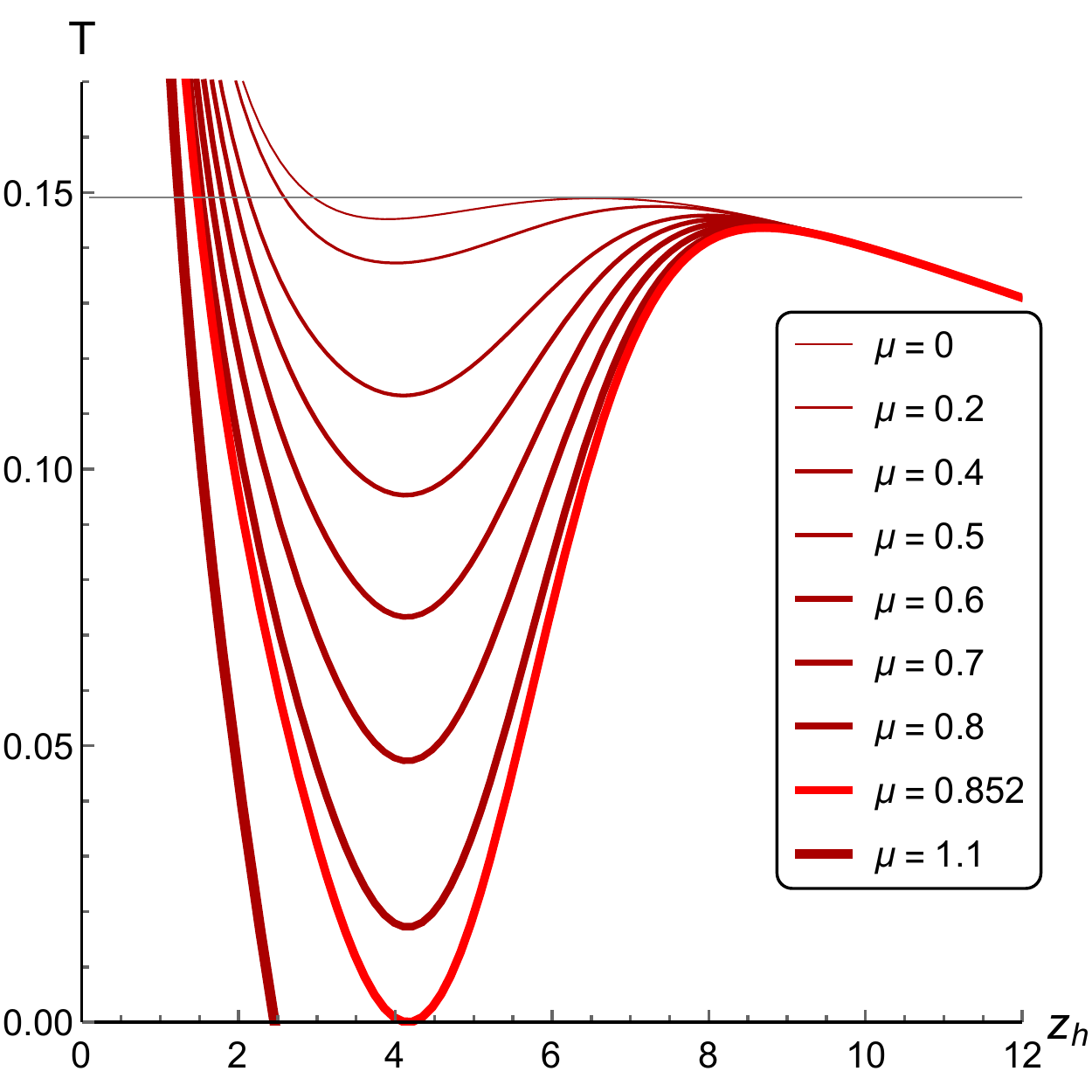} \\
  A \hspace{220pt} B \\
  \includegraphics[scale=0.5]{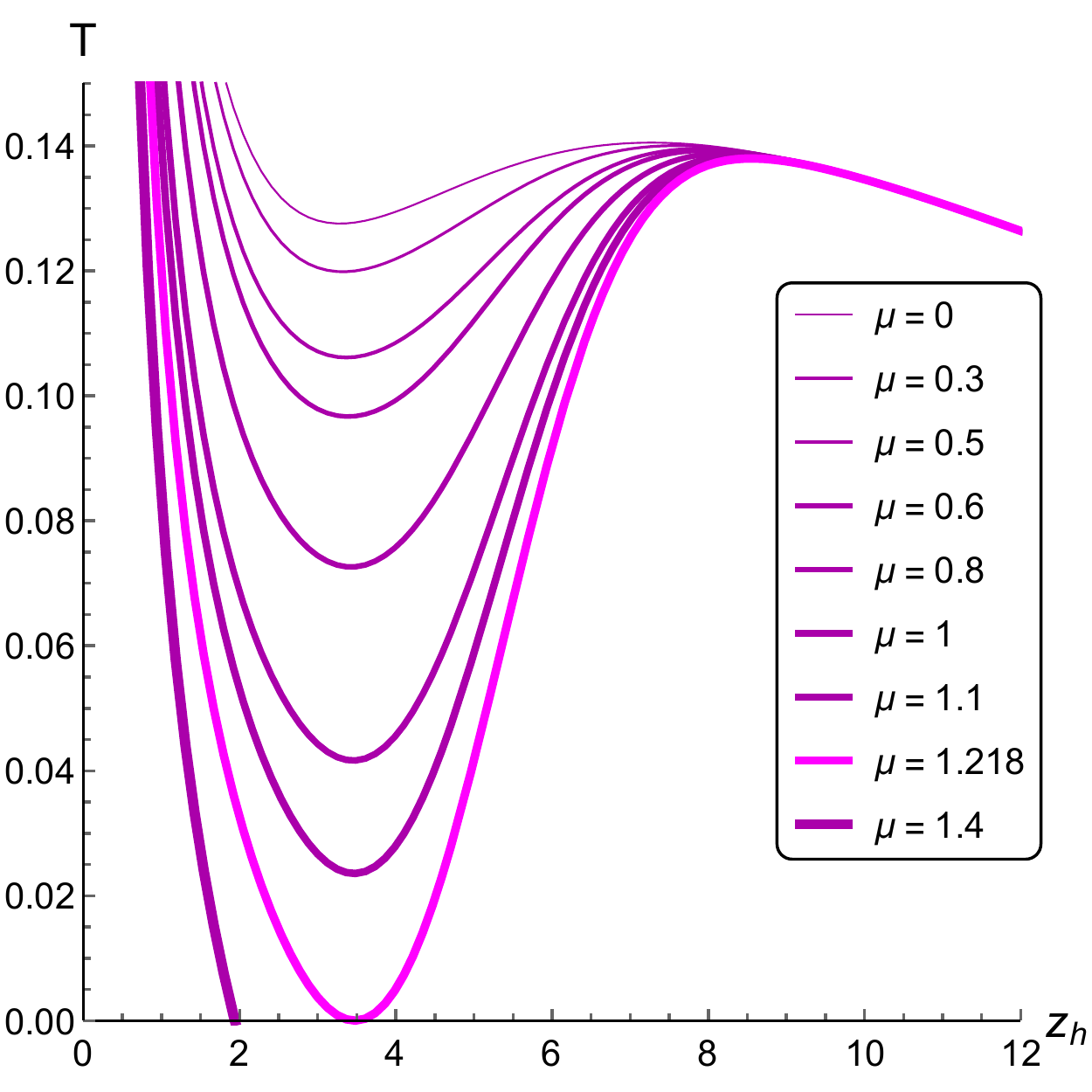} \qquad
  \includegraphics[scale=0.5]{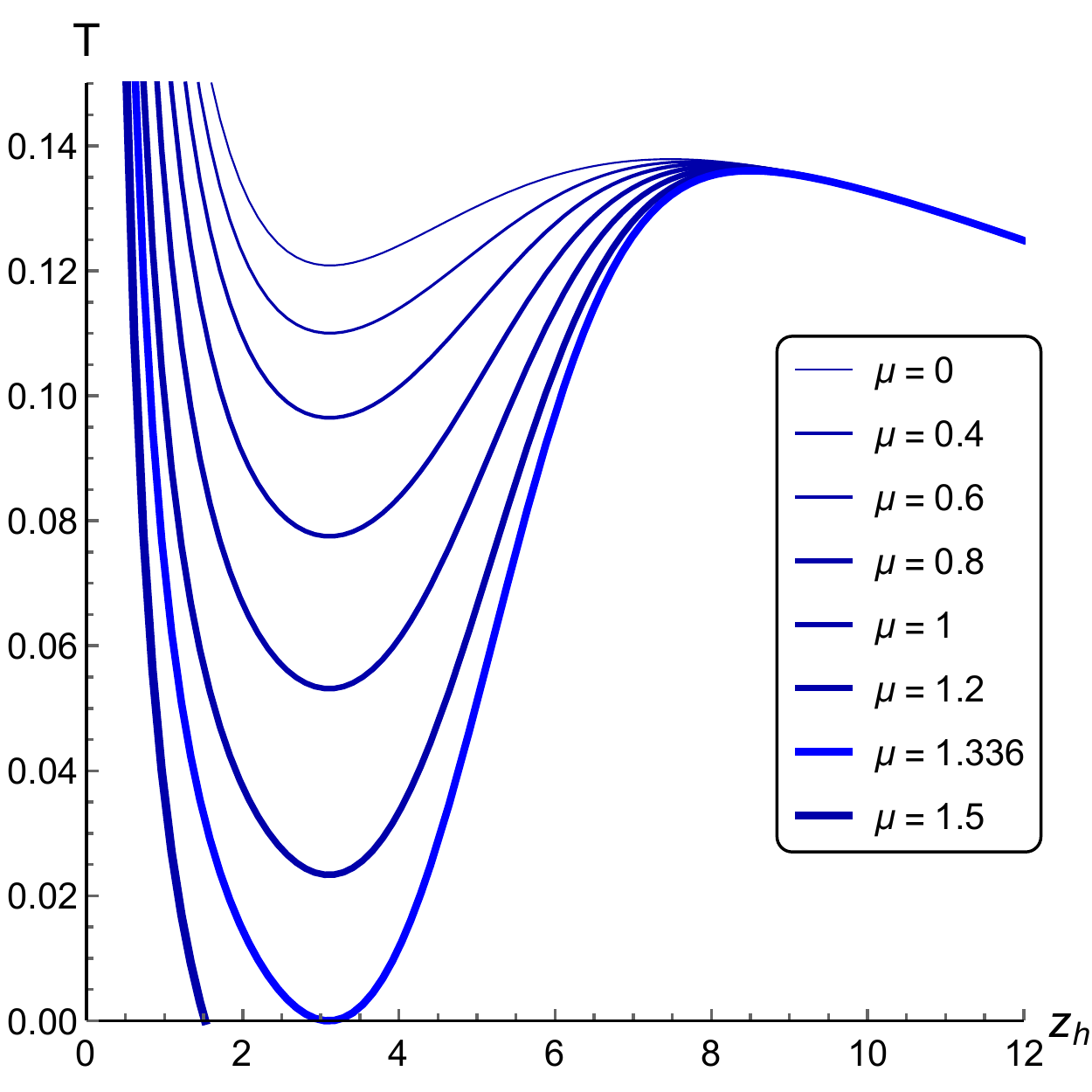} \\
  C \hspace{220pt} D
  \caption{Temperature as function of horizon for different $\mu$ in
    isotropic (A) and anisotropic cases for $\nu = 1.5$ (B), $\nu = 3$
    (C), $\nu = 4.5$ (D); $a = 4.046$, $b = 0.01613$, $c = 0.227$.}
  \label{Fig:TS}
\end{figure}

\section{Confinement/deconfinement phase transition}

\subsection{Temperature and entropy} \label{Thermodynamics}

For metric \eqref{eq:1.20} and the warp factor \eqref{eq:1.04}
temperature can be written as:
\begin{gather}
  \begin{split}
    T &= \cfrac{|g'|}{4 \pi} \, \Bigl|_{z=z_h} = \cfrac{1}{4 \pi}
    \left|
      - \ \cfrac{\left( 1 + b z_h^2 \right)^{3a}
        z_h^{1+\frac{2}{\nu}}}{
      \int_0^{z_h} \left(1 + b \, \xi^2 \right)^{3a}
      \xi^{1+\frac{2}{\nu}} \, d \xi} \left[1
      - \cfrac{2 \mu^2 c \, e^{2cz_h^2}}{L^2 \left( 1 - e^{cz_h^2}
        \right)^2} \right. \right. \times \\
      &\, \left. \left. \times
      \left( 1 - e^{-cz_h^2} \cfrac{
          \int_0^{z_h} e^{c\xi^2} \left(1 + b \, \xi^2 \right)^{3a}
          \xi^{1+\frac{2}{\nu}} \, d \xi}{
          \int_0^{z_h} \left(1 + b \, \xi^2 \right)^{3a}
          \xi^{1+\frac{2}{\nu}} \, d \xi} \right)
      \int_0^{z_h} \left(1 + b \, \xi^2 \right)^{3a}
      \xi^{1+\frac{2}{\nu}} \, d \xi
    \right] \right|.
  \end{split}\label{eq:2.03}
\end{gather}

Fig.\ref{Fig:TS}.A shows that in isotropic case for $\mu = 0$
temperature is a monotonically decreasing function of
horizon. Increasing chemical potential makes $T(z_h)$-function
three-valued at some interval, and local minimum appeares. As we will
see below, this is directly related to the Hawking-page-like
confinement/deconfinement phase transition. Indeed, in the isotropic
case first-order phase transition for light quarks shouldn't exist
near zero chemical potential and we should see a crossover
(Fig.\ref{Hybrid2}.B). The larger chemical potential is the lesser
temperature value at this local minimum becomes. For $\mu \approx
0.557$ local minimum temperature $T_{min} = 0$ and second horizon
appears.

\begin{figure}[b!]
  \centering
  \includegraphics[scale=0.46]{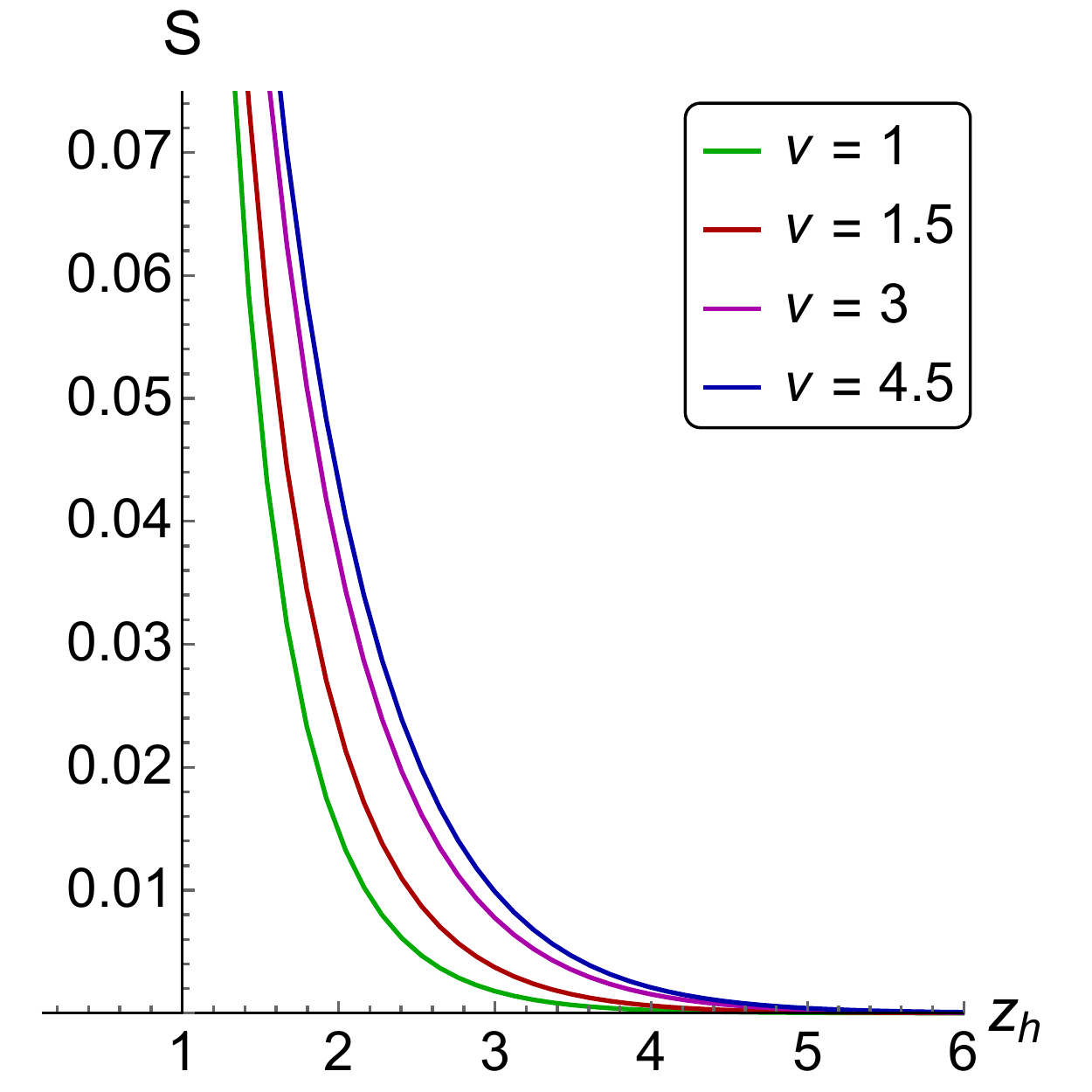} \
  \includegraphics[scale=0.71]{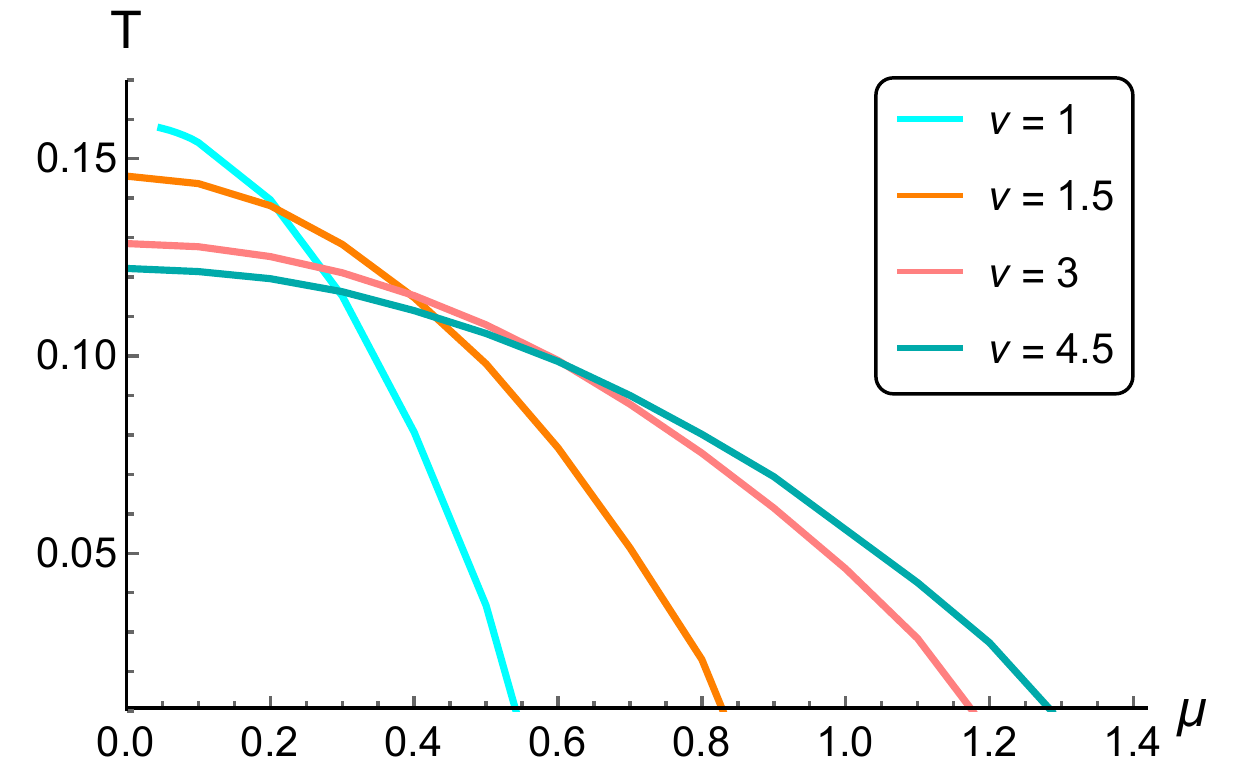} \\
  A \hspace{230pt} B
  \caption{Entropy as function of horizon (A) and
    Hawking-Page-like phase transition lines $T(\mu)$ for isotropic
    ($\nu = 1$) and anisotropic ($\nu = 1.5, \ 3, \ 4.5$) cases (B);
    $a = 4.046$, $b = 0.01613$, $c = 0.227$.}
  \label{Fig:ETmu}
\end{figure}

In the anisotropic case global behavior of temperature persists, but
it is a three-valued function for $\mu = 0$ already and the second
horizon appears at about $\mu \approx 0.852$ for $\nu = 1.5$
(Fig.\ref{Fig:TS}.B), $\mu \approx 1.218$ for $\nu = 3$
(Fig.\ref{Fig:TS}.C). and $\mu \approx 1.336$ for $\nu = 4.5$
(Fig.\ref{Fig:TS}.D). This indicates that the Hawking-Page-like phase
transition line should exist even in the absence of chemical
potential, and Fig\ref{Fig:ETmu}.B confirms this.

\begin{figure}[t!]
  \centering
  \includegraphics[scale=0.54]{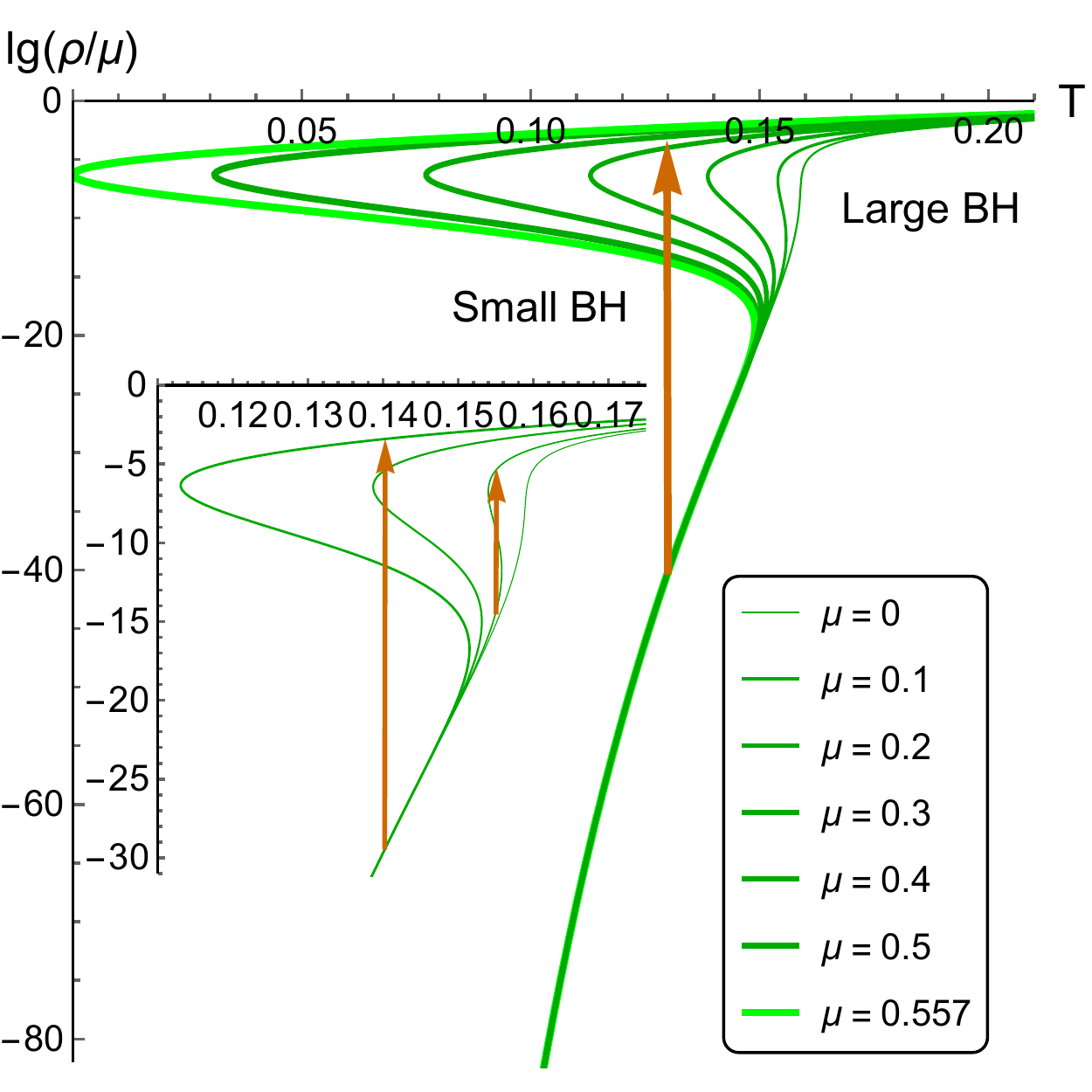} \qquad
  \includegraphics[scale=0.54]{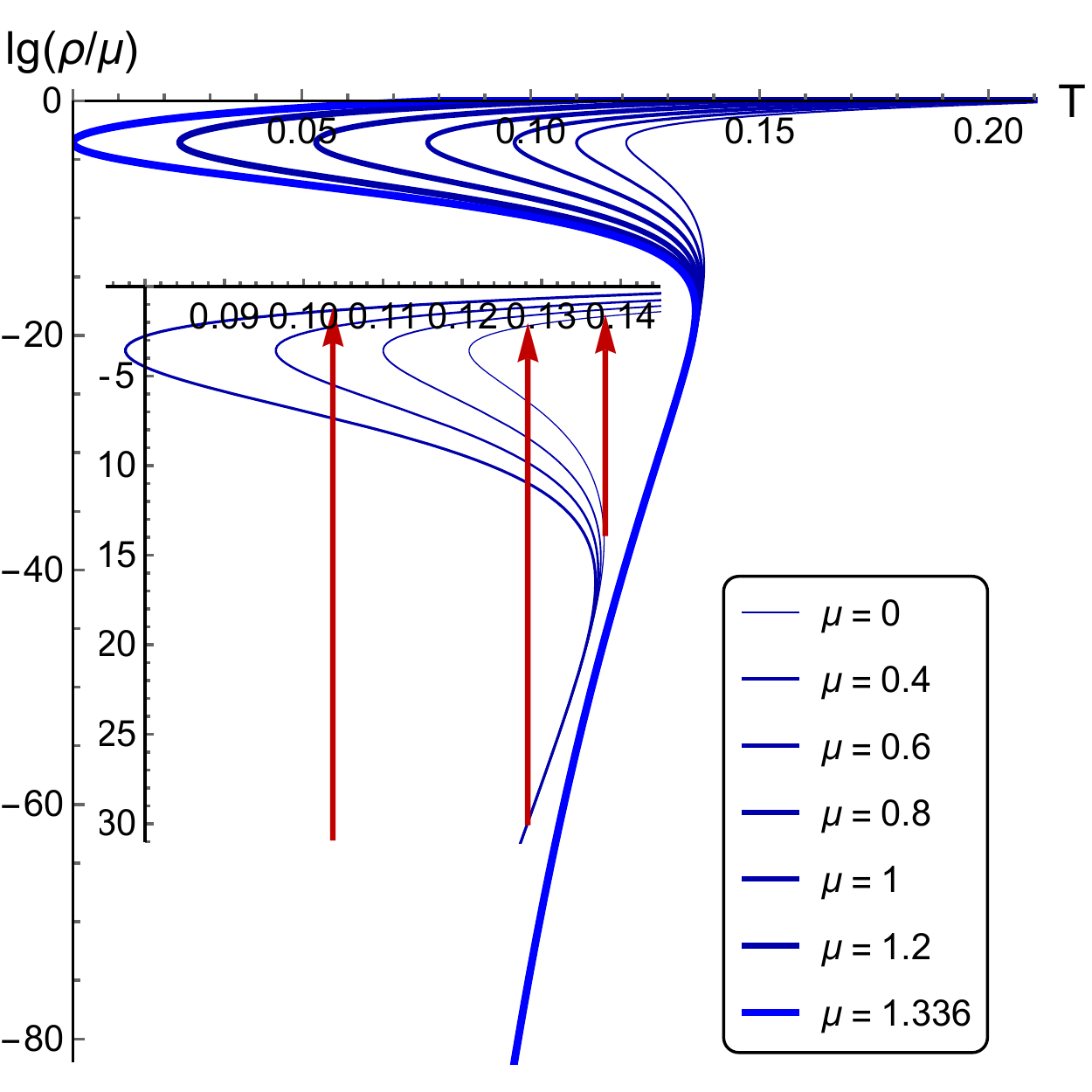} \\
  A \hspace{230pt} B
  \caption{Density $\rho/\mu(T)$ in logarithmic scale for different
    $\mu$ for $\nu = 1$ (A) and $\nu = 4.5$ (B); $a = 4.046$, $b =
    0.01613$, $c = 0.227$. Inner plots show the fragments of main
    plots zoomed.}
  \label{Fig:rhoT}
\end{figure}

For metric \eqref{eq:1.20} and the warp factor \eqref{eq:1.04} entropy
becomes
\begin{gather}
  s = \left( \cfrac{L}{z_h} \right)^{1+\frac{2}{\nu}} \cfrac{\left( 1
      + b z_h^2 \right)^{-3a}}{4} \, . \label{eq:2.04}
\end{gather}
It decreases monotonocally and quickly with horizon growth
(Fig.\ref{Fig:ETmu}.A).

The BH-BH phase transition caused by three-valued temperature function
produces a  jump of density $\rho$, that is a coefficient in $A_t$
expansion:
\begin{gather}
  A_t = \mu - \rho z^2 + \dots 
  = \mu - \cfrac{c \mu z^2}{1 - e^{c z_h^2}} + \dots, \qquad
  \rho = - \, \cfrac{c \mu}{1 - e^{c z_h^2}} \, . \label{eq:2.041}
\end{gather}
On Fig.\ref{Fig:rhoT} $\rho/\mu$ ratio as a function of temperature
for primary isotropic solution ($\nu = 1$, Fig.\ref{Fig:rhoT}.A) and
anisotropic solution ($\nu = 4.5$, Fig.\ref{Fig:rhoT}.B) are
plotted in logarithmic scale. Vertical red arrows show the BH-BH
transition direction. Function $\rho/\mu$ is a three-valued function
of $T$ as expected, and we can see that collapse from small black
holes (larger $z_h$) to the large ones (smaller $z_h$) is accompanied
by a sharp rise of the density for any appropriate chemical
potential.

\begin{figure}[b!]
  \centering
  \includegraphics[scale=0.65]{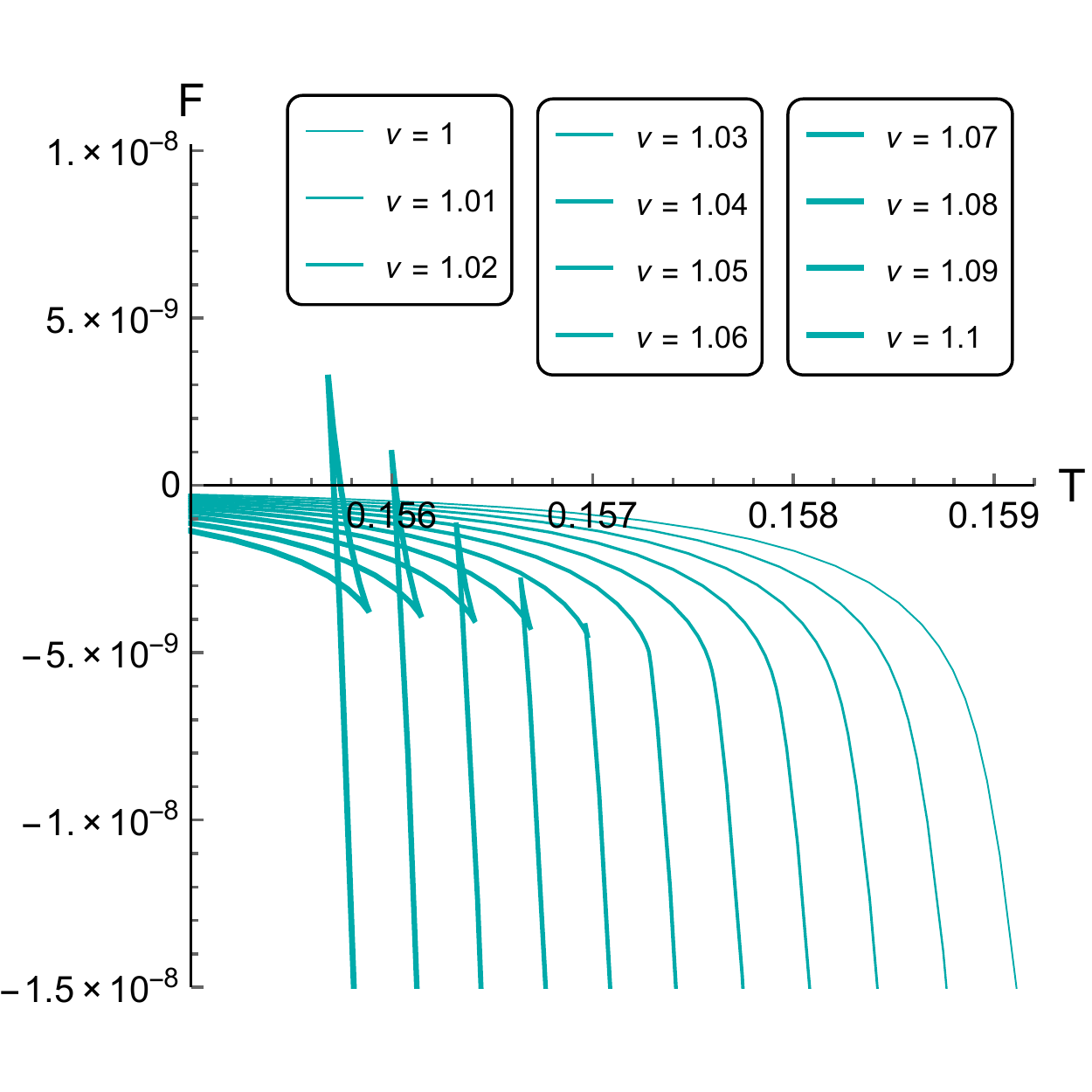}
  \caption{Free energy as function of temperature $F(T)$ for $\mu = 0$
    in isotropic ($\nu = 1$) and slightly anisotropic ($\nu = 1.01, \
    1.02, \ 1.03, \ 1.04, \ 1.05, \ 1.06, \ 1.07, \ 1.08, \ 1.09, \
    1.1$) cases; $a = 4.046$, $b = 0.01613$, $c = 0.227$.}
  \label{Fig:FTnu}
\end{figure}

To get Hawking-Page-like transition line (BH-BH phase transition) we
need to consider free energy as a function of temperature:
\begin{gather}
  F =  \int_{z_h}^{z_{h_2}} s \, T' dz. \label{eq:2.05}
\end{gather}
While $T \ge 0$, i.e for small chemical potentials, we integrate to
$z_{h_2} = \infty$. When second horizon where $T = 0$ appears, one
should integrate to it's value, i.e. to $z_{h_2} = 4.609$ for $\nu =
1$, $\mu = 0.557$, to $z_{h_2} = 4.163$ for $\nu = 1.5$, $\mu = 0.852$
and to $z_{h_2} = 3.102$ for $\nu = 4.5$, $\mu = 1.336$. These
conditions determine the end-point of the phase diagram, i.e. maximum
permissible chemical potential $\mu_{max}$ for chosen $\nu$. Thus
increasing anisotropy parameter $\nu$ allows larger chemical
potentials, but reduces the temperature.

On Fig.\ref{Fig:ETmu}.B Hawking-Page-like phase transition for $\nu =
1, \, 1.5, \, 3, \, 4.5$ is depicted. In isotropic case BH-BH phase
transition starts from a critical point $\mu_{c} = 0.04779$, $T_{c} =
0.1578$ that fully coincides with previous result in
\cite{Yang-2017}.

For the Hawking-Page-like phase transition the free energy should be a
multi-valued function of temperature. Graphically it is displayed as a
``swallow-tail''. The point where the free energy curve intersects
itself determines the Hawking-Page-like phase transition
temperature. On Fig.\ref{Fig:FTnu} the free energy as a function of
temperature for different values of $\nu$ in the absence of chemical
potential is plotted. We see a smooth free energy curve for $1 \le \nu
\le 1.04$, therefore no self-intersection and no Hawking-Page-like
phase transition for $\mu = 0$ exists. For $\nu = 1.05$ an obtuse
angle appears on the curve -- this is a germ of the
``swallow-tail''. The larger $\mu$ becomes the more pronounced the
``swallow-tail'' is. So turning the anisotropy on causes the gap
between $\mu = 0$ and the starting point of the Hawking-Page-like
line on the confinement/deconfinement phase diagram to close. Slight
anisotropy with $\nu = 1.05$ is enough to make this type of phase
transition exist for all chemical potential values $0 \le \mu \le
\mu_{max}$.


\subsection{Temporal Wilson loops} \label{Wilson loop}

Following \cite{ARS-2019plb} we consider temporal Wilson loops in
anisotropic background to calculate the parameters of Cornell
potential and find the conditions of confinement/decon\-fine\-ment
phase transition. Calculations for the Wilson loops were done in the
string frame. To determine the confinement/deconfinement condition and
the string tension $\sigma_{DW}$ let us study the asymptotics of the
Nambu-Goto action for the test string $S$ at large character length of
the string $\ell$. At $\ell \to \infty$ one has 
\begin{gather}
  S \sim \sigma_{DW}\ \ell.
\end{gather}
Like it was in \cite{ARS-2019plb}, we take the world sheet
parameterized as
\begin{gather*}
  X^{0} \equiv t, \
  X^{1} \equiv x = \xi \cos\theta, \
  X^{2} \equiv y_{1} =\xi\sin\theta, \
  X^{3} \equiv y_{2} = const, \
  X^{4} \equiv z = z(\xi). 
\end{gather*}
Angle $\theta$ defines orientation of the Wilson loop in the
considered background. The string tension
\begin{gather}
  \sigma(z_{DW},z_0) = \cfrac{ b(z_{DW})}{z^2 } \
  e^{\sqrt{\frac{2}{3}}\phi(z_{DW},z_0)}\sqrt{g(z_{DW})
    \left(z^{2-\frac{2}{\nu}} \sin^2(\theta) + \cos^2(\theta)\right)},
\end{gather}
obviously depends on the dilation field boundary conditions. Point
$z_{DW}$ is the position of the dynamical wall, where $\partial
\sigma/\partial z = 0$. In isotropic case this result coinsides with
\cite{Yang-2017}.
\begin{figure}[t!]
  \centering
  \includegraphics[scale=0.4]{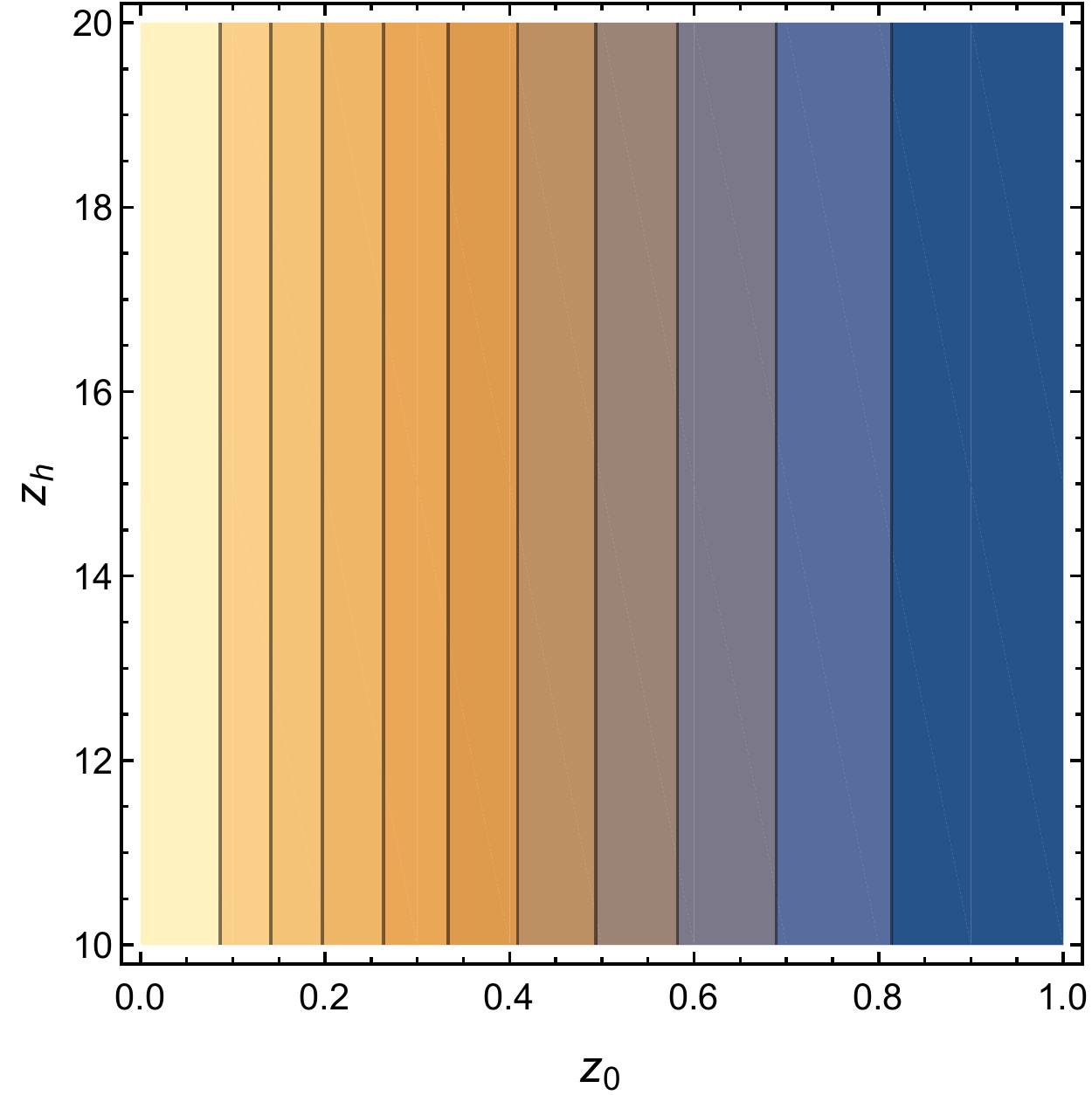} \
  \includegraphics[scale=0.4]{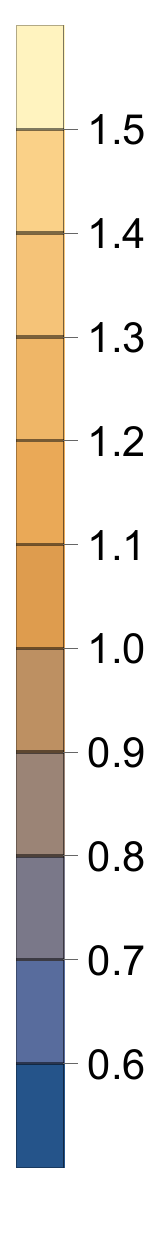} 
  \begin{picture}(0,0) \put(-20,150){$\sigma$}
    \put(-115,-25){$\sigma$} \put(83,-25){$ \sigma $}
  \end{picture}\\
  A \\ \ \\
  \includegraphics[scale=0.4]{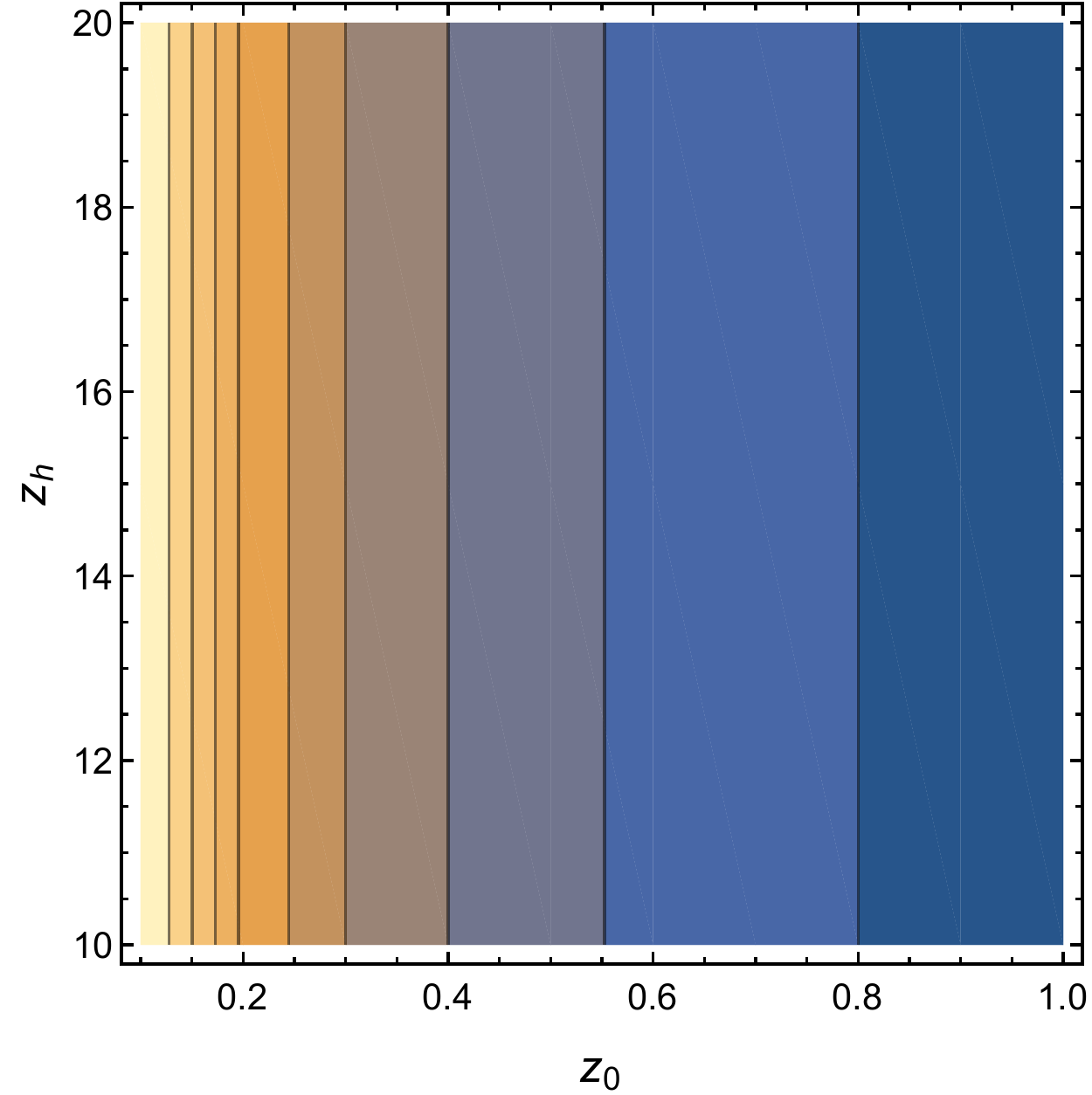} \
  \includegraphics[scale=0.4]{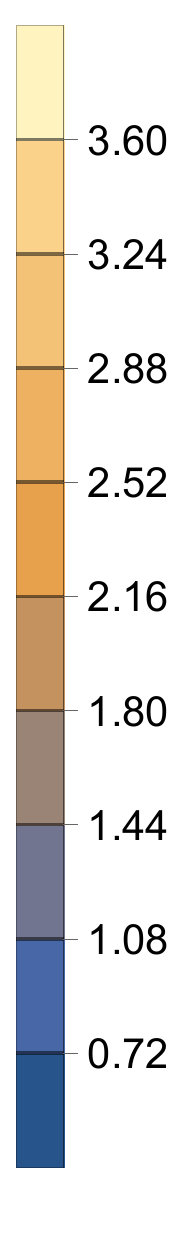} \qquad 
  \includegraphics[scale=0.4]{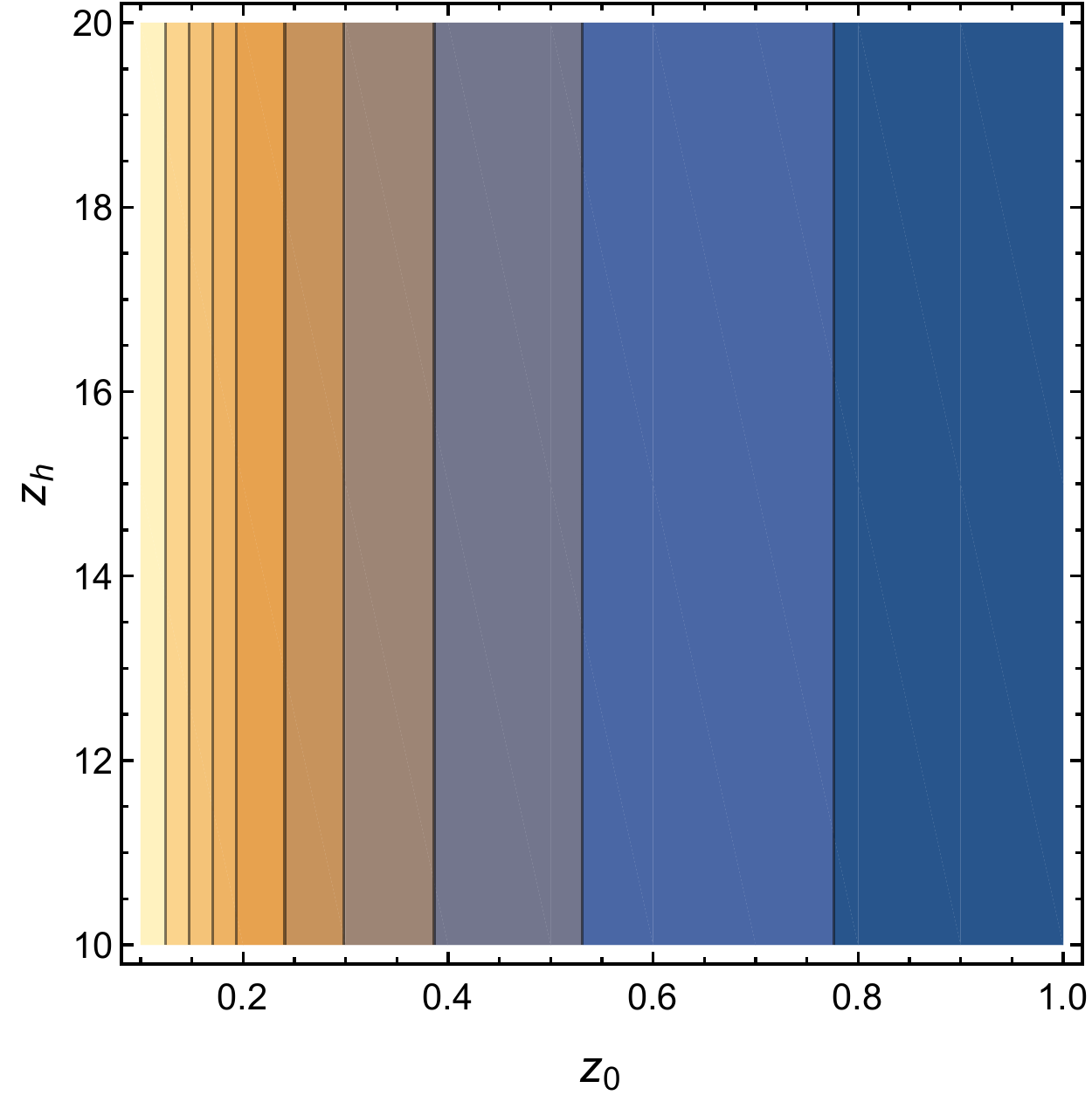} \
  \includegraphics[scale=0.4]{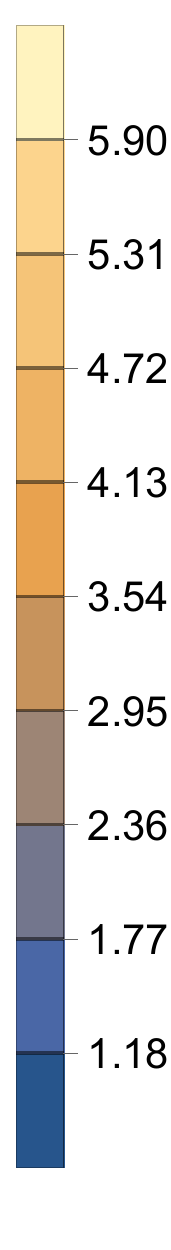}\\
  B \hspace{180pt} C
  \caption{Contour plot of the string tension as function of $z_{0}$
    and $z_h$ for $\mu = 0$, $\nu = 1$ (A), and for $\mu = 0.2$, $\nu
    = 4.5$, longitudinal (B) and transversal (C); $a = 4.046$, $b =
    0.01613$, $c = 0.227$.}
  \label{Fig:sigmazz0}
\end{figure}
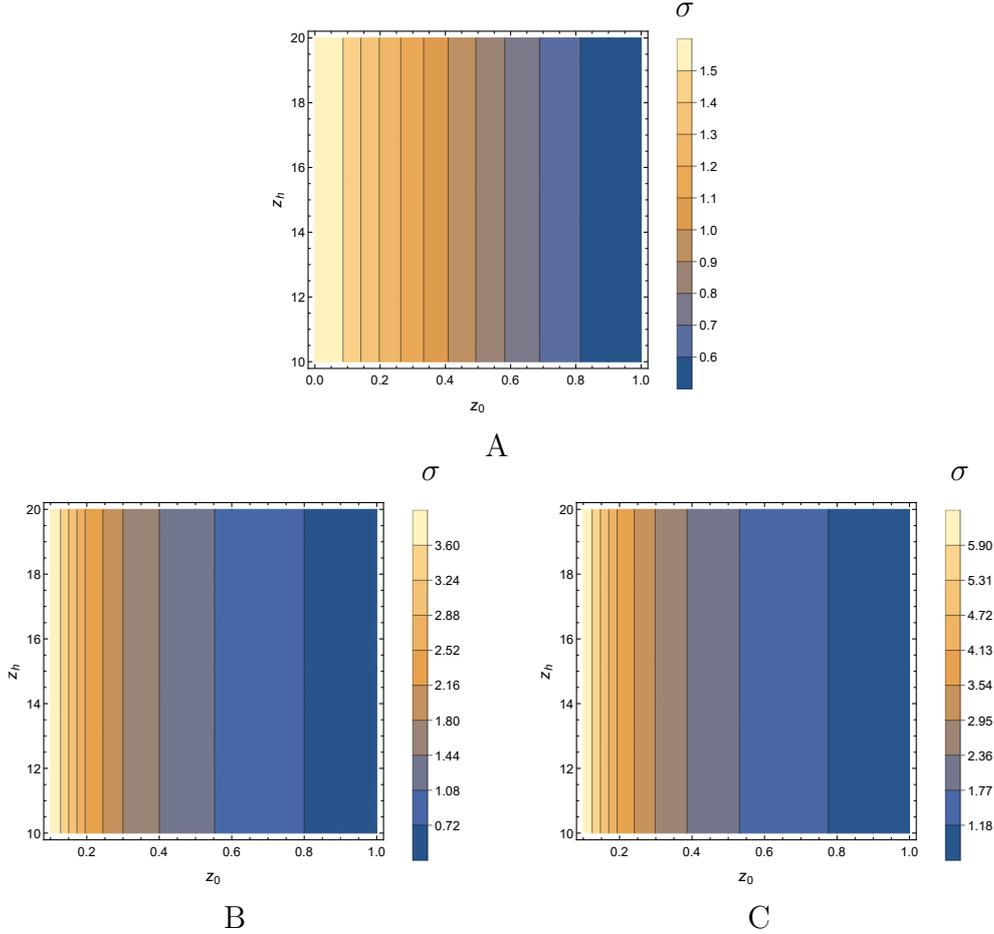

\begin{figure}[t!]
  \centering
  \includegraphics[scale=0.35]{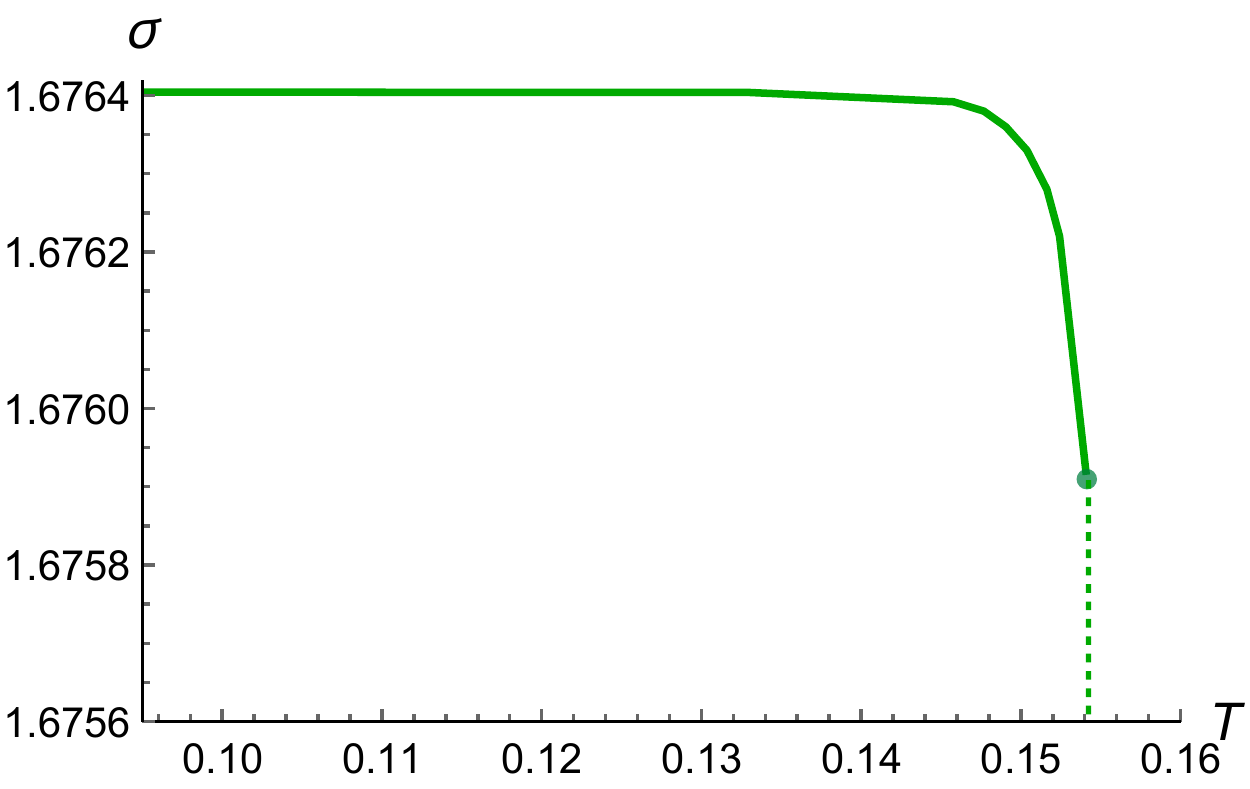}\qquad
  \includegraphics[scale=0.35]{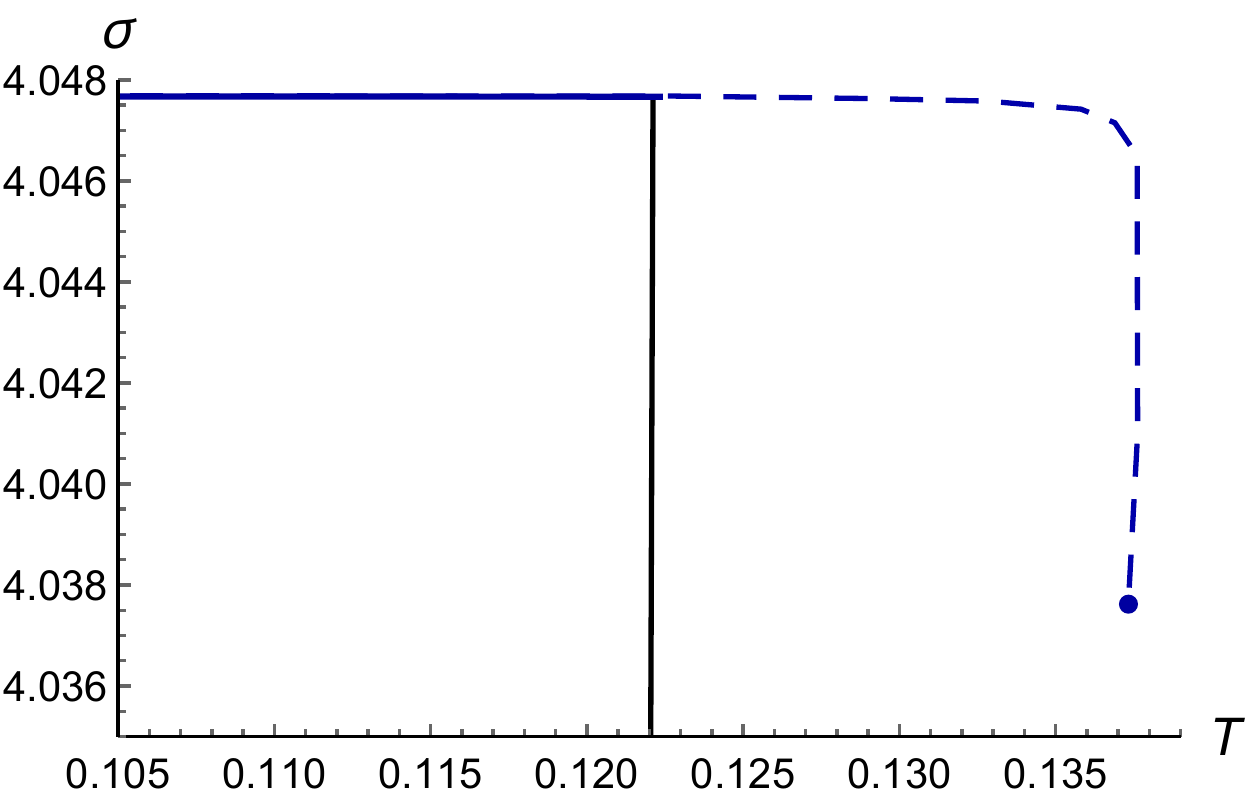}\qquad
  \includegraphics[scale=0.35]{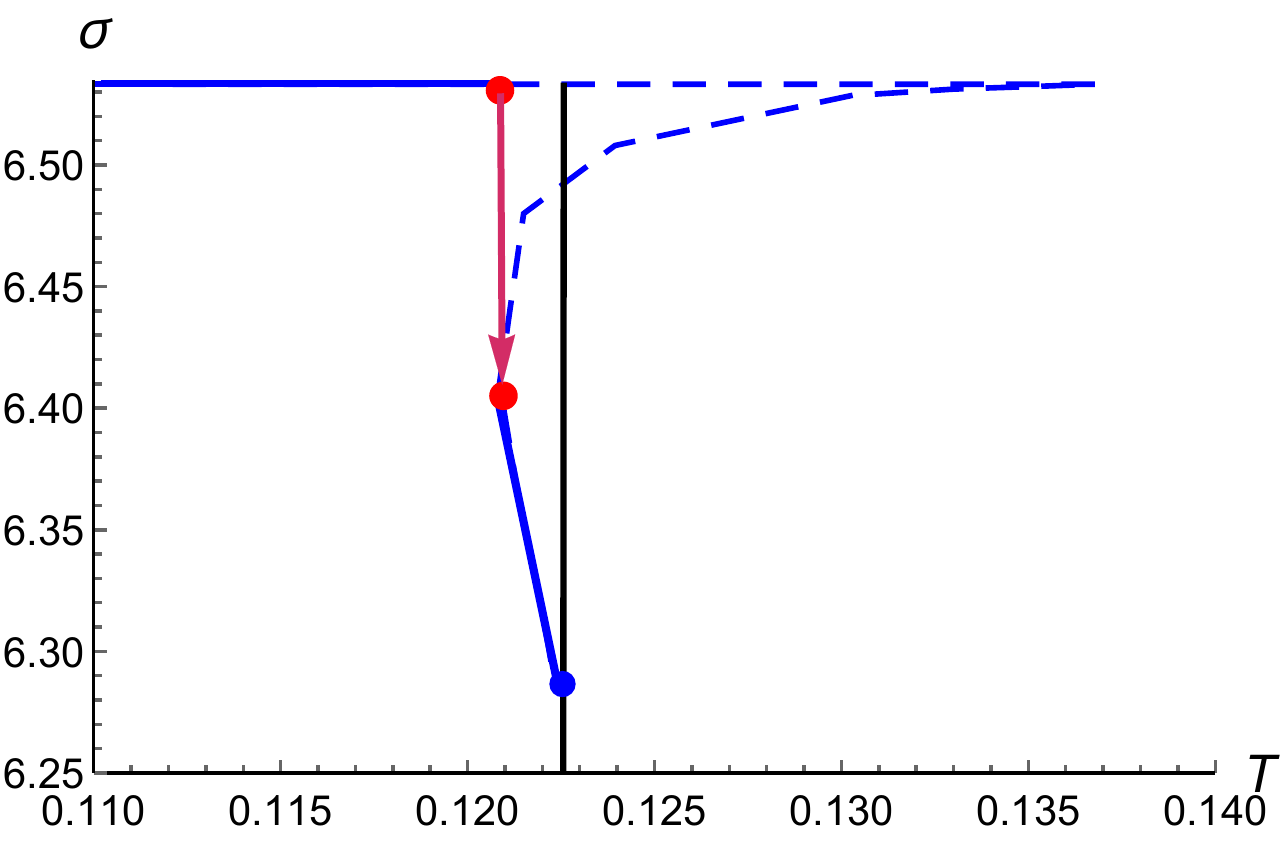} \\
  A \hspace{125pt} B \hspace{125pt} C
  \caption{String tension as function of temperature for $z_{0} = 0$,
    $\mu = 0$, $\nu = 1$ (A) and $z_{0} = 0.1$, $\mu = 0$, $\nu =
    4.5$, longitudinal (B) and transversal (C); $a = 4.046$, $b =
    0.01613$, $c = 0.227$. Solid lines -- the realized values of
    string tension; dotted green line (A) -- the WL phase transition;
    black solid lines (B,C) --- BH-BH phase transition; dashed blue
    lines -- string tension for temperature higher than the
    temperature of BH-BH phase transition, $T > T_{BH-BH}$. In the
    transversal (C) case the string tension is three-valued function
    and the string tension has phase transition that happens for less
    temperatures than the temperature of BH-BH phase transition. Red
    arrow shows jumps of the string tension.}
  \label{sigmaT}
\end{figure}

String tension dependence on the boundary point $z_0$ and horizon
$z_h$ for isotropic case (Fig.\ref{Fig:sigmazz0}.A) and both
orientations of anisotropic case (Fig.\ref{Fig:sigmazz0}.B,C) is
presented. The behavior of $\sigma(T)$ is similar for $\nu = 1$ and
$\nu \ne 1$. Adding chemical potential doesn't change the main picture
as well. For fixed $z_h$ value larger $z_0$ leads to lesser
$\sigma$. However, the string tension for fixed $z_0$ depends on $z_h$
weakly in all cases (Fig.~\ref{sigmaT}). Due to the boundary condition
$z_0 = const < z_h$ the string tension slowly decreased with
temperature till the very end, where it drops sharply to zero. This
behavior persists for any $\nu$, $\mu$ and $\theta$ values.

\begin{figure}[t!]
  \centering
  \includegraphics[scale=0.4]{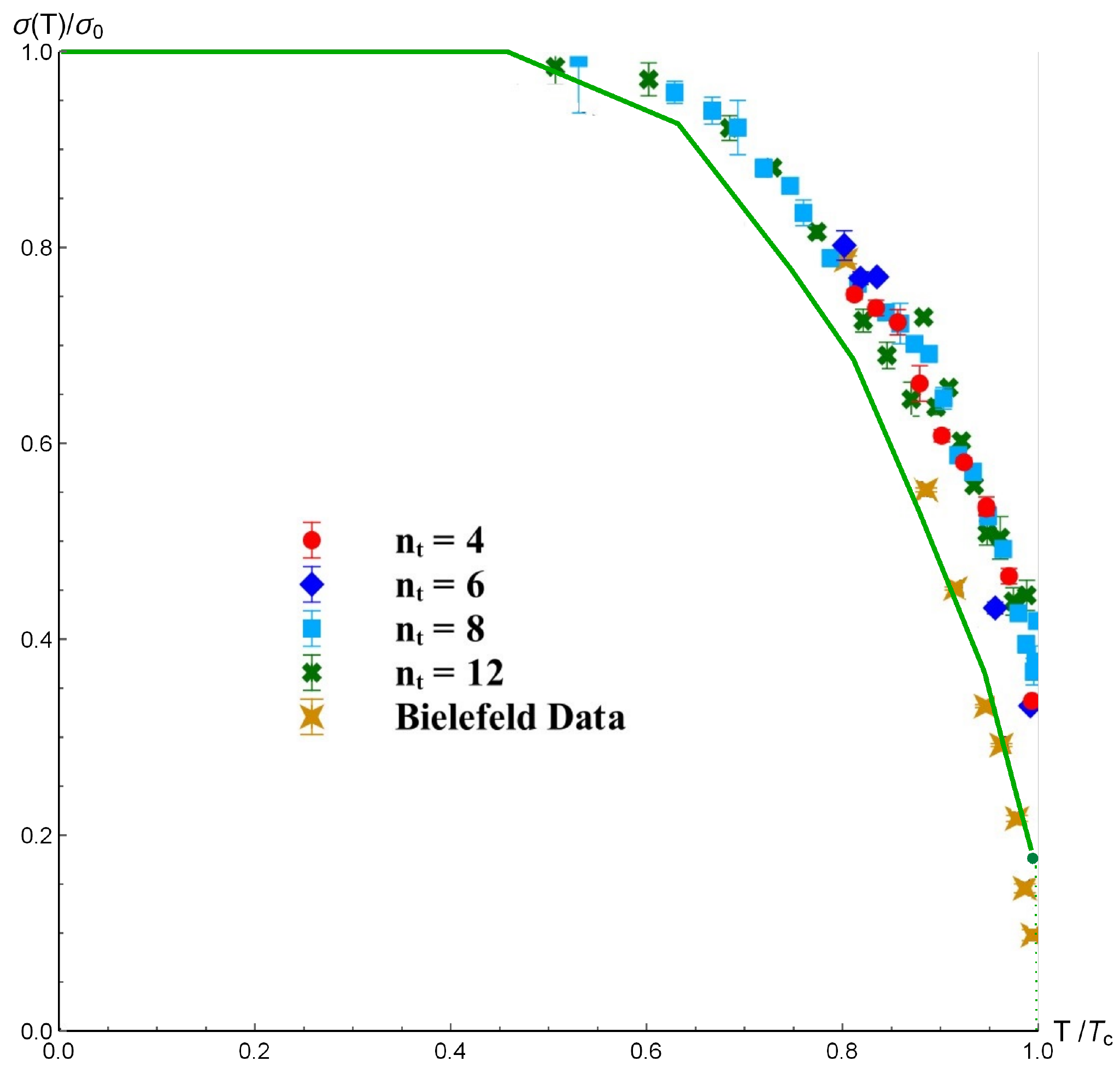}
   \caption{The green curve shows the string tension as function of
     temperature for $z_{0} = 10 \exp(-z_h/4)+0.1$, $\mu = 0$, $\nu =
     1$, $a = 4.046$, $b = 0.01613$, $c = 0.227$. The dots with
     different decorations show results of lattice calculations
     obtained in \cite{Cardoso:2011hh}.}
  \label{sigma}
\end{figure}
\begin{figure}[t!]
  \centering
  \includegraphics[scale=0.45]{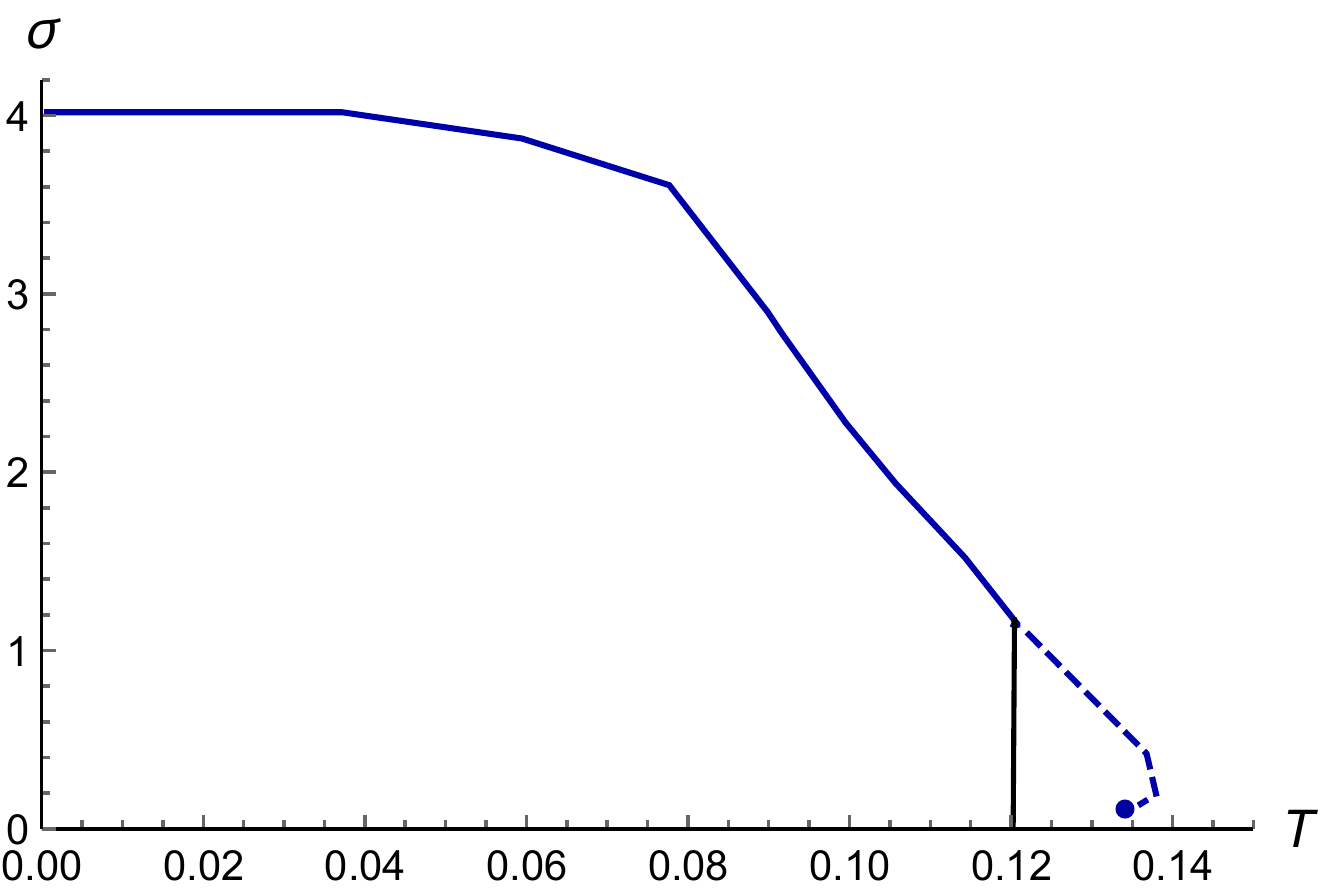}\qquad\qquad
  \includegraphics[scale=0.45]{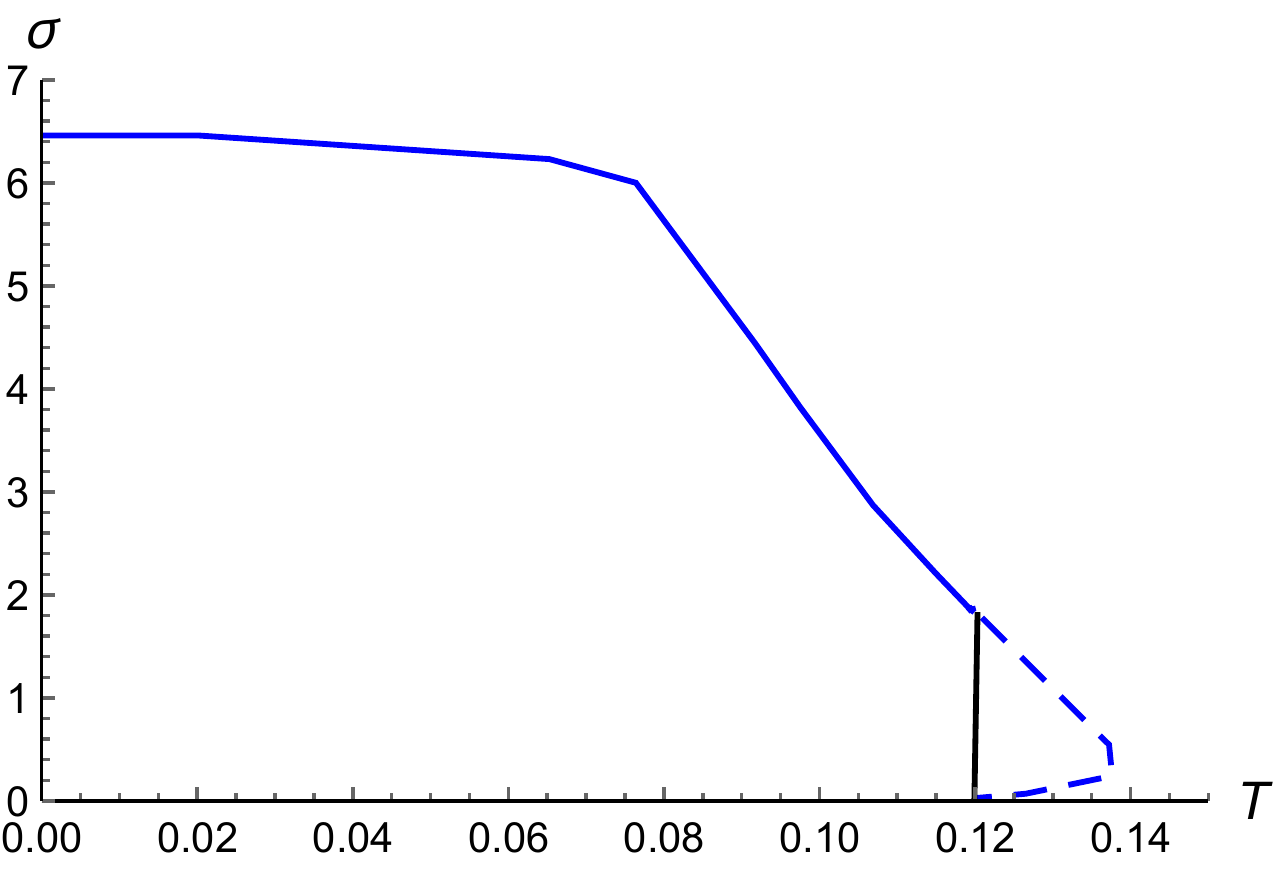} \\
  A \hspace{200pt} B 
  \caption{String tension as function of temperature for $z_{0} = 10
    \exp(-z_h/4)+0.1$,   $\mu = 0.2$, $\nu
    = 4.5$, longitudinal (A) and transversal (B); $a = 4.046$, $b =
    0.01613$, $c = 0.227$. Solid lines -- the realized values of
    string tension; dotted line (A) -- the WL phase transition; black
    solid lines (B,C) --- BH-BH phase transition; dashed blue lines --
    string tension for temperature higher than the temperature of
    BH-BH phase transition, $T > T_{BH-BH}$.}
  \label{sigmaTcond}
\end{figure}

The string tension behavior in Lattice QCD was discussed in
\cite{Laermann:2003cv, Digal:2005ht, Cardoso:2011hh, Bicudo:2010hg,
  J.P.:2020xmq}. It was shown that $\sigma(T)$ is a decreasing
function, but different factors  can influence the particular form of
the curve. Therefore fitting the experimental data could help to
specify the model's parameters. In particular, it is interesting to
consider the boundary $z_0$ as a function $z_0 = f(z_h, z_{DW},
\dots)$ and use it to fit the lattice results.

To get the best matching of isotropic $\sigma(T)$  with   the lattice
calculations \cite{Cardoso:2011hh, Bicudo:2010hg} we use the function 
\be
  z_{0} = 10 \exp(-z_h/4)+0.1. \label{z0}
\ee
In this case the string tension decreases significantly faster than
for $z_{0} = 0$ (Fig.\ref{sigma}). For anisotropic cases, where we do
not have lattice data, we use  the same function \eqref{z0} and  get
the temperature dependences presented in Fig.\ref{sigmaTcond}.A and B.

\begin{figure}[t!]
  \centering
  \includegraphics[scale=0.35]{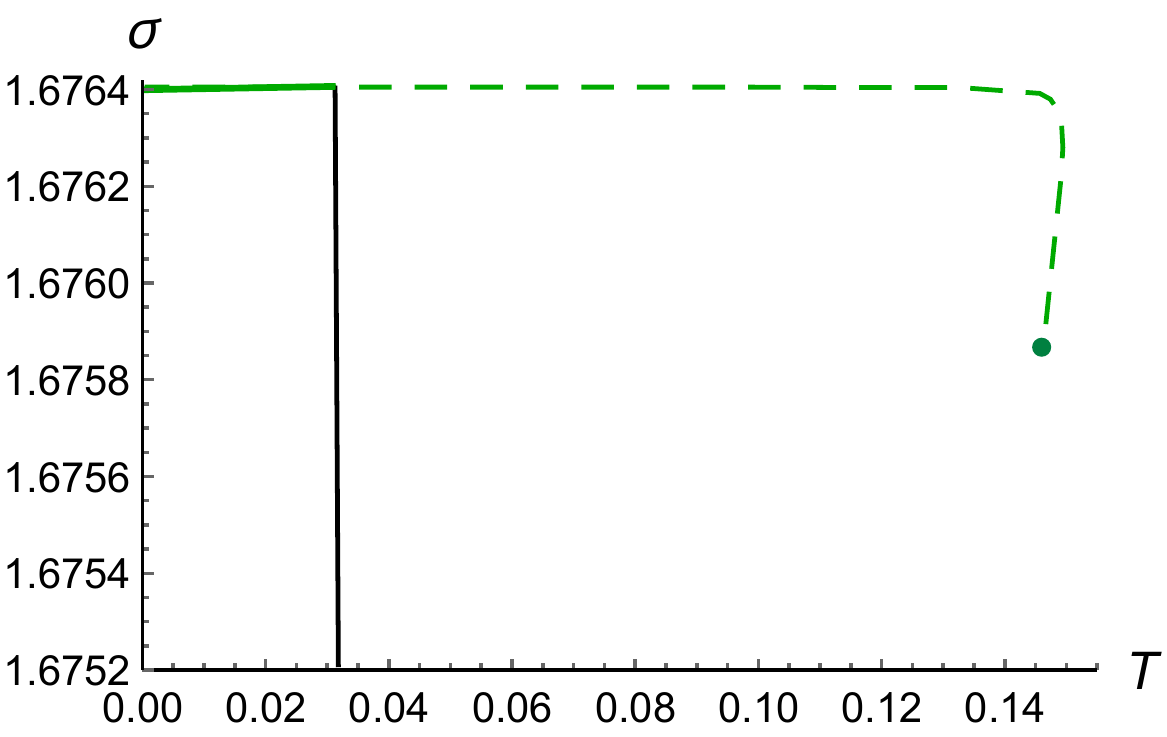}\qquad
  \includegraphics[scale=0.32]{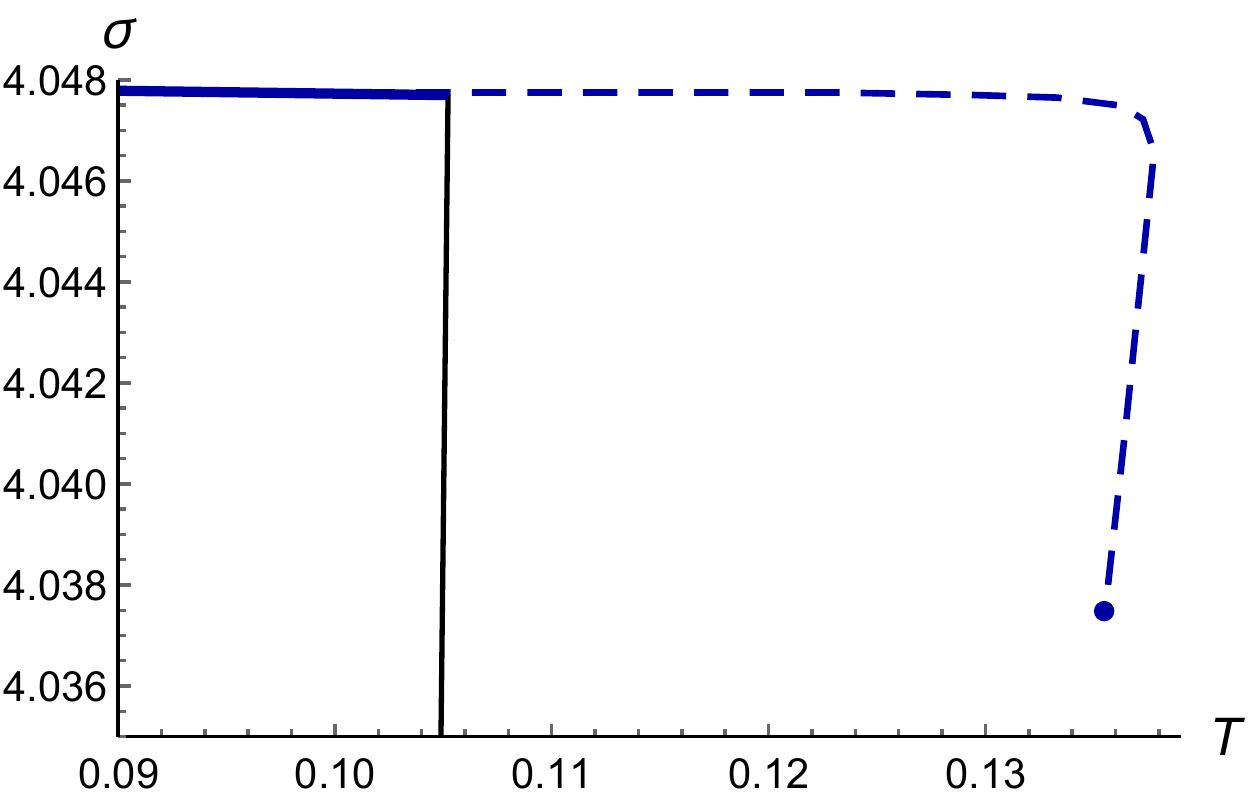}\qquad
  \includegraphics[scale=0.32]{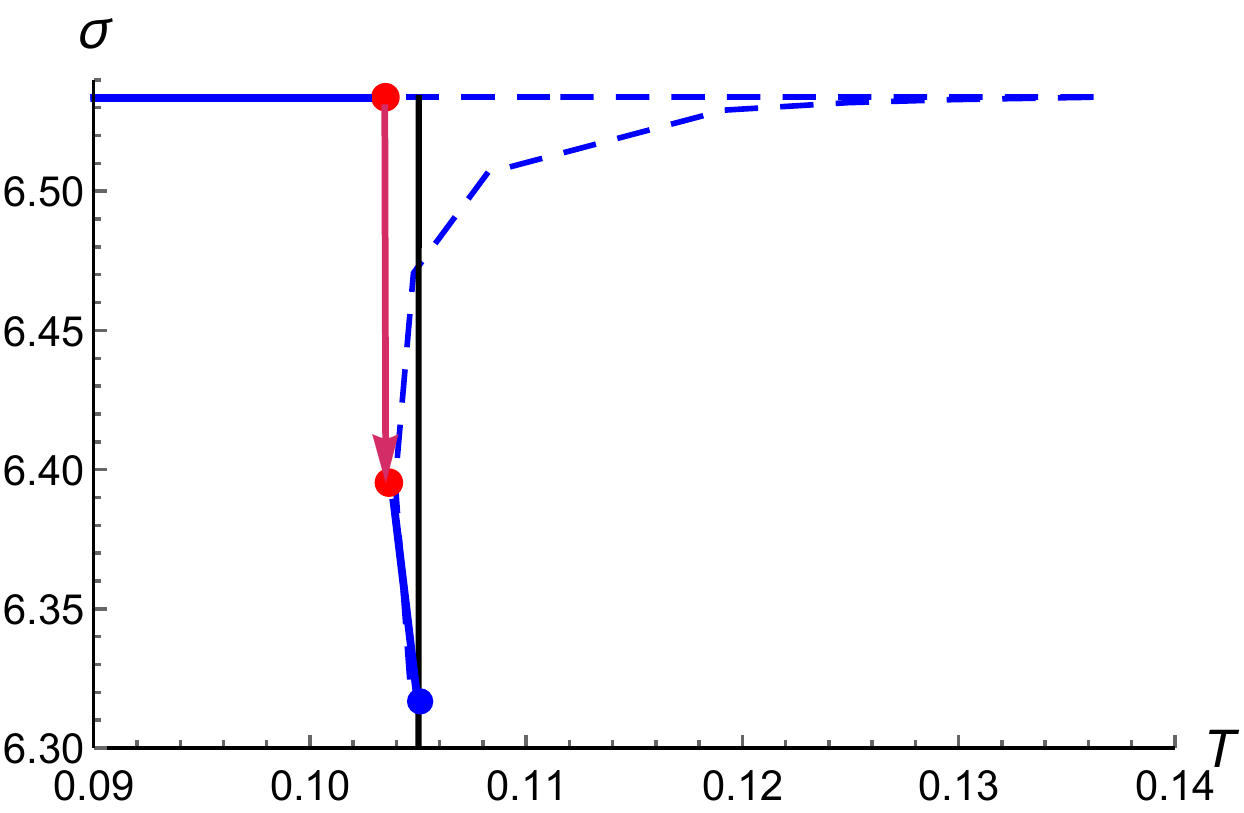} \\
  A \hspace{125pt} B \hspace{125pt} C
  \caption{String tension as function of temperature for $z_{0} = 0$,
    $\mu = 0.5$, $\nu = 1$ (A) and $z_{0} = 0.1$, $\mu = 0.5$, $\nu =
    4.5$, longitudinal (B) and traversal (C); $a = 4.046$, $b =
    0.01613$, $c = 0.227$. Solid lines -- the realized values of
    string tension; dotted line (A) -- the WL phase transition; black
    solid lines (B,C) --- BH-BH phase transition; dashed blue lines --
    string tension for temperature higher than the temperature of
    BH-BH phase transition, $T > T_{BH-BH}$, i.e.  string tensions in
    unstable phase. In the transversal (C) case the string tension is
    three-valued function and the string tension has phase transition
    that happens for less temperatures than the temperature of BH-BH
    phase transition. Red arrow shows jumps of the string tension.}
  \label{sigmaTmu05}
\end{figure}

On Fig.\ref{sigmaTcond}--\ref{sigmaTmu05} string tension as function
of temperature for different boundary limits $z_0$ and different
chemical potential $\mu$ values are presented. Dotted lines on A-plots
indicate the WL phase transition, when the connected string
configuration changes to the disconnected one and the string tension
drops to zero. Black solid lines on B (longitudinal orientation) and C
(transversal orientation) plots show the BH-BH phase transition, when
the connected string configuration in the first thermodynamic phase
changes to disconnected string configuration in the second
thermodynamic phase; dashed blue lines depict the string tension
values for the temperatures higher than the temperature of the BH-BH
phase transition. In anisotropic cases $\nu \ne 1$ (B,~C) string
tension $\sigma(T)$ can be multi-valued function for some values of
temperature.


In all plots of Fig.\ref{sigmaT} and Fig.\ref{sigmaTmu05} $\sigma(T)$
can be multi-valued function due to the multi-valued function of
temperature $T$ on the size of horizon $z_h$. To understand the
behavior of the string tension on temperature we should use the
knowledge about the BH-BH phase transitions structure.

In the considered case we have two thermodynamic phases -- small black
holes and large black holes -- and a phase transition between
them. The end point of $\sigma(T)$ in all plots is indicated by blue
or green dot. After this point dynamical wall (DW) for the effective
potential does not exist anymore and the connected string
configuration disappears, that indicates the WL phase transition.

On Fig.\ref{sigmaT}, \ref{sigmaTmu05} solid black (vertical) lines show
the BH-BH phase transition for the chosen set of parameters and
orientation. For isotropic case (Fig.\ref{sigmaTmu05}.A) and
longitudinal orientation of anisotropic case (Fig.\ref{sigmaT}.B,
Fig.\ref{sigmaTmu05}.B) multi-valued (dashed) branch lies rather far
from the BH-BH transition and does not interfere with it. Transversal
orientation has a specific feature: multi-valued branch intersects the
BH-BH transition line and thus starts to play significant role in the
general phase transition process. The string tension has a phase
transition at lower temperatures than the temperature of the BH-BH
phase transition. At this temperature the connected string
configuration with first string tension value  $\sigma_1$ in first
thermodynamic phase is changed by  the connected string configuration
with second string tension value $\sigma_2$ in second thermodynamic
phase, $\sigma_1 > \sigma_2$. Red arrows on Fig.\ref{sigmaT}.C and
\ref{sigmaTmu05}.C show these jumps of the string tension. We also see
that non-zero chemical potential suppresses this effect, making the
interval where the lowest part of the multi-valued $\sigma(T)$-branch
is realized, narrower (compare Fig.\ref{sigmaT}.C for $\mu = 0$ and
Fig.\ref{sigmaTmu05}.C for $\mu = 0.5$).

To get full picture we need to consider lines corresponding to Wilson
loops and depending on quarks pair orientation. As the current model
differs from the previous one by the form of the warp factor only, all
the reasoning in \cite{AR-2018, ARS-2019plb} remains applicable
here. Therefore the dynamical wall equations become:
\begin{gather}
  - \ \cfrac{4 a b z}{1 + b z^2} + \sqrt{\cfrac{2}{3}} \ \phi'
  + \cfrac{g'}{2g} = \cfrac{2}{z} \, , \label{eq:2.06} \\
  - \ \cfrac{4 a b z}{1 + b z^2} + \sqrt{\cfrac{2}{3}} \ \phi'
  + \cfrac{g'}{2g} = \cfrac{\nu + 1}{\nu z} \, \label{eq:2.07}
\end{gather}
for longitudinal ($x$) and transversal ($y$) direction
correspondingly.

\begin{figure}[b!]
  \centering
  \includegraphics[scale=0.55]{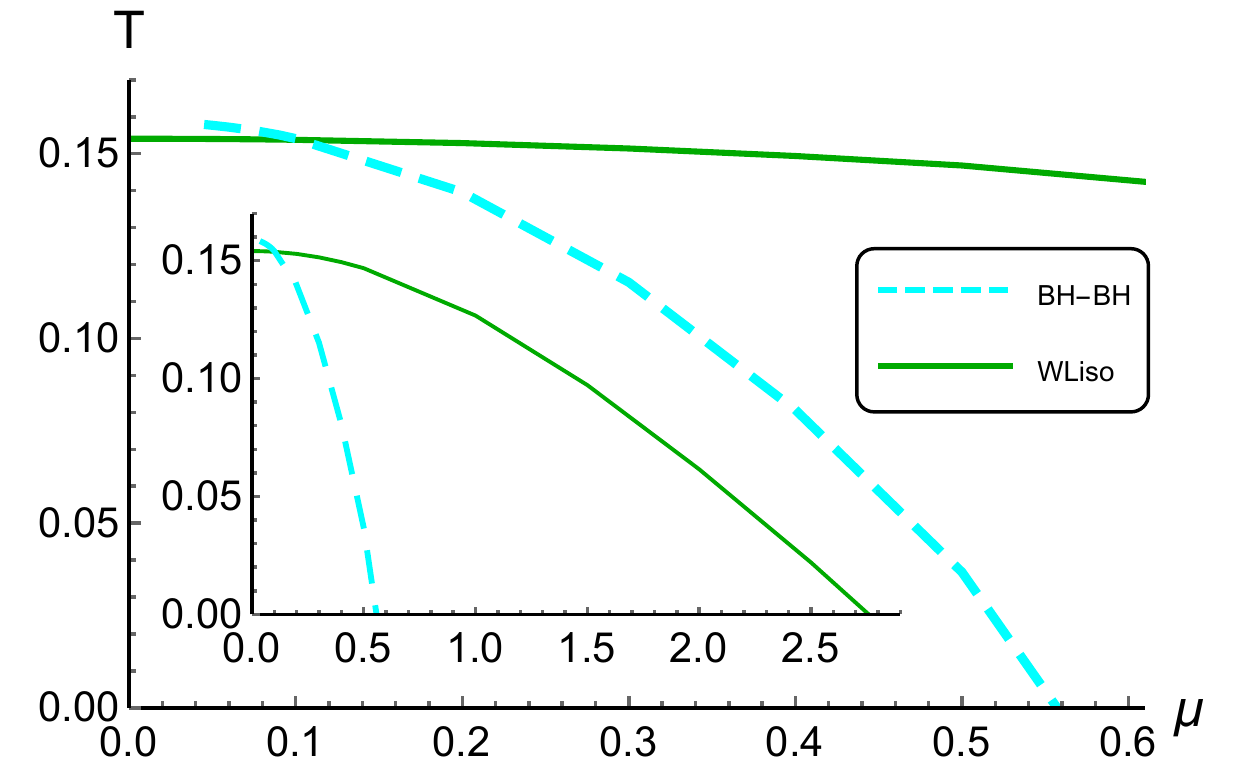} \qquad
  \includegraphics[scale=0.55]{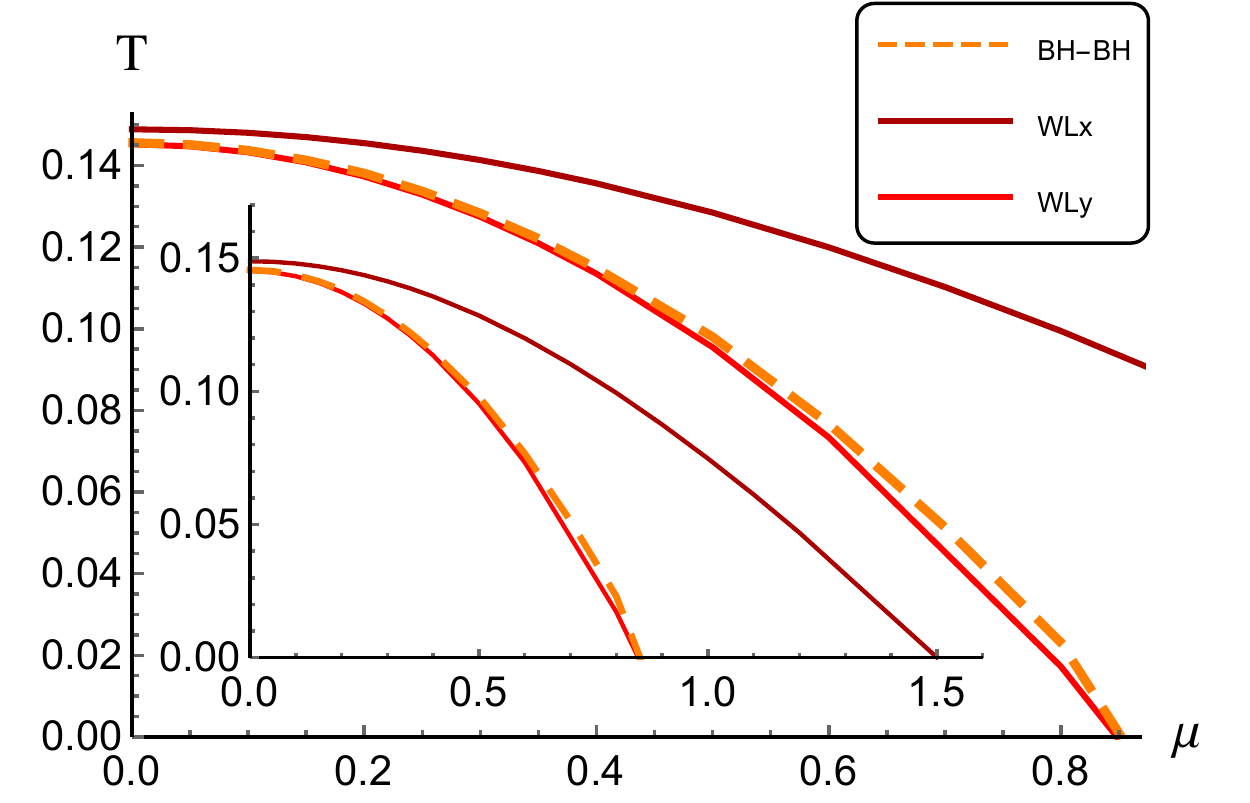} \\
  A \hspace{220pt} B \\
  \includegraphics[scale=0.55]{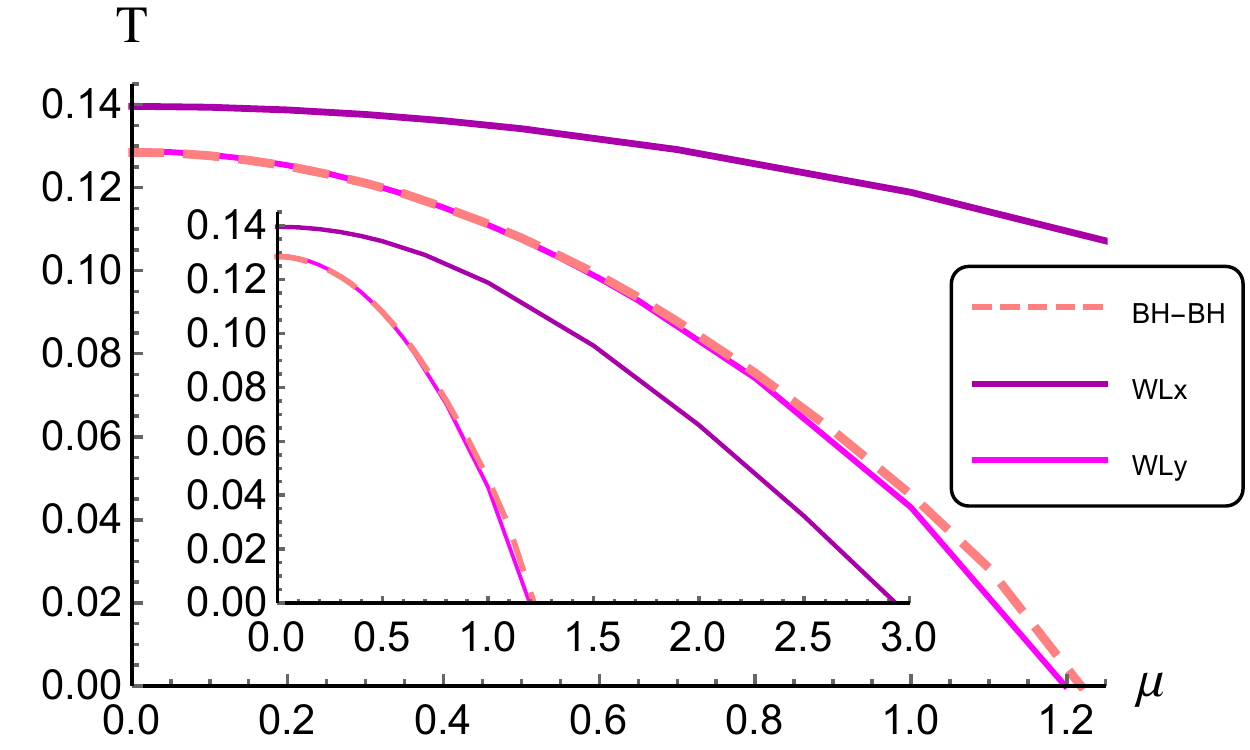} \qquad
  \includegraphics[scale=0.55]{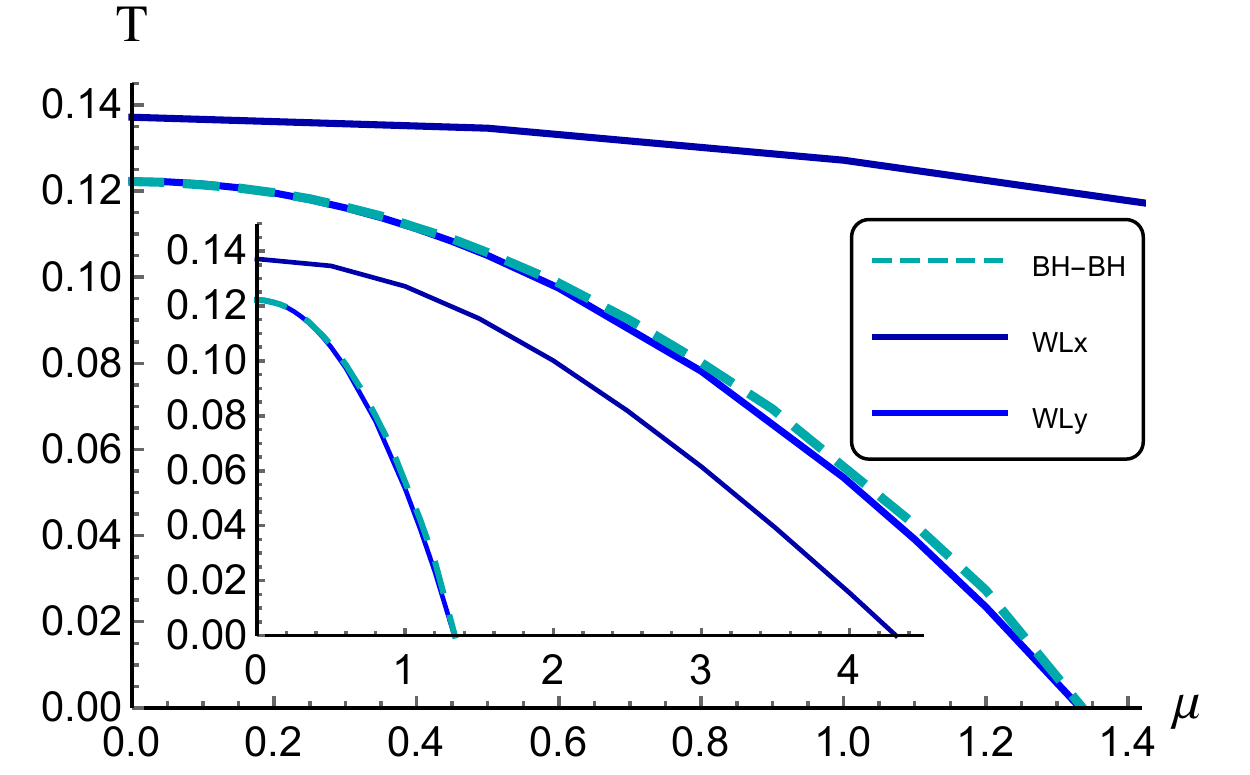} \\
  C \hspace{220pt} D
  \caption{Confinement/deconfinement phase diagram $T(\mu)$ in
    isotropic (A) and anisotropic cases for $\nu = 1.5$ (B), $\nu = 3$
    (C), $\nu = 4.5$ (D); $a = 4.046$, $b = 0.01613$, $c = 0.227$. Dashed lines show  Hawking-Page-like phase transitions (BH-BH).}
  \label{Fig:CDPD}
\end{figure}

The resulting phase transition lines, determined by the Wilson loops
(along with the Hawking-Page-like phase transition lines) for the
isotropic and anisotropic cases are showed on
Fig.~\ref{Fig:CDPD}.

On Fig.~\ref{Fig:CDPD}.A the isotropic case is depicted. The
confinement/deconfinement phase transition is mostly determined by the
Hawking-Page-like transition (BH-BH transition). Wilson loop is
sufficient in a small region of crossover for $0 <
  \mu < 0.104$, i.e. till the point $(0.104, 0.153)$, where two phase
transition lines intersect.

In the anisotropic case the isotropic Wilson line evolves into the
line corresponding to longitudinal Wilson loop. Starting from $\mu =
0$ the longutudinal Wilson line lies above the Hawking-Page-like and
doesn't actually influence the phase transition. For larger anisotropy
longitudinal Wilson line has lower temperature values, but the
difference between it and the Hawking-Page-like line increases with
$\nu$ (Fig.~\ref{Fig:CDPD}.A-D). Phase transition line corresponding
to the transversal Wilson loop almost coincides with the
Hawking-Page-like line scaled, so there is no evident crossover region
as it was in anisotropic case. Therefore influence of the transversal
Wilson line and the Hawking-Page-like line on the
confinement-deconfinement phase transition could be hardly 
distinguished from each other.


\section{Conclusion}

We have considered the anisotropic holographic model for light quarks. This model
is  invariant in the transversal directions with the unique anisotropy
scaling factor supported by the Einstein-Dilaton-two-Maxwell
action. The analogous model for heavy quarks was presented in
\cite{AR-2018}. 
We have found characteristic features inherent in the description of
light quarks within the holographic approach. Thermodynamical peculiar properties
and their influence on the confinement/deconfinement phase diagram
are considered.

\begin{figure}[b!]
  \centering
  \includegraphics[scale=0.56]{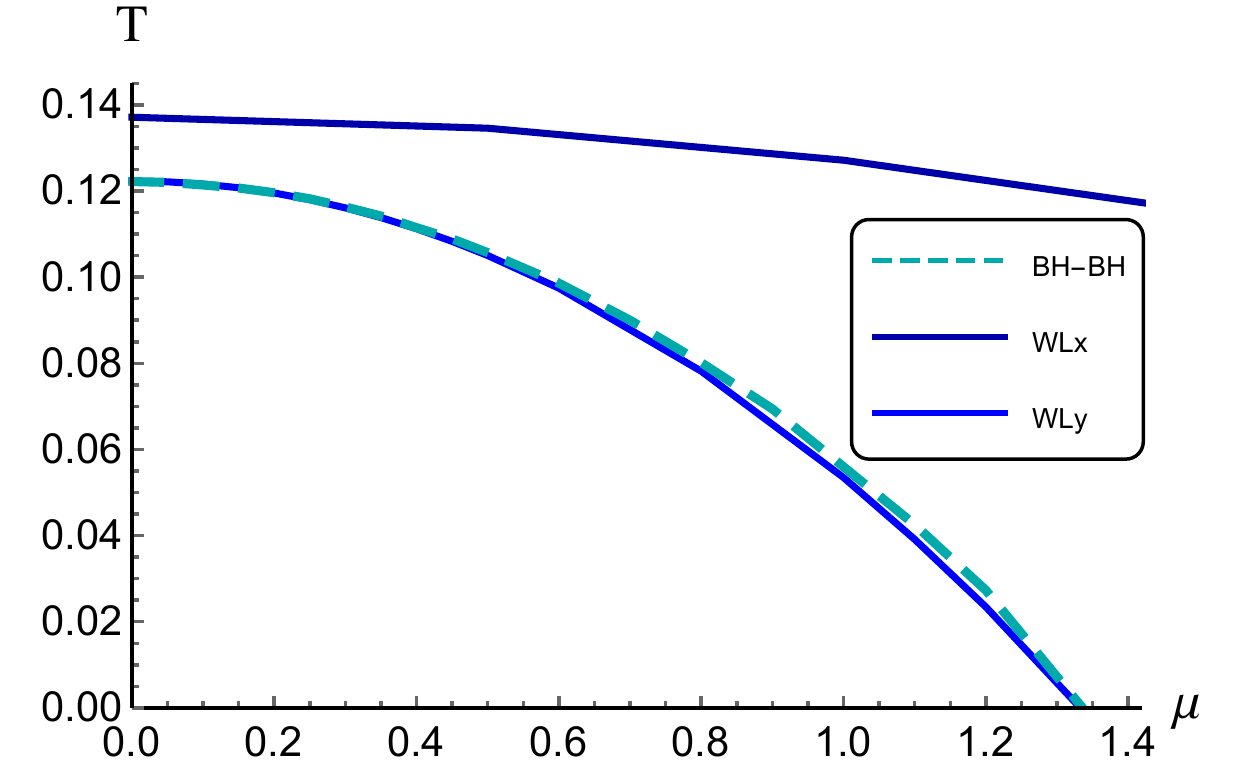} \quad
  \includegraphics[scale=0.56]{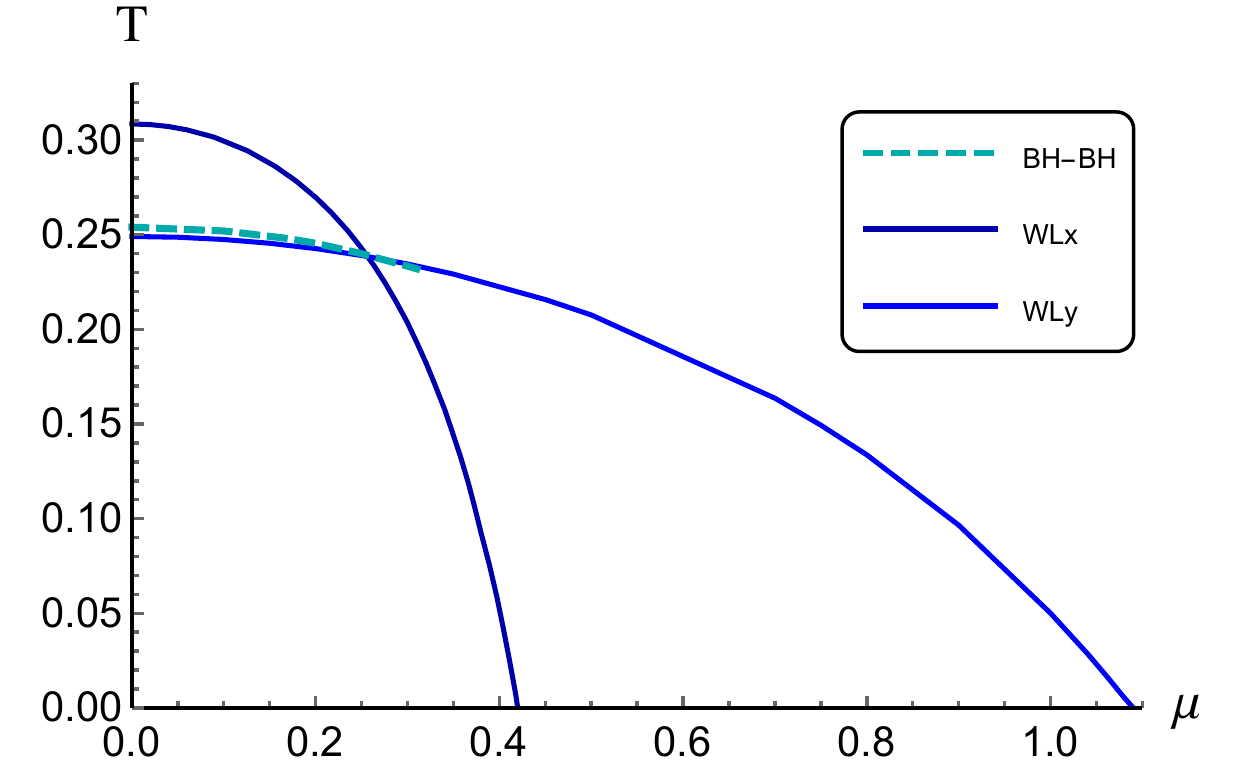} \\
  A \hspace{220pt} B
  \caption{Holographic QCD phase diagrams for light
        quarks (A) and for heavy quarks (B) in the anisotropic case
        \cite{AR-2018}. Here  Hawking-Page-like phase transitions
        (BH-BH) are indicated by dashed lines. Wilson loop phase transitions for different
        orientations (WLx and WLy) are shown by solid lines.}
  \label{Hybrid2} 
\end{figure}

\begin{figure}[h!]
  \centering
  \includegraphics[scale=0.55]{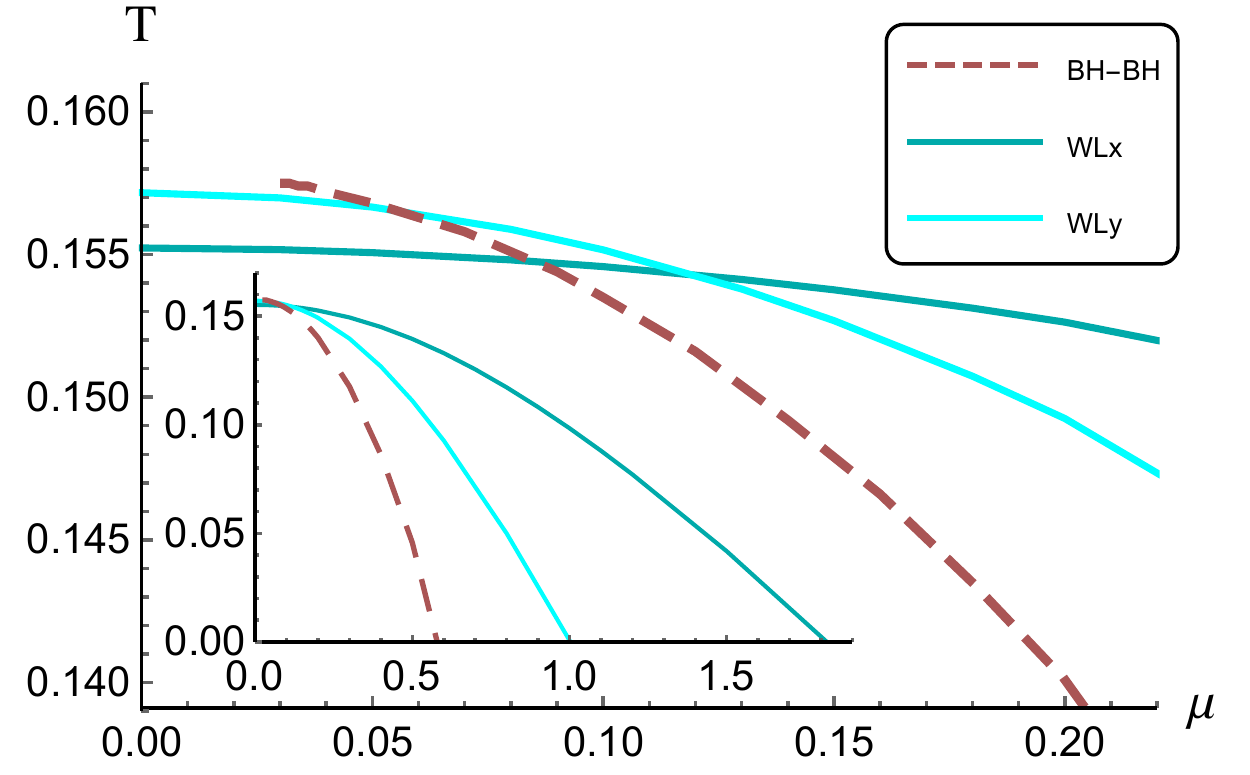} \quad
  \includegraphics[scale=0.55]{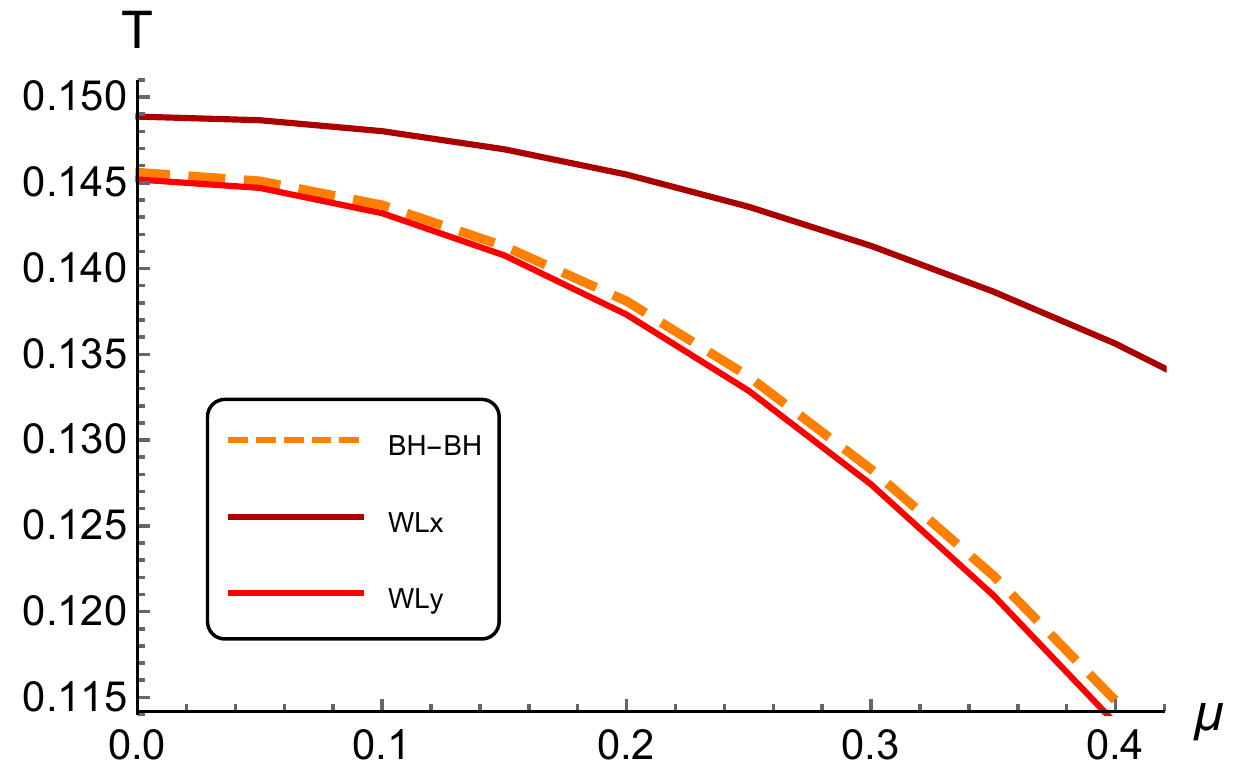} \\
  A \hspace{220pt} B \\
  \includegraphics[scale=0.55]{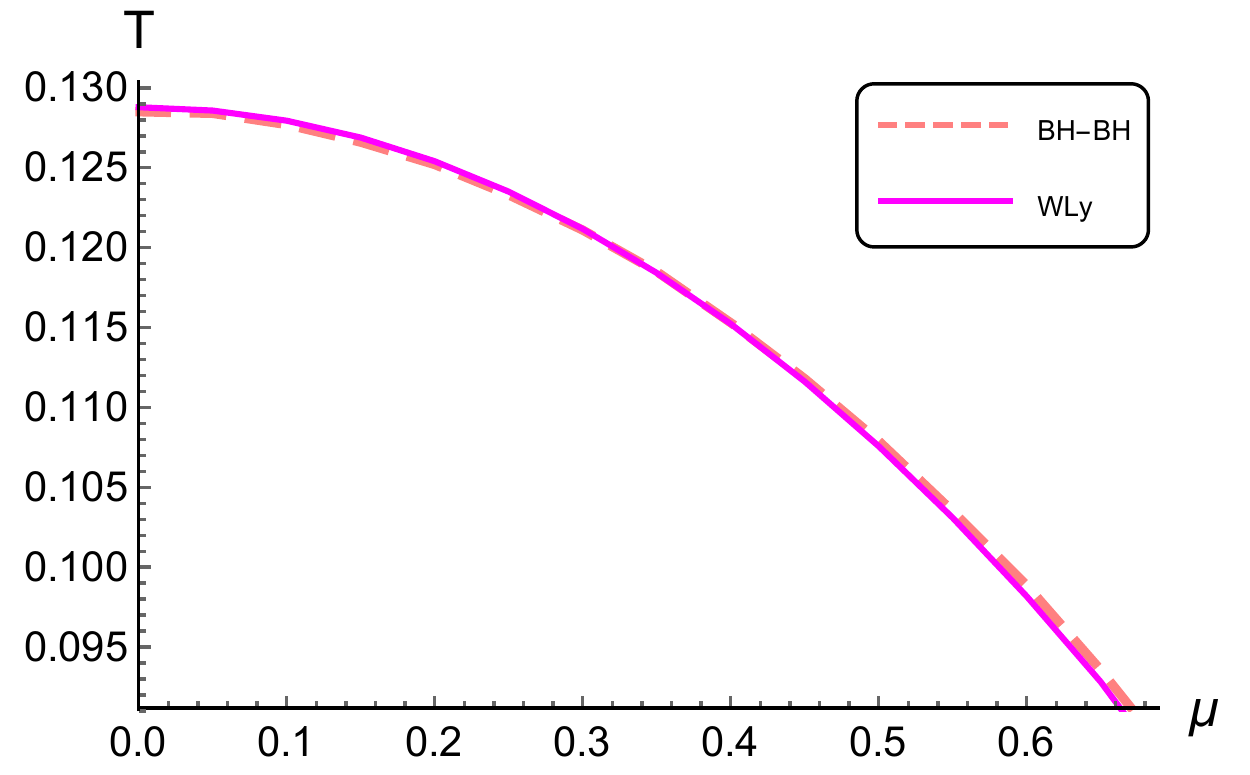} \quad
  \includegraphics[scale=0.55]{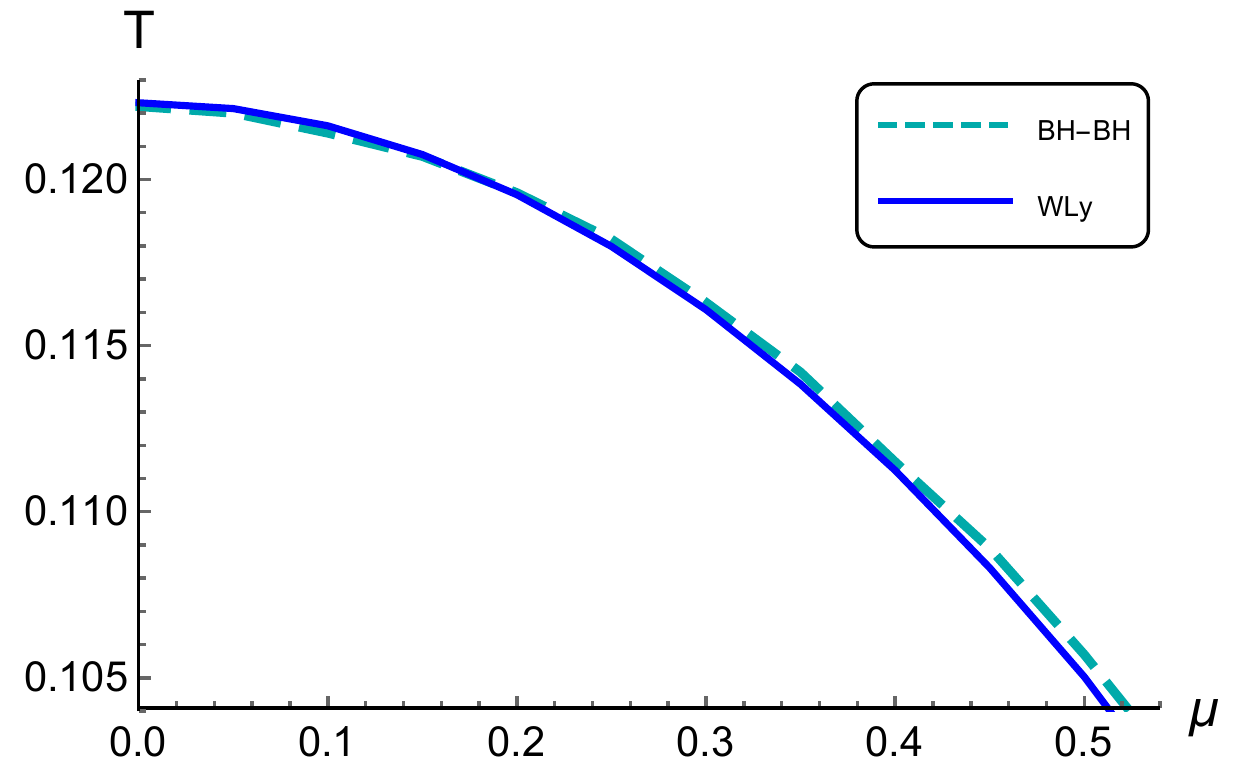} \\
  C \hspace{220pt} D
  \caption{Confinement/deconfinement phase diagram $T(\mu)$ for $\nu =
    1.03$ (A), $\nu = 1.5$ (B), $\nu = 3$ (C), $\nu = 4.5$ (D); $a =
    4.046$, $b = 0.01613$, $c = 0.227$.}
  \label{Fig:CDPDs}
\end{figure}

Unlike the heavy quarks model (Fig.\ref{Hybrid2}.B) \cite{AR-2018},
the Hawking-Page-like phase transition line does not break at a
relatively high temperature, but lasts till $T = 0$
(Fig.\ref{Hybrid2}.A). Also longitudinal orientation of quarks pairs
does not contribute to confinement/deconfinement phase transition, so
the influence is shared by the Hawking-Page-like and the transversal
Wilson loop. Transfer of the main role in the phase transition looks
rather smooth and simple and is not accompanied by jumps as it was in
heavy quarks model \cite{AR-2018}.

Plots on Fig.\ref{Fig:CDPDs} show the features of the phase diagram in
more details. For $\nu = 3$ (Fig.~\ref{Fig:CDPDs}.C) and $\nu = 4.5$
(Fig.~\ref{Fig:CDPDs}.D) the Hawking-Page-like phase transition
dominates for small chemical potentials (till $\mu = 0.3$ and $\mu =
0.2$ correspondingly), then the transversal Wilson loop takes over
control providing narrow strip of the crossover for larger chemical
potentials. However these effects seem to be rather weak, so
practically the crossover region is unlikely to be registered for the
light quarks confinement/deconfinement phase transition. For $\nu =
1.5$ the transversal Wilson line lies right under the
Hawking-Page-like line, so, strictly speaking, it is the transversal 
component that determines the confinement-deconfinement phase
transition (Fig.~\ref{Fig:CDPDs}.B). On the other hand the
Hawking-Page-like line is located too close to make the crossover
tangible.

Let us note that some interesting features appear for light quarks
model for small anisotropy already. For example weak anisotropy with
$\nu = 1.03$ was considered (Fig.~\ref{Fig:CDPDs}.A). Generally the
phase transition picture for such a slight anisotropy is the same as
in the isotropic case. When the anisostropy is turned on the crossover
region narrows. For $\nu = 1.03$ it end at the point $(0.085;
0.155)$. Actually the length of the crossover seems to be the most
essential manifestation of anisotropy. It's width shouldn't be large
as longitudinal and transversal Wilson lines are rather close to each
other and for $\mu > 0.073$ both of them lie above the
Hawking-Page-like curve and do not affect the
confinement/deconfinement phase transition.

Within the change of the boundary conditions of the dilaton field the
form of the string tension dependence on temperature $\sigma(T)$ in
isotropic case can be qualitatively fit by the lattice results
(Fig.\ref{sigmaTcond}.A). Keeping the same boundary condition for the
anisotropic case we obtain realistic string tension behavior. In both
models for heavy and light quarks $\sigma(T)$ can be a multi-valued
function for high temperatures. The appearance of the multi-valued
$\sigma(T)$ is the consequence the multi-valued temperature $T$ on the
size of horizon $z_h$ for some set of parameters. The end point of
$\sigma(T)$ is interpreted as a point of the phase transition
associated with the Wilson loop (i. e. $\sigma$ undergoes a jump to
zero and the connected configuration is replaced by the disconnected
one). Also $\sigma$ undergoes a jump to zero due to the BH-BH phase
transition, when the phase transition happens from first thermodynamic
phase with connected string configuration to second thermodynamic
phase with disconnected string configuration. As it was shown on
Fig.\ref{sigmaT}.C and Fig.\ref{sigmaTmu05}.C, the string tension
$\sigma$ has the phase transition -- a jump of the string tension
value $\sigma_1$ to the string tension value $\sigma_2$, $\sigma_1 >
\sigma_2$. This transition happens for lower temperatures than the
BH-BH phase transition temperature and can be interpreted as the
transition between quarkyonic and hadronic phases of QGP.

In both models (for heavy and light quarks) $\sigma(T)$ can be a
multi-valued function for high temperatures. The appearance of
multi-valued behavior $\sigma(T)$ is interpreted as a transition
associated with the Wilson loop (i. e. $\sigma$ undergoes a jump to
zero and the connected configuration is replaced by the disconnected
one). For weak anisotropization ($\nu = \ 1.01 \div 1.05$), the BH-BH
transition competes with the transition for the Wilson loop and for
small chemical potentials ($\mu = \ 0 \div 0.1$) the transition for
the Wilson loop is dominant. The result plots  of the
confinement/deconfinement phase transition on $(\mu,T)$-plane for
light and heavy  quarks' mass in anisotropic media are displayed in
Fig.~\ref{Hybrid2} and  Fig.~\ref{Fig:CDPDs}. We can see that the
phase transitions structure is more complex in the anisotropic case
than is isotropic one (Fig.~\ref{Hybrid2} and  Fig.~\ref{Fig:CDPDs}
can be compared with Fig.~\ref{Hybrid1}).

Let us remind that the choice of the model \cite{AG}, that is a
starting point of our consideration of the anisotropic models, was
motivated by agreement of the energy dependence of the produced
entropy with the experimental data for the energy dependence of the
total multiplicity of particles produced in HIC \cite{ALICE,ATLAS}. It
would be interesting to study the change in the produced entropy under
deformations of the anisotropic model \cite {AG} with which we are
dealing in this article. Also it would be interesting to study
modification of the entanglement entropy, compare with
\cite{Dudal:2018ztm,APS}.


This leads us to the next step for obtaining more realistic model --
to investigate some kind of a mixed model, where both heavy and light
quarks would be included. Study of such a mix should be rather
instructive for better understanding of confinement/deconfinement
phase transition and futher interpretations of experimental data. This
model should also be fully anisotropic because of presence of external
magnetic field. Analagous consideration inspired by
\cite{Gursoy:2018ydr} was already perfomed for heavy quarks only in
\cite{2011.07023}, see also \cite{Gursoy:2020}.

The holographic entanglement entropy (HEE) can be related to the phase
transitions in quark matter, therefore it is interesting to calculate
the HEE for the considered light quarks anisotropic model and compare
the results with \cite{APS, Arefeva:2019dvl, Slepov:2019guc}. Also
drag forces and tensions for spatial Wilson loops can be compared
following \cite{IADrag}. Main considerations for heavy quarks can be
found in \cite{2012.05758}.

We hope that the results presented in this paper and their further
possible adjustment to the phenomenological data can be of interest
for experiments at the future facilities of FAIR, NICA, for RHIC's BES
II program and CERN, III run.


\section*{Acknowledgments}

This work is supported by RFBR Grant 18-02-40069 and partially
(I.A. and P.S.) by the ``BASIS'' Science Foundation grant
No. 18-1-1-80-4.

\end{document}